\documentclass[%
 twocolumn,
 10pt,
 superscriptaddress,
 longbibliography,
 amsmath,amssymb,
 aps,
 pra
]{revtex4-1}

\usepackage{graphicx,amsthm}% Include figure files
\usepackage{bm}% bold math
\usepackage{times}
\usepackage[caption=false]{subfig}
\usepackage{epstopdf}
\usepackage{mathtools,enumerate}

\usepackage[colorlinks=true,linkcolor=blue,citecolor=red, linktocpage=true,breaklinks=true]{hyperref}

\newcommand{\eq}{\begin{equation}}
\newcommand{\en}{\end{equation}}
\newcommand{\eqa}{\begin{eqnarray}}
\newcommand{\ena}{\end{eqnarray}}
\newcommand{\re}{\mathrm{Re}}
\newcommand{\tr}{\mathrm{Tr}}
\newcommand{\im}{\mathrm{Im}}

\allowdisplaybreaks[1]
\begin{document}

\title{Influence of equilibrium and nonequilibrium environments on macroscopic realism through the Leggett-Garg inequalities}% Force line breaks with \\

\author{Kun \surname{Zhang}}
%\email{Email: kun.h.zhang@stonybrook.edu}
\affiliation{Department of Chemistry, State University of New York at Stony Brook, Stony Brook, New York 11794, USA}
\author{Wei \surname{Wu}}
\affiliation{State Key Laboratory of Electroanalytical Chemistry, Changchun Institute of Applied Chemistry, Chinese Academy of Sciences, Changchun 130022, China}
\author{Jin \surname{Wang}}
\email{Email: jin.wang.1@stonybrook.edu}
\affiliation{Department of Chemistry, State University of New York at Stony Brook, Stony Brook, New York 11794, USA}
\affiliation{Department of Physics and Astronomy, State University of New York at Stony Brook, Stony Brook, New York 11794, USA}

\date{\today}

\begin{abstract}

We study the macroscopic realism (macrorealism) through the two- and three-time Leggett-Garg inequalities (LGIs) in a two interacting qubits system. The two qubits are coupled either with two bosonic (thermal or photonic) baths or fermionic (electronic) baths. We study both how the equilibrium and nonequilibrium environments influence the LGIs. One way to characterize the nonequilibrium condition is by the temperature difference (for the bosonic bath) or the chemical potential difference (for the fermionic bath). We also study the heat or particle current and the entropy production rate generated by the nonequilibrium environments. Analytical forms of LGIs and the maximal value of LGIs based on the quantum master equation beyond the secular approximation are derived. The LGI functions and the corresponding maximal value have separated contributions, the part describing the coherent evolution and the part describing the coupling between the system and environments. The environment-coupling part can be from the equilibrium environment or the nonequilibrium environment. The nonequilibrium dynamics is quantified by the Bloch-Redfield equation which is beyond the Lindblad form. We found that the nonequilibriumness quantified by the temperature difference or the chemical potential difference can lead to the LGIs violations or the increase of the maximal value of LGIs, restoring the quantum nature from certain equilibrium cases where LGIs are preserved. The corresponding nonequilibrium thermodynamic cost is quantified by the nonzero entropy production rate. The violation enhancement increases with the increase of the entropy production rate under certain nonequilibrium conditions. Therefore, the LGIs violation enhancement can be realized by the thermodynamic nonequilibrium cost. Our results shed light on the nature of the macrorealism and the relationship between the nonequilibriumness and the quantum temporal correlation. Our finding of the nonequilibrium promoted LGIs violations suggests a new strategy for the design of quantum information processing and quantum computational devices to maintain the quantum nature and quantum correlations for long.
	
\end{abstract}

\maketitle

\section{Introduction}

Quantum correlations, which distinguish the quantum world from the classical world, not only are rooted in the fundamental nature of quantum mechanics, but also become valuable resources for quantum information processing tasks \cite{NC10}. The spatial quantum correlations, known as the entanglement \cite{HHHH09} or the discord \cite{MBCPV12}, show perhaps the most spooky phenomenon in the nature. The well-known Bell inequalities \cite{Bell64} were proposed in 1964 to distinguish the classical correlations with the quantum ones, and the violation is inconsistent with the local hidden variable theory (the classical correlation). The genuine nonlocality, also called Bell nonlocality (violation of Bell inequalities),  has been demonstrated in many experiments since 1972 \cite{FC72,Hensen15,BCPSW14}.

The temporal quantum correlations, generated by sequential non-commuting measurements on a single system at different times, is also different from the classical probabilistic descriptions. Such a discrepancy (between the classical and the quantum) on temporal correlations can be distinguished by the correlation inequalities called Leggett-Garg inequalities (LGIs) \cite{LG85,ELN13}. LGIs are temporal analog version of Bell inequalities. Bell inequalities and LGIs have the same spirit: the joint probability distribution can not be assigned to all measurement results, regardless if the measurements are performed on the separated space or time \cite{MKTLSPK14}. LGIs were firstly motivated to demonstrate the macroscopic coherence, namely how to justify the existence of the macroscopic superposition state. LGIs test the realism of physical states: the system is in definite states with distinct observable values, which is also the essence of the hidden variable theory. Violation of LGIs implies that the system is undergoing the quantum mechanical time evolution which is beyond the classical probabilistic description. 

Although Bell inequalities and LGIs share the same spirit, Fine's theorem \cite{Fine82} can not be applied to the original LGIs, since LGIs require measuring the noncommuting observables at different times and the no-signaling condition fails. LGIs are only the sufficient condition (not the necessary condition) for testing the macrorealism \cite{CK16}. There are different proposals to bring the no-signaling condition in testing the macrorealism for suggesting the sufficient and necessary condition. For examples, the no-signaling in time condition \cite{JB13} has a direct analogy to the no-signaling condition in Bell inequalities. Variants of LGIs, such as the Wigner's form of LGIs \cite{SMPH15} and the augmented set of LGIs \cite{Halliwell16}, can also remedy the sufficient condition issue. Halliwell recently showed that these necessary and sufficient conditions are not equivalent, instead they are testing different degrees (notions) of the macrorealism \cite{Halliwell17,Halliwell19}. 

Although the LGIs are devoted to demonstrate the macroscopic coherence, the violation of LGIs can be attributed from the violation of the macrorealism description or/and the violation of the noninvasive measurability (NIM) \cite{ELN13}. There have been intensive theoretical and experimental studies on how to achieve the NIM to demonstrate the violation of the macrorealism assumption. There are four major strategies to exclude the invasive measurement issue. First, the NIM assumption, in some circumstances, can be replaced by the stationary assumption \cite{HMS95,HMS96} (see the experiments based on the stationary assumption \cite{WNHJW11,GGC15,FKMW16}). Second, experiments can be performed by the weak measurement to diminish the effects of measurements on the system \cite{WHWSHPLM00,AMNBVEK10,GABLOWP11}. Third, the original LGIs argued that the ideal negative measurement (INM) in principle can detect the system without disturbing it \cite{LG85}. The quantum version of the INM has been realized in many experiments \cite{KSGMRABPITBB12,KSRM13,KBLL17,MHGJR19}. Wilde and Mizel almost closed the loophole by designing the control experiments to show that the violation of LGIs does not come from the invasive measurements \cite{WM12} (see the experiments \cite{Knee16}).

Quantum coherence is the reason why the macrorealism notion has to be rejected. However, quantum coherence is notoriously fragile due to the coupling with the environments. This leads to the decoherence \cite{Leggett87,BP02,Wojciech03}. Decoherence has to be included when studying the violation of LGIs in the real world. When the system is coupled with the environments (also called reservoirs or baths in this paper), the dynamics between the system and the environments can be classified as Markovian (memoryless) \cite{BP02,Scully97} or non-Markovian (with the memory effect) \cite{VA17}. Both Markovian \cite{Emary13,CCLC13,LLD15,FL17,CDMSS18,CRQ13} and non-Markovian \cite{CA14,DMM18} effects on violations of LGIs have been studied before. The non-Markovian case requires special care since the quantum coherent evolution can be rewritten as a non-Markovian rate equation which can violate the LGIs \cite{ELN13,LECN10}. 

The effects of the environments to the system can be classified as equilibrium and nonequilibrium, where the nonequilibrium condition is quantified as the temperature difference or the chemical potential difference of the environments or baths. The nonequilibrium condition represents the degree of the energy or matter exchange between the system and the reservoirs, respectively. Coherence \cite{ABHM14,ZHXYF15,GKF18} or entanglement \cite{PHBK99,ABV01,Wang01,KS02,Braun02,BFP03,VWC09,ZJCY11,HRP12} generated and controlled from the equilibrium environment has been intensely studied in last 20 years, due to their potential applications in the quantum information processing. Nonequilibrium environments \cite{EHM09} draw more attentions in recent years. Nonequilibrium environments have their own significance for maintaining and enhancing the long time coherence \cite{ZW14,LCS15,WWCW18} or entanglement \cite{LAB07,QRRP07,SPB08,WS11,LHK11,DSH12,BA13,ZW15,WWW18,THHBB18,TASG18,WW19}. At the equilibrium scenario, coupling with the environment has only negative influence on the violations of LGIs \cite{CCLC13,LLD15,FL17}. However, the steady-state coherence generated under the nonequilibrium condition \cite{ZW14,LCS15,WWCW18} seems to suggest that the nonequilibriumness may contribute to or enhance the violations of LGIs.  Only very limited numbers of studies have been devoted to the issue of how the nonequilibriumness (the energy or particle exchange with the system) influences the quantum dynamical nature of the system \cite{CRQ13,MGRQ19}.

To address the question on how equilibrium and nonequilibrium environments influence the LGIs, we study the following setup: the two-coupled-qubit system (may have different frequencies) immersed into two individual reservoirs respectively. The two reservoirs can have the same or different temperatures (bosonic baths) or chemical potentials (fermionic baths). The weak coupling between the system and the environments and the Markovian dynamics are assumed in the model. The dynamics of the system is described by the Bloch-Redfield equation \cite{Bloch57,Redfield57}, without the secular approximation made in the Lindblad equation. The Lindblad equation is not used here because the nonequilibrium steady state coherence is neglected by the secular approximation in the Lindblad equation \cite{ZW14,LCS15,WWCW18,GKF18,WWW18}. Specifically, if the secular approximation has been applied, the population space and the coherent space will be decoupled. Moreover, the LGIs are dependent on the local observables (local measurements performed on one qubit). Furthermore, if the secular approximation is applied, LGIs will have the symmetric response to the nonequilibrium environments \cite{CRQ13}. However, the Bloch-Redfield equation is criticized by non-positivity of the density matrix evolution in certain parameter regimes. The positivity of Bloch-Redfield equation may be recovered by the initial conditions \cite{SSO92} or further approximations \cite{JIPHC15,FG19}. The validity of Bloch-Redfield equation is beyond the scope of the study in this paper. Instead, we circumvent the positivity issue by concentrating on the parameter regimes which give the positive density matrix.

In our study, we consider the equilibrium and nonequilibrium steady state as the initial state for the LGIs. Therefore, only the time interval of sequential measurements matters. We adopt the augmented LGIs (a set of two- and three-time LGIs) as the necessary and sufficient condition for the macrorealism. We give analytical results about the steady state at both equilibrium and nonequilibrium scenarios. In nonequilibrium case, the two reservoirs are sustained with a constant temperature difference or chemical potential difference if the two baths are bosonic or fermionic, respectively. Sequentially measuring one local qubit gives the temporal correlations of the local observable. We analytically find the augmented set of LGIs based on the weak-coupling assumption. We approximate the augmented set of LGIs up to the first order coupling strength between the system and the environments. The zeroth order (in terms of the coupling strength) can be viewed as the coherent evolution part and the first-order part describes the effects on LGIs from the dynamical coupling between the system and the environments. The steady-state coherence in the energy basis, which is only nonzero in the nonequilibrium setups, has the contribution to the two-time LGI functions. However, such contribution is not significant enough for the violation of two-time LGIs. Therefore, we concentrate on the three-time LGIs (the original proposed LGIs) in our study. The maximum of LGI (MLGI) functions can be used to quantify the degree of the LGIs violations.

In the equilibrium cases, the MLGI function has the non-monotonic relation with the common temperature or the chemical potential. In the low temperature regime, increasing the temperature can increase the population of the excited states, which are superposition of local states. In the high temperature regime, increasing the temperature leads to a stronger thermal effect (decoherence). In the nonequilibrium scenario, the MLGI function can be enhanced by the nonequilibrium conditions: the temperature difference or the entropy production rate in the bosonic environments; the chemical potential difference or the entropy production rate in the fermionic environments. The LGIs violation enhancement has a thermodynamic cost quantified by the nonzero entropy production rate. The bosonic enhancement is only realized at the low mean temperature regime. If we choose to measure the qubit 1, then the MLGI function has greater enhancement if the qubit 1 is coupled with the lower temperature bath. Note that the Lindblad gives symmetric results: the qubit 1 coupled with the bath with a lower temperature $T_1$ and the qubit 2 coupled with the bath with a higher temperature $T_2$ gives the same result when the qubit 1 is coupled with the bath with a higher temperature $T_2$ and the qubit 2 is coupled with the bath with a lower temperature $T_1$. In other words, the Lindblad does not characterize well the nonequilibrium dynamics. We obtained the MLGI function enhancement from the chemical potential difference (fermionic baths) when the mean chemical potential is away from the resonant point. The resonance occurs when the mean chemical potential equals to the mean energy of the two-qubit system. We have also studied the LGIs violations when the two-qubit system has detuned energy splitting. We got a larger MLGI function (than the equilibrium case) when the low-frequency qubit is coupled with the high-temperature bath or high chemical potential bath and the high-frequency qubit is coupled with the low-temperature bath or high chemical potential. Although the model in our study is far from macroscopic, the results suggest that the nonequilibrium environments may be beneficial to the real macroscopic coherence.

The rest of the paper is organized as follows. Section \ref{sec: LGI} reviews the time correlation functions and the two- and three-time LGIs. We also briefly review how the ideal negative measurement (as the noninvasive measurement) can be realized on the qubit system. Section \ref{sec:model} presents the dynamic quantum master equation of the system and the analytical form of the steady state. We study the LGI functions and MLGI function given by the system coupled with the equilibrium and nonequilibrium environments in Secs. \ref{sec:equ_LGI} and \ref{sec:nonequ_LGI} respectively. The results are based on analytical expressions and are demonstrated numerically.  The last section gives the conclusion. The detailed expressions for the Bloch-Redfield equation and the heat or particle current are presented in the Appendix.

\section{The two- and three-time Leggett-Garg inequalities}

\label{sec: LGI}

\subsection{The correlation function}

In classical probabilistic theory, the results of measurements performed at different times $t_j$ and $t_l$ can be described by the joint probability $P(Q_j,t_j;Q_l,t_l)$. {Here $Q_j$ is the measurement value at $t_j$.} The correlation function {characterizing the measurement results at $t_j$ and $t_l$} is defined as
\eq
\label{C ij}
 C_\text{cl}(t_j,t_l)=\sum_{Q_j,Q_l}Q_jQ_lP(Q_j,t_j;Q_l,t_l)
\en
The subscript ``cl'' distinguishes the correlation functions from the quantum case. 

In the quantum mechanical description, there is no unique analog of classical correlation function defined in (\ref{C ij}), because of the operator ordering. We can define the quantum correlation function in the perspective of the measurement results. Suppose we have dichotomic observables 
\begin{equation}
    \hat Q = \sum_m a_m\hat\Pi_m
\end{equation}
with $a_m=\pm1$. Operators $\hat\Pi_m$ are the corresponding projections. Since we only consider the quantum cases, we omit the hat notation on the operator for simplicity. Suppose that the system has the time evolution
\begin{equation}
\label{def w}
    \frac{d\rho}{dt} = \mathcal W \rho
\end{equation}
where $\mathcal W$ is the superoperator generating the dynamics. Note that we do not restrict ourselves in the unitary evolution only. Based on the perspective of the classical measurement values, the correlation function in the quantum case has the form
\begin{align}
\label{def Cq}
    C_\text{q}(t_j,t_l) = &\sum_{m,n}a_ma_n \tr\left(\Pi_m e^{\mathcal W (t_j-t_l)}\left(\Pi_n \rho(t_l)\Pi_n\right)\right) \nonumber \\
    =& \tr\left(Q(t_j)\left(\sum_n a_n \Pi_n\rho(t_l)\Pi_n\right)\right)
\end{align}
The subscript q means quantum. Here, the operator $Q(t_j)$ is defined in the Heisenberg picture. The correlation function $C_\text{q}(t_j,t_l)$ has the same interpretation as the classical correlation $C_\text{cl}(t_j,t_l)$ in Eq. (\ref{C ij}).

Studies regarding the LGIs usually take the other form of the quantum correlation function \cite{ELN13}:
\begin{align}
\label{def C'q}
    C'_\text{q}(t_j,t_l) = &\frac 1 2 \tr\left(\left\{Q(t_j),Q(t_l)\right\}\rho(t_l)\right) \nonumber \\
    =& \tr\left(Q(t_j)\frac 1 2\left\{Q(t_l),\rho(t_l)\right\}\right) 
\end{align}
which is also the real part of the ``naive'' quantum correlation function
\begin{equation}
    C''_\text{q}(t_j,t_l) = \tr(Q(t_j)Q(t_l)\rho(t_l))
\end{equation}
The quantum correlation function $C''_\text{q}(t_j,t_l)$ is in general a complex function. The imaginary part of it is a measure of noncommutativity for observables $Q(t_j)$ and $Q(t_l)$ (without classical analog).  

Note that the anticommutator in Eq. (\ref{def C'q}) has the explicit form
\begin{multline}
    \frac 1 2\left\{Q(t_l),\rho(t_l)\right\}\\
     =\sum_n a_n \Pi_n \rho(t_l)\Pi_n +\frac 1 2 \sum_{n\neq n'}(a_n+a_{n'})\Pi_n\rho(t_l)\Pi_{n'}
\end{multline}
Since we set that $a_n=\pm 1$, the second term in the above equation is zero. If we substitute the above expression into Eq. (\ref{def C'q}) (with the setting $a_n=\pm 1$), then the correlation function $C_\text{q}(t_j,t_l)$ in Eq. (\ref{def Cq}) has the same form of $C'_\text{q}(t_j,t_l)$ in Eq. (\ref{def C'q}). Based on the setting $a_n=\pm 1$, we reach the conclusion that the commonly used quantum correlation function $C'_\text{q}(t_j,t_l)$ in Eq. (\ref{def C'q}) has the same interpretation with the classical correlation $ C_\text{cl}(t_j,t_l)$ in Eq. (\ref{C ij}). Note that such interpretation is only valid with the dichotomic observables with eigenvalues $\pm1$. Some literatures using the measurement values $0,1$ \cite{HMS96,WNHJW11} do not have such interpretation.

%We can define the correlation functions by taking the real part of the observables:

%It can be interpreted exactly the same as the classical correlation function: the correlation between the two measurement results at time $t_j$ and $t_l$. Such interpretation is only valid for dichotomic observables (with values $\pm 1$). Note that the imaginary part of {$\tr(Q(t_j)Q(t_l)\rho)$} is a measure of non-communitativity for observables $Q(t_j)$ and $Q(t_l)$ (without classical analog).

\subsection{The three-time Leggett-Garg inequalities}

Originated from the macrorealism test, the original LGIs are based on the following assumptions:
\begin{enumerate}[(a)]
    \item Macrorealism per se: the physical object is in one of the distinct states at any time.
    
    \item NIM: it is in principle possible to reveal the state of the object without disturbing the following dynamics.
    
    \item Induction: the present state can not be affected by the future measurements.
\end{enumerate}
The induction assumption is also assumed. The NIM assumption is based on the macrorealism per se assumption. Quantum systems do not admit either macrorealism per se assumption or the NIM assumption. However, a macrorealist can always claim that the violation of LGIs is from the failure to perform the NIM. This is the ``clumsiness loophole'' in testing the macrorealism. Although the loophole free test is not proposed yet, multiple different strategies have been suggested to exclude the violation of the NIM assumption (see the review \cite{ELN13} for more comments). Nevertheless, the aim of this paper is {\it not} to address the NIM assumption in the macrorealism tests. Instead, we assume that the INM protocol can always be perfectly realized. How to realize the INM in the quantum system is briefly reviewed in Sec. \ref{subsec:INM}. Then, we focus on the influence of the equilibrium and nonequilibrium environments on the violations of LGIs.

The essence of deriving the LGIs is to assign the probability distribution over the three-time measurements $P(Q_3,t_3;Q_2,t_2;Q_1,t_1)$. Then, the two-time probability $P(Q_j,t_j;Q_l,t_l)$ can be obtained from the marginal of the three-time probability. Then, the two-time correlations are bounded by the following inequalities \cite{LG85}
\begin{subequations}
\begin{align}
\label{def LGI 1}
 &C_\text{cl}(t_1,t_2)+C_\text{cl}(t_2,t_3)-C_\text{cl}(t_1,t_3)\leq 1 \\
 \label{def LGI 2}
 &C_\text{cl}(t_1,t_2)-C_\text{cl}(t_2,t_3)+C_\text{cl}(t_1,t_3)\leq 1 \\
 \label{def LGI 3}
-&C_\text{cl}(t_1,t_2)-C_\text{cl}(t_2,t_3)-C_\text{cl}(t_1,t_3)\leq 1 \\
\label{def LGI 4}
-&C_\text{cl}(t_1,t_2)+C_\text{cl}(t_2,t_3)+C_\text{cl}(t_1,t_3)\leq 1 
\end{align}
\end{subequations}
The above inequalities only concern the constraints on the two-time
probability distribution given by the three-time probability
distribution. The three-time LGIs can be generalized to the multi-time cases \cite{ELN13}. If the initial time is irrelevant, then the correlation function only depends on the time interval. In practice, we can keep the two-time intervals in the three-time measurements to be the same, i.e., $t_2-t_1=t_3-t_2=t$. Then, the four LGIs in Eqs. (\ref{def LGI 1})-(\ref{def LGI 4}) can be reduced into two inequalities
\begin{equation}
\label{def LGI ss}
 \pm2 C_\text{cl}(t)-C_\text{cl}(2t)\leq 1 \\
\end{equation}
where $C_\text{cl}(t) = C_\text{cl}(t_1,t_2)=C_\text{cl}(t_2,t_3)$ and $C_\text{cl}(2t) = C_\text{cl}(t_1,t_3)$.

Since quantum mechanics rejects either the microrealism or the macrorealism, the quantum correlation functions do not satisfy the inequalities (\ref{def LGI ss}). We define the corresponding three-time LGI functions 
\begin{equation}
\label{LGI steady state}
\mathcal I_\pm(t,\rho^\text{ss}) =
\pm2C_\text{q}(t) - C_\text{q}(2t)
\end{equation}
where the quantum correlation function $C_\text{q}(t)$ is given by Eq. (\ref{def Cq}) or Eq. (\ref{def C'q}). Note that the functions $\mathcal I_\pm(t,\rho^\text{ss})$ depend on the initial density matrix $\rho^\text{ss}$ and the equal interval time $t$ once the dichotomic observable $Q$ is fixed. Here, the superscript ``ss'' in $\rho^\text{ss}$ represents the steady state of the system (therefore, the initial time is irrelevant). The LGI functions $\mathcal I_\pm(t,\rho^\text{ss})$ can break the classical bound but are limited with the value \cite{ELN13}:
\begin{equation}
    \mathcal I_\pm(t,\rho^\text{ss})\leq \frac 3 2
\end{equation}
Only the evolution defined in Eq. (\ref{def w}) is unitary, the LGI functions $\mathcal I_\text{a,b}(t,\rho^\text{ss})$ can be saturated.  

\subsection{The two-time Leggett-Garg inequalities}

The original three-time LGIs are only the sufficient condition for testing the macrorealism \cite{CK16}. There are several different proposals for the necessary and sufficient conditions \cite{JB13,SMPH15,Halliwell16}. Different proposals are not exactly equivalent \cite{Halliwell17,Halliwell19}. Here we follow a approach which remains closely with the original three-time LGIs. A new set of two-time LGIs combined with the three-time LGIs in Eqs. (\ref{def LGI 1})-(\ref{def LGI 4}) form the necessary and sufficient condition \cite{Halliwell16,Halliwell17,Halliwell19}. 

Since both the single-time probability and the two-time probability distributions can be obtained from the marginals of the three-time probability, the averages and the two-time correlation functions also satisfy the inequalities
\begin{equation}
\label{def LGI two}
    -a_j\langle Q(t_j)\rangle-a_l\langle Q(t_l)\rangle-a_ja_l C_\text{cl}(t_j,t_l)\leq 1
\end{equation}
where $a_j,a_l=\pm1$. Here, the time pair $(j,l)$ takes the values $(1,2),(2,3),(1,3)$. Therefore, there are 12 inequalities in total. The averages and the correlation functions can be measured in three different experiments. The averages do not have the invasive measurement issue since only one measurement is performed. Similar with the three-time LGIs, we assume that the correlation function $C_\text{cl}(t_j,t_l)$ satisfies the NIM assumption. Note that the inequalities in Eq. (\ref{def LGI two}) are the necessary and sufficient conditions for the macrorealism in the two-time level. In other words, if the two-time inequalities are satisfied, there are no contradictions between the joint probability $P(Q_j,t_j;Q_l,t_l)$ and the single-time probability $P(Q_j,t_j)$. Based on a simpler proof of Fine's theorem \cite{Halliwell14}, Halliwell showed that if the three-time LGIs in Eqs. (\ref{def LGI 1})-(\ref{def LGI 4}) combined with the two-time LGIs in Eq. (\ref{def LGI two}) are satisfied, the three-time joint probability can always be constructed, therefore giving the necessary condition for the macrorealism.

If the system reaches the steady state, the initial time is not concerned and the correlation function only depends on the time interval between the two measurements. Surprisingly, only 2 of the 12 inequalities are nontrivial and the two remaining inequalities have the unified form
\begin{equation}
\label{def classical LGI 2}
    2\langle Q\rangle_\text{ss} -C_\text{cl}(t)\leq 1
\end{equation}
We define the corresponding two-time LGI function with the steady state condition:
\begin{equation}
\label{def LGI function 2}
    \mathcal I_2(t,\rho^\text{ss}) = 2\langle Q\rangle_\text{ss}-C_\text{q}(t)
\end{equation}
where the quantum correlation function $C_\text{q}(t)$ is given by Eq. (\ref{def Cq}) or Eq. (\ref{def C'q}). Instead of the classical bound shown in Eq. (\ref{def classical LGI 2}), the two-time LGI function is bounded by
\begin{equation}
    \mathcal I_2(t,\rho^\text{ss}) \leq 3
\end{equation}
See examples in Sec. \ref{sec:equ_LGI}. Recent experiments on nuclear spins showed that different initial states can separately violate either the three-time LGIs or the two-time LGIs \cite{MHGJR19}.

\subsection{The ideal negative measurements}

\label{subsec:INM}

\begin{figure}
	\includegraphics[width=\columnwidth]{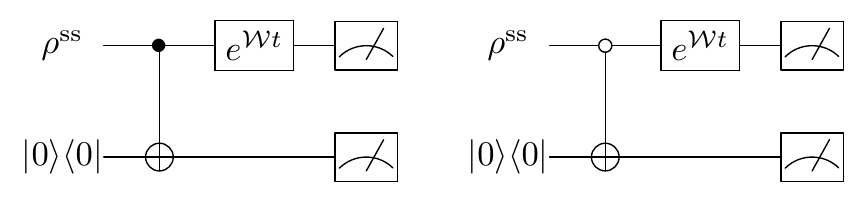}
	\caption{\label{fig_INM} The quantum INM based on the classical INM protocol. The CNOT and anti-CNOT gates are defined in Eqs. (\ref{def CNOT}) and (\ref{def antiCNOT}), respectively. The measurement results are discarded of the ancilla qubit $|0\rangle\langle 0|$ is flipped. The evolution of the density matrix is described by the superoperator $\mathcal W$ defined in (\ref{def w}).}
\end{figure}

To demonstrate the violation of the macrorealism assumption via the two-time and three-time LGIs, the experiments performed need to rule out the possibility that the violations of LGIs do not come from the invasive measurements. Leggett and Garg originally argued that the noninvasive measurement can always be applied via the INM \cite{LG85}. Suppose that the measurement apparatus only clicks (interacting with the system) when the system is at the state assigned with the value $1$. Therefore, if the measurement device remains unclicked, we can learn that the system stays at the state with the assigned value $-1$. And we discarded the measurement results when the measurement device is triggered. 

In quantum mechanics, the NIM is rejected. However, we can design the ``INM'' with the same spirit of the classical INM described above \cite{KSGMRABPITBB12}. The interaction between the system and the measurement apparatus (an ancilla qubit) can be described by the controlled NOT (CNOT) gate or anti-CNOT gate, which are defined as 
\begin{subequations}
\begin{align}
\label{def CNOT}
    \text{CNOT} = & |0\rangle\langle 0|\otimes 1\!\!1_2+|1\rangle\langle 1|\otimes \sigma_x, \\
\label{def antiCNOT}
    \text{anti-CNOT} = & |1\rangle\langle 1|\otimes 1\!\!1_2+|0\rangle\langle 0|\otimes \sigma_x
\end{align}
\end{subequations}
where $1\!\!1_2$ is the identity operator and $\sigma_x$ is the Pauli-$X$ gate. Clearly, the CNOT gate (anti-CNOT gate) only flips the ancilla qubit if the control state is $|1\rangle$ ($|0\rangle$). Then we can design the quantum circuits to measure the correlation function $C_\text{q}(t)$ according to the classical noninvasive way (see FIG. \ref{fig_INM}). 

The classical INM suggests that we discard the measurement results when the ancilla qubit is flipped. To understand why the quantum circuits in FIG. \ref{fig_INM} work, we can have the intuitive understanding in the following way. Suppose that we know the probability for the measurement results $|00\rangle$ (with the CNOT gate). The measurement results $|00\rangle$ imply that the system is always in the state $|0\rangle$ during the time $t$, since the ancilla qubit is not flipped. If the measurement result is $|10\rangle$, we know that the initial state of the system is $|0\rangle$ and the system becomes $|1\rangle$ after the time evolution. In fact, the diagonal terms in the final state of the quantum circuits represent the two-time probability of the system (see the detail calculations in Ref. \cite{MHGJR19}). Note that we can apply a single-qubit gate conjugated at the CNOT gate to perform the INM on other observables of the system.  

The INM protocol does not exclude all interactions between the system and the measurement device, therefore, the loophole is not completely closed. One way to circumvent the NIM issue is to replace the NIM assumption by the stationary assumption \cite{HMS95,HMS96}. 
We assume the steady state of the system which seems to satisfy the stationary assumption. However, the essence of the stationary assumption is to prepare the initial state in some definite states, where the first-time measurement is not necessary (see more comments in Ref. \cite{ELN13}). The steady-state condition is proposed in this study due to the long-time limit reached from the interactions between the system and the environments. The steady state of the system can be conveniently measured in the experiments. Since we do not assume the specific form of the steady state, the initial measurement can not be omitted. The stationary condition proposed in our paper does not aim to replace the NIM assumption.

\section{Quantum master equation and steady state}

\label{sec:model}

\subsection{Model}

\begin{center}
\begin{figure}
	\includegraphics[width=0.8\columnwidth]{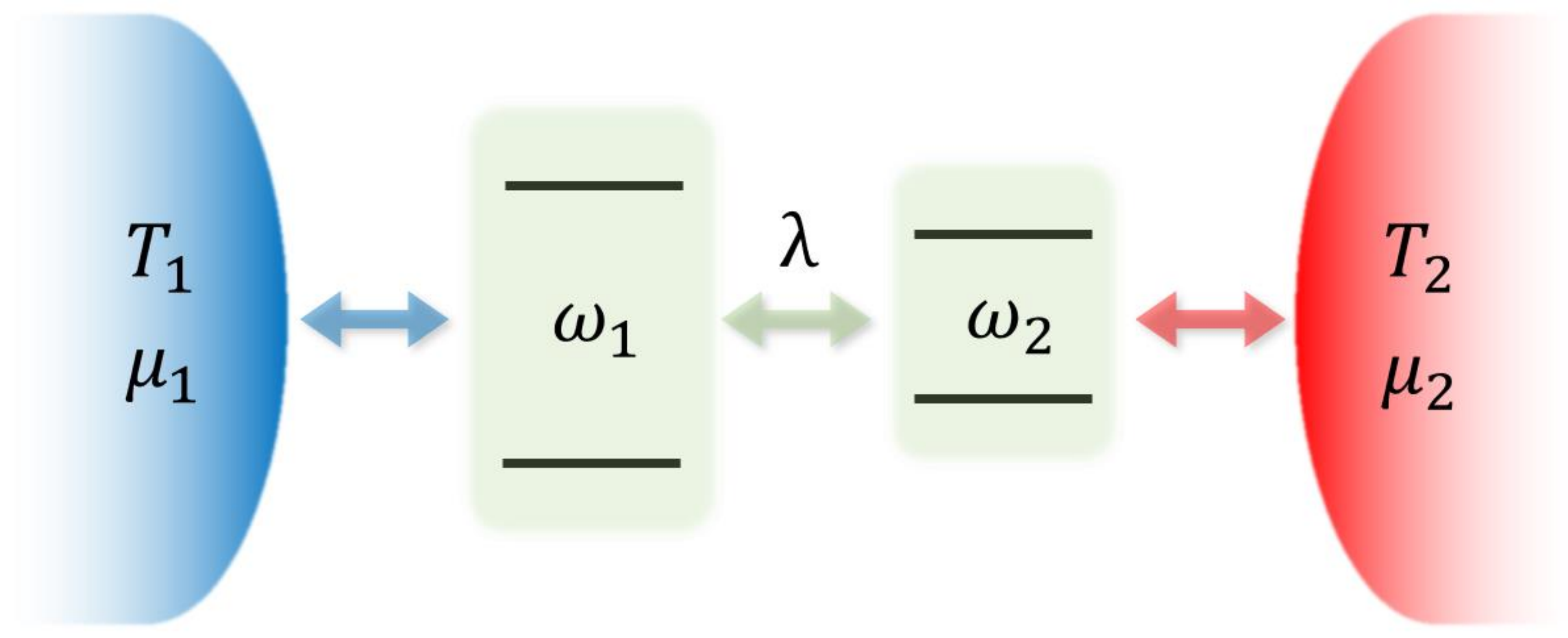}
	\caption{\label{fig_model} Two qubits with different frequencies weakly coupled with two environments. The two environments have different temperatures or chemical potentials. The inter-qubit coupling is characterized by $\lambda$. } 
\end{figure}
\end{center}

We study a two-coupled two-level (qubit) system (with different {transition} frequencies) individually coupled to two bosonic or fermionic baths, respectively, with different temperatures or chemical potentials (see FIG. \ref{fig_model}). The free Hamiltonian of the system (coupled two qubits) and the environments are given by
\begin{align}
\label{def H S}
&H_S=\omega_1|e_1\rangle\langle e_1|+\omega_2|e_2\rangle\langle e_2|+\frac \lambda 2\left(\sigma_1^+\sigma_2^-+\sigma_1^-\sigma_2^+\right),\\
\label{def H R}
&H_R=\sum_{k}\omega_{bk}b_{k}^{\dagger}b_{k}
+\sum_{k}\omega_{ck}c_{k}^{\dagger}c_{k}
\end{align}
where $\omega_1$ and $\omega_2$ represent the energy splittings {(transition frequencies)} of the first and second qubit respectively; state $|e_1\rangle$ or $|e_2\rangle$ is the excitation in the first or second qubit; the ground state has energy 0; the third term in $H_S$ describes the coupling between the two qubits and $\lambda$ is the coupling strength between qubit 1 and qubit 2; operators $\sigma_1^+=|e_1\rangle\langle g_1|$ and $\sigma_2^+=|e_2\rangle\langle g_2|$ are the raising (creation) operators for the qubit one and the qubit two respectively; The bosonic bath gives the commutative relation $[b_k,b_{k'}^\dag]=\delta_{kk'}$, $[b_k,b_{k'}]=0$ and the fermionic bath gives the anti-commutative relation $\{b_k,b_{k'}^\dag\}=\delta_{kk'}$, $\{b_k,b_{k'}\}=0$; the two baths have the energy spectral $\omega_{bk}$ and $\omega_{ck}$ for the $k$-th mode ($b$ denotes the first bath coupled with qubit number 1 and $c$ denotes the second bath coupled with qubit number 2). The constant $\hbar=1$ is set to 1 for convenience. 

The model having the Hamiltonian $H_S$ in (\ref{def H S}) can be understood as two spatial separated atoms coupled through the dipole-dipole interaction \cite{PK02}. Note that the effective Hamiltonian of the dipole-dipole interaction characterizes the resonant coupling of two identical atoms which have $\omega_1=\omega_2$ \cite{CT98}. In the language of spin chain, Hamiltonian $H_S$ in (\ref{def H S}) describes the Heisenberg XY model ($\sigma_x\sigma_x$ and $\sigma_y\sigma_y$ interactions) and the two spins are subjected to inhomogeneous magnetic field (different transition frequencies) \cite{LSM61,Wang01,KS02,QRRP07}. If the two-level system is understood as a (spin-degenerate) quantum dot: $|g\rangle$ as the empty site and $|e\rangle$ as the occupied site, the coupling between the two qubits is essentially the non-interaction tunneling \cite{LAB07,WW19}. The onsite energies characterized by $\omega_i$ can be controlled by the electrochemical potentials \cite{HFCJH03}. Note that the Hamiltonian $H_S$ in (\ref{def H S}) does not include the interdot Coulomb interaction of the two sites (as a toy model for double quantum dots). 

{Because of the dipole-dipole interaction, atomic basis is not the eigenstate. The system forms the dimer eigenstates (with the new eigenenergies) defined by:}
\begin{align}
\label{def nonlocal basis}&\omega_{gg}=0,   &|1\rangle=&|gg\rangle, \nonumber \\
&\omega_1'=\bar\omega -\frac 1 2 \sqrt{\Delta \omega^2+\lambda^2},  &  |2\rangle=&\cos\frac \theta 2|eg\rangle+\sin\frac \theta 2|ge\rangle, \nonumber \\
&\omega_2'=\bar\omega +\frac 1 2 \sqrt{\Delta \omega^2+\lambda^2}, & |3\rangle=&-\sin\frac \theta 2|eg\rangle+\cos\frac \theta 2|ge\rangle, \nonumber \\
&\omega_{ee}=\omega_1+\omega_2, &|4\rangle=&|ee\rangle
\end{align}
We use short notations: $|eg\rangle = |e_1\rangle\otimes |g_2\rangle$. Here, $\bar \omega=(\omega_1+\omega_2)/2$ is the mean energy splitting and $\Delta \omega=\omega_1-\omega_2$ is the degree of energy detuning. The interqubit interaction is relatively weak. We limit the coupling strength in the regime $\lambda<\sqrt{\omega_1\omega_2}$. The angle $\theta$ is defined by
\begin{align}
\label{def theta}
    \theta =
            \begin{cases}
                \arctan(\lambda/\Delta\omega), & \text{if~}\Delta\omega<0\\
                \arctan(\lambda/\Delta\omega)-\pi, & \text{if~}\Delta\omega>0
            \end{cases}
\end{align}
The identical two qubits ($\omega_1=\omega_2=\omega$) gives $\theta=-\pi/2$. And the eigenstates $|2\rangle$ and $|3\rangle$ are Bell-type states (maximal entangled two-qubit states). The weak interqubit coupling $\lambda<\sqrt{\omega_1\omega_2}$ gives the ground state $|gg\rangle$. We have quantum phase transition \cite{Sachdev07} at $\lambda=\sqrt{\omega_1\omega_2}$: state $|2\rangle$ will be ground state if $\lambda>\sqrt{\omega_1\omega_2}$. 

We can expand the lowering (annihilation) operators $\sigma_l^-$ (with $l=1,2$) and raising (creation) operators $\sigma_l^+$ into the basis of eigenstates $|j\rangle$ (with $j=1,2,3,4$):
\eq
\label{eqa sigma with eta xi}
\sigma_1^-=\eta_1+\xi_1,\quad\quad\sigma_2^-=\eta_2+\xi_2
\en
with the operators $\eta$ and $\xi$ defined in the energy basis
\begin{subequations}
\begin{align}
  \label{eta1}\eta_1=&\cos\frac\theta 2\left(|1\rangle\langle 2|+|3\rangle\langle 4|\right), \\
  \label{eta2}\eta_2=&\cos\frac\theta 2\left(|1\rangle\langle 2|-|3\rangle\langle 4|\right), \\
  \label{xi1}\xi_1=&\sin\frac\theta 2\left(|2\rangle\langle 4|-|1\rangle\langle 3|\right), \\
  \label{xi2}\xi_2=&\sin\frac\theta 2\left(|2\rangle\langle 4|+|1\rangle\langle 3|\right)
\end{align}
\end{subequations}
Similarly, the raising operators $\sigma_l^+$ can be reformulated with the operators $\eta^\dag_l$ and $\xi^\dag_l$ (with $l=1,2$).

{We adopt the rotation wave approximation (neglecting oscillations with high frequency) to describe the interaction between the system and the environments:}
\eq
\label{H SR}
H_{SR}=\sum_k g_k\left(\sigma^-_1b_k^\dag+\sigma^+_1b_k\right)+\sum_k h_k\left(\sigma^-_2c_k^\dag+\sigma^+_2c_k\right)
\en
where $g_k$ and $h_k$ are the coupling strengths between the system and the environments. We can assume that $g_k$ and $h_k$ are both real numbers without losing generality. When the environments are bosonic (operators $b_k$ and $c_k$ follow the commutative relations), the model describes two atoms or two 1/2-spin system interacting with photonic or thermal baths. When the  environments are fermionic (operators $b_k$ and $c_k$ follow the anti-commutative relations), the model describes double quantum dots coupled with two metal leads.

In the interaction picture, the Hamiltonian $H_{SR}$ (\ref{H SR}) takes the form
\begin{align}
\label{H SR interaction}
H_{SR}(t)=&\sum_kg_k\left(\eta_1e^{-i\omega'_{1}t}+\xi_1e^{-i\omega'_{2}t}\right)b_k^\dag e^{i\omega_{bk}t}+\text{H.c.} \nonumber \\
+&\sum_kh_k\left(\eta_2e^{-i\omega'_{1}t}+\xi_2e^{-i\omega'_{2}t}\right)c_k^\dag e^{i\omega_{ck}t}+\text{H.c.}
\end{align}
{where H.c. is short for Hermitian conjugate. The physical meaning of operators $\eta_l$ and $\xi_l$ (\ref{eta1})-(\ref{xi2}) is clear: operators $\eta_l$ are lowering the energy $\omega_1'$ and operators $\xi_l$ are lowering the energy $\omega'_2$.}

\subsection{Quantum Master Equation}

Based on the weak-coupling (between the system and environment) and Born-Markov approximations, the dynamics of system (in the interaction picture) is governed by the quantum master equation for the reduced density matrix (after tracing over the baths): \cite{BP02}
\eq
\frac{d\rho_I(t)}{dt}=-\int_0^\infty ds \text{Tr}_R \left[H_{SR}(t),\left[H_{SR}(t-s),\rho_I(t)\otimes\rho_R\right]\right]
\en
where $\rho_I(t)=\exp\left(-iH_S t\right)\rho_S\exp\left(iH_S t\right)$ and $\rho_R$ is the density matrix of the two baths in their thermal equilibrium states. {Here $i$ is the imaginary unit $i=\sqrt{-1}$.} {The above equation is called Bloch-Redfield equation \cite{BP02,Bloch57,Redfield57}. For our model described by the interaction Hamiltonian $H_{SR}$ defined in (\ref{H SR}), we have the Bloch-Redfield equation (back to Schrödinger's picture): } 
\eq
\label{QME}
\frac{d\rho_S}{dt}=i\left[\rho_S,H_S\right]+\sum_{l=1}^2\mathcal D_l[\rho]
\en
with the dissipators expressed as ($l=1,2$)
\begin{align}
\label{def dissipators}
\mathcal D_l[\rho] =& \alpha_l(\omega_1')\left(\eta_l^\dag\rho \eta_l+\eta_l^\dag\rho \xi_l-\eta_l\eta_l^\dag\rho-\xi_l\eta_l^\dag\rho+\text{H.c.}\right) \nonumber \\
+& \alpha_l(\omega_2')\left(\xi_l^\dag\rho \xi_l+\eta_l^\dag\rho \xi_l-\xi_l\xi_l^\dag\rho-\eta_l\xi_l^\dag\rho+\text{H.c.}\right) \nonumber \\
+& \beta_l(\omega_1')\left(\eta_l\rho \eta_l^\dag+\eta_l\rho \xi_l^\dag-\eta_l^\dag \eta_l\rho-\xi_l^\dag \eta_l\rho+\text{H.c.}\right) \nonumber \\
+& \beta_l(\omega_2')\left(\xi_l\rho \xi_l^\dag+\eta_l\rho \xi_l^\dag-\xi_l^\dag \xi_l\rho-\eta_l^\dag \xi_l\rho+\text{H.c.}\right)
\end{align}
Here, the coefficients are
\eq
\label{def alpha beta}
\alpha_l(\omega)=J_l(\omega)n_{l}(\omega),
\quad\quad  \beta_l(\omega)=J_l(\omega)(1\pm n_{l}(\omega))
\en
{with the coupling spectrum of the two baths defined as}:  
\eq
\label{def J}
J_1(\omega)=\pi\sum_k g_k^2\delta(\omega-\omega_{bk}),\quad J_2(\omega)=\pi\sum_k h_k^2\delta(\omega-\omega_{ck})
\en
Here the plus sign in $\beta_l(\omega)$ is for the case of the bosonic bath and the minus sign is for the fermionic bath. Variable $n_l(\omega)$ is the mean occupation number of particles with the energy $\omega$ at temperature $T_l$ and chemical potential $\mu_l$ for the $l$-th bath, namely 
\begin{equation}
\label{def n i omega}
     n_l(\omega)=\frac{1}{\exp\left((\omega-\mu_l)/T_l\right)\mp1}
\end{equation}
The minus sign is for bosonic bath {(Bose-Einstein distribution)} and plus sign is for fermionic bath {(Fermi-Dirac distribution)}. Boltzmann constant is set to be 1. {Photons or phonons have negligible self-interactions.} We set $\mu_1=\mu_2=0$ both for the equilibrium ($T_1=T_2=T$) and nonequilibrium ($T_1\neq T_2$) bosonic baths. And we consider the temperature equilibrium ($T_1=T_2=T$) for fermionic baths both for equilibrium chemical potential ($\mu_1=\mu_2=\mu$) and nonequilibrium chemical potential ($\mu_1\neq \mu_2$) cases, especially in the low-temperature regime.

Operators $\eta_l$ defined in (\ref{eta1}) and (\ref{eta2}) and $\xi_l$ defined in (\ref{xi1}) and (\ref{xi2}) characterize the transitions with different frequencies ($\eta_l$ is for $\omega_1'$ and $\xi_l$ is for $\omega_2'$). In the interaction picture, cross terms in the dissipators in (\ref{def dissipators}), such as $\eta_l^\dag\rho \xi_l$, are usually considered as oscillating processes and therefore often neglected under fast oscillations (secular approximation). After secular approximation, the Bloch-Redfield equation becomes the Lindblad form. Since the cross terms couple the population and coherent space of the density matrix, dropping the crossing terms gives zero steady state coherence (in the eigenstate representation) (see \cite{LCS15,WWCW18,WWW18}). Coherence is crucial for the violations of LGIs. Therefore, we keep the cross terms in our study and apply the master equation without the secular approximation, or the Bloch-Redfield equation for time evolution of the system. 

%{ Explanations for balance constant coupling $J$. Or the case which is valid?}.

\subsection{Steady State Solutions}

{The steady-state solution of the Bloch-Redfield equation in a similar setting (for the study of a different problem) had been obtained in \cite{WWW18}. Here, we follow the similar procedure.} We can reformulate the Bloch-Redfield equation with the form (\ref{QME}) in the Liouville space, where the system density matrix takes the vector form
\begin{equation}
    |\rho_S\rangle=\left(\rho_{11},\rho_{22},\rho_{33},\rho_{44},\rho_{23},\rho_{32}\right)^T
\end{equation}
with $T$ as the matrix transpose. Other coherence terms {(such as $\rho_{14}$ and $\rho_{41}$)} are decoupled with the population terms and therefore can be dropped in the steady-state solution. The Bloch-Redfield equation (\ref{QME}) has the matrix form:
\eq
\label{QME matrix}
\frac d {dt}|\rho_S\rangle=\mathcal M |\rho_S\rangle
\en
The expressions of the matrix elements $\mathcal M$ are given in Appendix \ref{Appendix:matrix}. 

The reduced density matrix of system can be grouped into two parts: $|\rho_S\rangle=\left(\rho_p,\rho_c\right)^T$ with population terms $\rho_p$ (diagonal terms) and coherence terms $\rho_c$ (off-diagonal terms). Under the same arguments, the dynamic matrix $\mathcal M$ has the block forms:
\eq
\mathcal M=\left(\begin{array}{cc}
  \mathcal M_{pp}	& \mathcal M_{pc} \\
  \mathcal M_{cp}	& \mathcal M_{cc}
\end{array}\right)
\en
{The steady state is given by}
\begin{equation}
     \mathcal M|\rho^\text{ss}_S\rangle=0
\end{equation}
The coherence terms can be substituted by
\begin{equation}
\label{def rho c}
    \rho_c=-\mathcal M_{cc}^{-1}\mathcal M_{cp} \rho_p
\end{equation}
with invertible block matrix $\mathcal M_{cc}$. Then, we define the steady-state population matrix $\mathcal A$ as
\eq
\label{def A}
\mathcal A= \mathcal M_{pp}-\mathcal M_{pc}\mathcal M_{cc}^{-1}\mathcal M_{cp}
\en
which satisfies 
\begin{equation}
\label{eqa A rho}
    \mathcal A|\rho^\text{ss}_p\rangle=0
\end{equation}
Note that the overall constant in $\mathcal A$ gives the same steady state solution. The matrix elements of the steady-state population matrix $\mathcal A$, both for the bosonic and fermionic baths, are presented in Appendix \ref{appen B}.

In the following, we consider the symmetric constant coupling spectrum:
\begin{equation}
\label{constant J}
     J_1(\omega)=J_2(\omega)=J
\end{equation}
To simplify the steady-state expressions $\rho^\text{ss}_S$, we introduce the following notations:
\begin{subequations}
\begin{align}
\label{def tilde n1}&\tilde n_1(\theta,\omega_1')=\cos^2\frac \theta 2 n_1(\omega'_1)+\sin^2\frac \theta 2n_2(\omega'_1), \\
\label{def tilde n2}&\tilde n_2(\theta,\omega_2')=\sin^2\frac \theta 2n_1(\omega'_2)+\cos^2\frac \theta 2n_2(\omega'_2), \\
\label{delta n 1}&\Delta n_1(\theta,\omega_1')= \frac 1 2 \sin\theta\left(n_2(\omega'_1)-n_1(\omega'_1)\right),\\
\label{delta n 2}&\Delta n_2(\theta,\omega_2')= \frac 1 2 \sin\theta\left(n_2(\omega'_2)-n_1(\omega'_2)\right)
\end{align}
\end{subequations}
{To avoid tedious notations, we set}
\begin{equation}
    \tilde n_l \equiv \tilde n_l(\theta,\omega_l'),\quad  \Delta n_l \equiv  \Delta n_l(\theta,\omega_l') 
\end{equation}
{with $l=1,2$.} Here $\tilde n_l$ can be viewed as the mean particle occupation number weighted by the mixing angle $\theta$ in the two baths with the same energy $\omega_l'$. And $\Delta n_l$ describes the difference in occupation number and therefore the degree of the nonequilibriumness which vanishes at the equilibrium case $T_1=T_2$ and $\mu_1=\mu_2$.

%\frac 1 2 \sin(\theta) \left(\alpha_2(\omega_1')-\alpha_1(\omega_1')\right)=\pm\frac 1 2 \sin(\theta) \left(\beta_2(\omega_1')-\beta_1(\omega_1')\right)=J\Delta n_1,

\subsubsection{Steady state solution for bosonic bath}

Directly solving the steady-state equation (\ref{eqa A rho}) for bosonic baths (the matrix $\mathcal A^\text{b}$ for bosonic baths has the elements in (\ref{A Bosonic})-(\ref{A Bosonic_end})) gives the steady solution {\cite{WWW18}}:
\begin{subequations}
\begin{align}
\label{Boson rho 11}\rho^\text{b}_{11}=&\frac 1 {\mathcal N^\text{b}}\left((1+\tilde n_1)(1+\tilde n_2)-\frac{\kappa^\text{b} s_1 s_2}{4(1+\tilde n_1+\tilde n_2)}\right), \\
\label{Boson rho 22}\rho^\text{b}_{22}=&\frac 1 {\mathcal N^\text{b}}\left(\tilde n_1(1+\tilde n_2)+\frac{\kappa^\text{b} s_2 s_3}{4(1+\tilde n_1+\tilde n_2)}\right), \\
\label{Boson rho 33}\rho^\text{b}_{33}=&\frac 1 {\mathcal N^\text{b}}\left(\tilde n_2(1+\tilde n_1)+\frac{\kappa^\text{b} s_1 s_4}{4(1+\tilde n_1+\tilde n_2)}\right), \\
\label{Boson rho 44}\rho^\text{b}_{44}=&\frac 1 {\mathcal N^\text{b}}\left(\tilde n_1\tilde n_2-\frac{\kappa^\text{b} s_3 s_4}{4(1+\tilde n_1+\tilde n_2)}\right)
\end{align}
\end{subequations}
{Here, $\kappa^\text{b}$ is defined as}
\begin{equation}
    \label{kappa b}\kappa^\text{b}=\frac{\tilde n_1+\tilde n_2+1}{(\tilde n_1+\tilde n_2+1)^2+(\Omega/2J)^2}
\end{equation}
{with the transition frequency given as}
\begin{equation}
    \label{def Omega}
    \Omega = \omega_2'-\omega_1'=\sqrt{\Delta\omega^2+\lambda^2}
\end{equation}
{Note that $\kappa^\text{b}$ is also defined in the population matrix $\mathcal A^\text{b}$ [see (\ref{def A aa bb}]. The other parameters are
\begin{subequations}
\begin{align}
\label{def a 1}&s_1=\Delta n_2-\Delta n_1(3+2\tilde n_1+2\tilde n_2), \\
&s_2=\Delta n_1-\Delta n_2(3+2\tilde n_1+2\tilde n_2), \\
&s_3=\Delta n_2+\Delta n_1(1+2\tilde n_1+2\tilde n_2), \\
\label{def a 4}&s_4=\Delta n_1+\Delta n_2(1+2\tilde n_1+2\tilde n_2),
\end{align}
\end{subequations}
{The normalization $\mathcal N^\text{b}$ is }
\eq
\mathcal N^\text{b}=(1+2\tilde n_1)(1+2\tilde n_2)-4\kappa_b\Delta n_1\Delta n_2(1+\tilde n_1+\tilde n_2)
\en
The superscript ``b'' stands for the bosonic reservoir setup. We omit the superscript ``ss'' for the steady-state notation.

{The steady state coherence is given by (\ref{def rho c}). We have}
\eq
\label{def rho 23}
\rho^\text{b}_{23}={\rho^\text{b}_{32}}^*=\frac 1 {\mathcal N_b}\left(\frac{\Delta n_1(1+2\tilde n_2)+\Delta n_2(1+2\tilde n_1)}{2(1+\tilde n_1+\tilde n_2)-i\Omega/J}\right)
\en
which vanishes if $\Delta n_1=\Delta n_2=0$, namely at the equilibrium case, off-diagonal terms {of the reduced density matrix at steady state} in the energy basis are always zero. The asterisk represents the complex conjugate. 

{Parameters $s_j$ with $j=1,2,3,4$ defined in (\ref{def a 1})-(\ref{def a 4}) solely characterize the nonequilibrium effects, since they all vanish at equilibrium cases.} We can view the second terms in (\ref{Boson rho 11})-(\ref{Boson rho 44}) as the nonequilibrium corrections which are proportional to {the square of coupling strength $J^2$ (since $\kappa^\text{b}$ (\ref{kappa b}) is proportional to $J^2$).} Although $J^2$ is negligible (weak coupling assumption), the nonequilibrium correction terms such as $J^2\Delta\tilde n^2_l$ are not bounded (only in the case of bosonic baths). {The first term of the steady-state population in (\ref{Boson rho 11})-(\ref{Boson rho 44}) can only reveal part of the nonequilibrium effects (the mean properties of the two baths) if we have intermediate temperature difference. In other words, the Lindblad can be used to characterize \textit{some} nonequilibrium results \cite{SPB08,WS11,CRQ13,WW19}. However, we will see later that the LGI function and MLGI function obtained by the Bloch-Redfield equation can have significant deviations from those characterized by the Lindblad, due to the dynamic differences.

{Equilibrium case ($T_1=T_2=T$) gives vanishing $\Delta n_i$ with $i=1,2$ (defined in (\ref{delta n 1}) and (\ref{delta n 2})). Therefore, we do not have coherence $\rho_{23}^\text{b}$ in the energy basis. We have} the equilibrium steady state:
\begin{align}
\label{Boson ss rho}\rho^\text{b,e}_{11}=&\frac 1 {\mathcal N^\text{b,e}}(1+\tilde n_1)(1+\tilde n_2),
& \rho^\text{b,e}_{22}=&\frac 1 {\mathcal N^\text{b,e}}\tilde n_1(1+\tilde n_2), \nonumber \\
\rho^\text{b,e}_{33}=&\frac 1 {\mathcal N^\text{b,e}}\tilde n_2(1+\tilde n_1),
& \rho^\text{b,e}_{44}=&\frac 1 {\mathcal N^\text{b,e}}\tilde n_1\tilde n_2
\end{align}
with the {normalization}
\begin{equation}
    \label{def N b e}
     \mathcal N^\text{b,e}=(1+2\tilde n_1)(1+2\tilde n_2)
\end{equation}
The superscript ``e'' reminds the equilibrium situation. The equilibrium steady state satisfies the canonical ensemble distribution. {At low temperatures, the system stays at the ground state with high probability: %$\rho^\text{b,e}_{11}\approx 1$.
\begin{equation}
    \lim_{T\rightarrow 0} \rho^\text{b,e}_{11}=1
\end{equation}
At high temperatures, we have the equally mixed state:}
\begin{equation}
\label{eqa rho high T}
    \lim_{T\rightarrow \infty} \rho^\text{b,e}_{jj}=\frac 1 4
\end{equation}
{with $j=1,2,3,4$.} 

%$\rho^\text{b,e}_{jj}\approx1/4$ with $j=1,2,3,4$.

%{ Steady state in extreme parameters, both equilibrium and non-equilibrium}.

\subsubsection{Steady state solution for fermionic bath}

{When the system is coupled with two fermionic baths (double quantum dots as the system), we have the steady-state population matrix $\mathcal A^\text{f}$ with matrix elements given in (\ref{A Fermionic})-(\ref{A Fermionic_end}). We can solve the reduced steady-state density matrix {\cite{WWW18}:}}
\begin{subequations}
\begin{align}
\label{Fermi rho 11}\rho^\text{f}_{11}=&(1-\tilde n_1)(1-\tilde n_2)-\frac{\kappa^\text{f}}{4}\left(\Delta n_1+\Delta n_2\right)^2, \\
\label{Fermi rho 22}\rho^\text{f}_{22}=&\tilde n_1(1-\tilde n_2)+\frac{\kappa^\text{f}}{4}\left(\Delta n_1+\Delta n_2\right)^2, \\
\label{Fermi rho 33}\rho^\text{f}_{33}=&\tilde n_2(1-\tilde n_1)+\frac{\kappa^\text{f}}{4}\left(\Delta n_1+\Delta n_2\right)^2, \\
\label{Fermi rho 44}\rho^\text{f}_{44}=&\tilde n_1\tilde n_2-\frac{\kappa^\text{f}}{4}\left(\Delta n_1+\Delta n_2\right)^2
\end{align}
\end{subequations}
with $\kappa^\text{f}$ defined as
\begin{equation}
\label{kappa f}
\kappa^\text{f}=\frac{1}{1+(\Omega/2J)^2}
\end{equation}
The superscript ``f'' means the fermionic bath setup. The corresponding coherence terms of the reduced density matrix are
\eq
\label{def rho 23 f}
\rho^\text{f}_{23}={\rho^\text{f}_{32}}^*=\frac{\Delta n_1+\Delta n_2}{2-i\Omega/J}
\en
which is zero if $\Delta n_1=\Delta n_2=0$. Nonequilibrium corrections (second terms in (\ref{Fermi rho 11})-(\ref{Fermi rho 44})) are proportional to $J^2\Delta n^2_l$. Unlike the bosonic environments, the particle occupation number difference $|\Delta n_l|<1/2$ is bounded in the fermionic case. Therefore, for $J\ll\Omega$, the Lindblad form can give reasonable description for effects of the nonequilibrium fermionic environments.

The equilibrium steady state ($T_1=T_2=T$ and $\mu_1=\mu_2=\mu$) is the special case of (\ref{Fermi rho 11})-(\ref{Fermi rho 44}) with $\Delta n_1=\Delta n_2=0$:
\begin{align}
\label{Fermi ss rho}\rho^\text{f,e}_{11}=&(1-\tilde n_1)(1-\tilde n_2),
& \rho^\text{f,e}_{22}=&\tilde n_1(1-\tilde n_2)\nonumber \\
\rho^\text{f,e}_{33}=&\tilde n_2(1-\tilde n_1),
& \rho^\text{f,e}_{44}=&\tilde n_1\tilde n_2
\end{align}
which satisfies the grand canonical ensemble distribution ($\mu\neq 0$). {At low chemical potentials, we reach
\begin{equation}
    \lim_{\mu\rightarrow 0} \rho^\text{f,e}_{11}=1
\end{equation}
The double dots are both empty. When we have high chemical potentials (with low temperatures), we reach
\begin{equation}
    \lim_{\mu\rightarrow \infty} \rho^\text{f,e}_{44}=1
\end{equation}
We have two occupied sites.}

%{ Steady state in extreme parameters, both equilibrium and non-equilibrium}.

\section{Leggett-Garg inequalities violations in equilibrium cases}

\label{sec:equ_LGI}

The two- and three-time LGIs defined in (\ref{def classical LGI 2}) and (\ref{def LGI ss}) impose constraints on the two-time and one-time probability distributions if they are obtained from the marginals of the joint three-time probabilities. In this section, we explore the LGIs for our two-qubit system with the equilibrium environments (two bosonic baths have the same temperatures or two fermionic baths have the same temperatures and chemical potentials). First we numerically show that not all LGIs are violated. We define the maximum of LGIs (MLGI) function. Note that the dynamics of the system is described by the Bloch-Redfield equation (\ref{QME}). It is difficult to give the closed expression of the time-evolution operator given any time $t$. In the following, we calculate the time-evolution operator in a perturbative way, namely the time-evolution operator in the zeroth and the first order of the coupling $J$ defined in (\ref{def J}). Such perturbation is valid if the system and bath are weakly coupled and the equilibrium temperature is relatively low. The weakly coupling condition is also the assumption for deriving the Bloch-Redfield equation (\ref{QME}). We study the MLGI function both analytically and numerically.

%We will check our analytical expressions with numerical results.

%LGI defined in (\ref{def LGI}) states the constraints on the probability distributions (simultaneously) for observables at different times. Quantum mechanical description is not equivalent to the classical probabilistic description. 

%The degree of LGI violation is characterized by the maximal value of LGI (MLGI). In this section, we study the LGI function $\mathcal I(t,\rho^\text{ss})$ (quantum version of LGI) defined in (\ref{LGI steady state}) and its maximal value $\mathcal I_\text{max}(\rho^\text{ss})$ defined in (\ref{def I max}) when the system is coupled with the equilibrium environments.

\subsection{Maximum of the Leggett-Garg Inequalities} 

\label{subsec MLGI}

The time-evolution operator for the reduced density matrix (the two qubit system) is generated by the superoperator $\mathcal W$, which is given by the Bloch-Redfield equation in Eq. (\ref{QME}). The superoperator $\mathcal W$ has two parts: the coherent evolution $\mathcal W_0$ and the dissipator $\mathcal W_d$. The coherent evolution is the unitary part defined by the von Neumann equation:
\begin{equation}
\label{def W 0}
    \mathcal W_0\rho_S=i[\rho_S,H_S]
\end{equation}
where $H_S$ is the system Hamiltonian defined in (\ref{def H S}). The dissipator originates from the interaction between the system and the environments, defined in (\ref{def dissipators}). Then the quantum correlation function $C_\text{q}(t)$ in (\ref{def Cq}) is calculated from
\begin{equation}
    C_\text{q}(t) = \re\tr\left(Qe^{\mathcal W t}Q\rho^\text{ss}\right)
\end{equation}
Note that to exclude the invasiveness on the macrorealism tests, the correlation function $C_\text{q}(t)$ is measured according to the INM protocol described in Sec. \ref{subsec:INM}.

The states $|g\rangle$ and $|e\rangle$ represent the local realism of the system. Although we can optimize the choice of local observables in order to maximize the violations of the LGIs \cite{Emary13}, we fix the observables to compare the violation of the LGIs with different environmental contexts. We choose the standard single-qubit dichotomic observable 
\begin{equation}
\label{def Q}
    Q=\sigma_{z,1}=(|g\rangle\langle g|-|e\rangle\langle e|)\otimes 1\!\!1_2
\end{equation}
to testify the LGIs. In the energy basis (\ref{def nonlocal basis}), the observable $Q$ has the matrix form
\eq
Q=\left(\begin{array}{cccc}
   1	& 0 & 0 & 0 \\
0	& -\cos\theta & \sin\theta & 0 \\
0	& \sin\theta & \cos\theta & 0 \\
0	& 0 & 0 & -1
\end{array} \right)
\en
with $\theta$ defined in (\ref{def theta}). 

First, we give an numerical results on the LGI functions $\mathcal I_{\pm}(t,\rho^\text{ss})$ and $\mathcal I_2(t,\rho^\text{ss})$ defined in (\ref{LGI steady state}) and (\ref{def LGI function 2}) respectively, see FIG. \ref{fig_LGIs}. We consider the case that two-qubit system is coupled with the equilibrium environments, where the two baths have the same temperatures and the same chemical potentials. Since the bosonic and fermionic baths have different statistics, they give different steady states (and different dynamics), see Eq. (\ref{Boson ss rho}) and (\ref{Fermi ss rho}), we draw the LGI functions for the bosonic and fermionic environments separately in FIG. \ref{fig_LGIs}. Clearly we can see that all the three LGI functions are damped due to the non-unitary evolution of the system. However, only the LGI function $\mathcal I_+(t,\rho^\text{ss})$ exceeds the classical limit either in the bosonic case or the fermionic case. We will give the analytical argument that only the LGI function $\mathcal I_+(t,\rho^\text{ss})$ is violated irrespective to the initial steady state in Sec. \ref{subsec LGI zero}.

\begin{figure}
	\includegraphics[width=0.85\columnwidth]{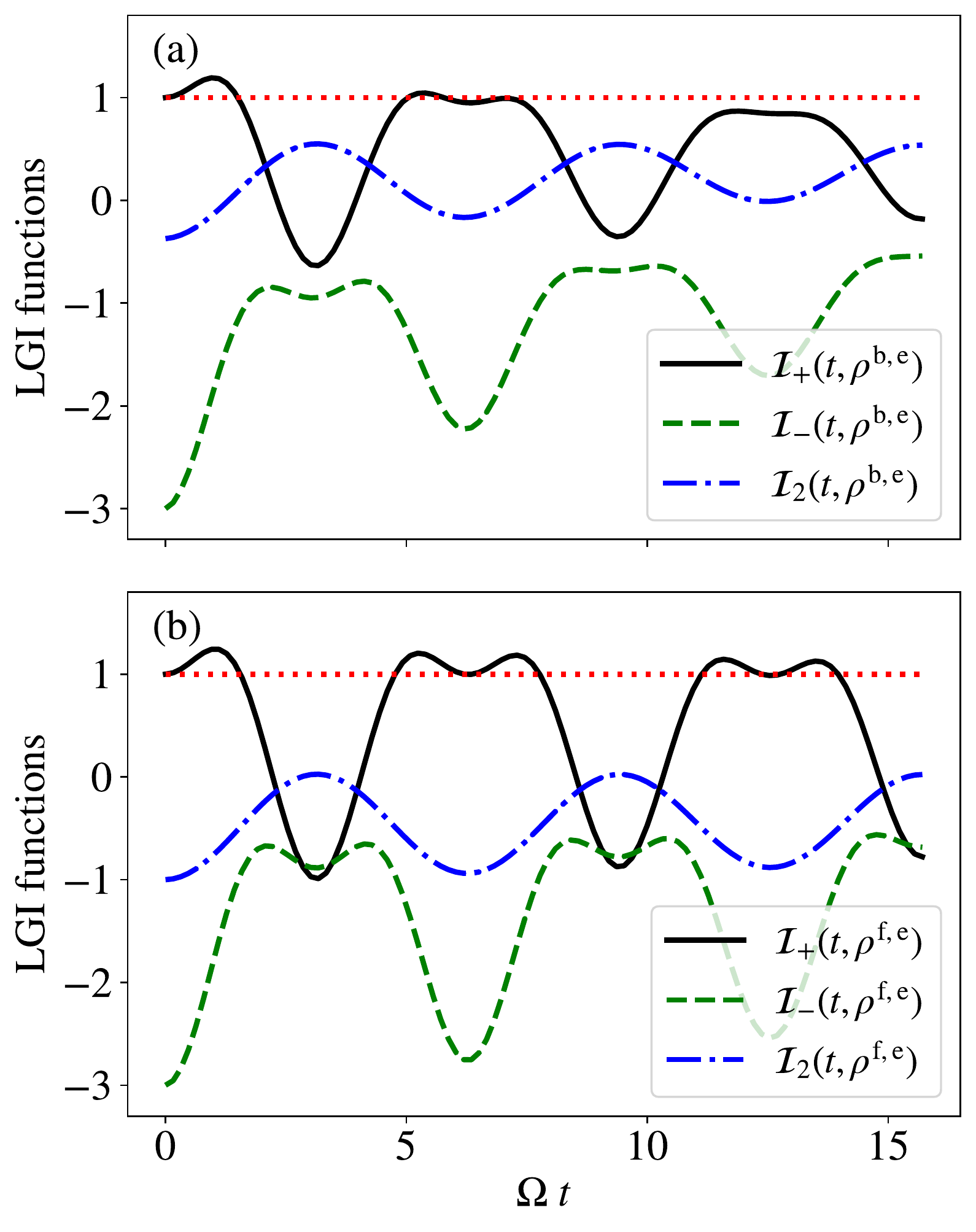}
	\caption{\label{fig_LGIs} LGI functions $\mathcal I_{\pm}(t,\rho^\text{ss})$ in (\ref{LGI steady state}) and $\mathcal I_2(t,\rho^\text{ss})$ in (\ref{def LGI function 2}) for the two-qubit system coupled with equilibrium environments. The oscillating frequency $\Omega$ is defined in Eq. (\ref{def Omega}). The parameters are set as $\lambda=\bar\omega$, $\theta=-\pi/2$ and $J=0.005\bar\omega$. (a) The two-qubit system is coupled with two bosonic baths with the same temperature $T=1.5 \bar\omega$. (b) The two-qubit system is coupled with two fermionic baths with the same temperature $T=1.5 \bar\omega$ and the same chemical potentials $\mu=\bar\omega$. The above red line suggests the violations of LGIs (beyond the classical description).} 
\end{figure}

To characterize the violations of the LGIs, we define the MLGI function by
\eq
\label{def I max}
\mathcal I_\text{max}(\rho^\text{ss}) =\max_{t\geq 0}\left\{\mathcal I_\pm(t,\rho^\text{ss}),\frac 1 2\mathcal I_2(t,\rho^\text{ss})\right\}-1
\en
The overall $1/2$ constant before the LGI function $\mathcal I_2(t,\rho^\text{ss})$ is for normalizing to the same maximal value with the LGI functions $I_\pm(t,\rho^\text{ss})$. The minimal value of $\mathcal I_\text{max}(\rho^\text{ss})$ is 0 since we have 
\begin{equation}
     \mathcal I_+(0,\rho^\text{ss}) = 1
\end{equation}
Violations of the LGIs in (\ref{LGI steady state}) and (\ref{def LGI function 2}) give $\mathcal I_\text{max}(\rho^\text{ss})>0$, which suggests that the quantum evolution (of the system) is beyond the classical descriptions (no joint probability distribution for observables at different times) \cite{ELN13}. Unitary evolution (full quantum description) of quantum states gives the maximal value of $\mathcal I_\text{max}(\rho^\text{ss})$. We have the range:
\begin{equation}
     0\leq\mathcal I_\text{max}(\rho^\text{ss})\leq \frac 1 2
\end{equation}
MLGI function is a natural quantity characterizing the maximal degree of LGIs violations. MLGI function is also the witness of the quantum evolution, which is beyond the classical probability evolution.

\subsection{Leggett-Garg Inequalities in the Zeroth Order of Coupling}

\label{subsec LGI zero}

The analytical form of the time-evolution operator $e^{\mathcal{W} t}$ given by the Bloch-Redfield equation in Eq. (\ref{QME}) is complicated. We consider the perturbation method. The time-evolution operator to the zeroth order of the coupling is the superoperator with the coherent evolution only. In zeroth order of the coupling $J$, the correlation function $C_\text{q}(t)$ defined in (\ref{def Cq}) is obtained from the coherent evolution $\mathcal W_0$. We have
\begin{equation}
\label{def c q 0}
     C^{(0)}_\text{q}(t,\rho) = 1-(1-\cos(\Omega t))\sin^2\theta(\rho_{22}+\rho_{33})
\end{equation}
The frequency $\Omega$ is defined in (\ref{def Omega}). The zeroth order correlation function $C^{(0)}_\text{q}(t)$ oscillates with the period $\Omega/2\pi$. We have the perfect oscillation (without decay) because the coupling to the environments is turned off. If we turn off the interqubit coupling, i.e., $\lambda=0$, which means the two-qubit systems are decoupled, we have
\begin{equation}
     C^{(0)}_\text{q}(t,\rho(\lambda=0))=1
\end{equation}
We have the perfect correlation because the observable $Q$ defined in Eq. (\ref{def Q}) commutes with the Hamiltonian $H_S$ when $\lambda=0$.

Let us firstly look up the zeroth order of two-time LGI function $\mathcal I_2(t,\rho^\text{ss})$ defined in Eq. (\ref{def LGI function 2}), which has the form
\begin{multline}
    \mathcal I^{(0)}_2(t,\rho^\text{ss}) = 2(\rho^\text{ss}_{11}-\rho^\text{ss}_{44})+2\cos\theta (\rho^\text{ss}_{33}-\rho^\text{ss}_{22})  \\
    +4\sin\theta \re\rho^\text{ss}_{23}+(1-\cos(\Omega t))\sin^2\theta(\rho^\text{ss}_{22}+\rho^\text{ss}_{33})-1
\end{multline}
Obviously, the maximum of $\mathcal I^{(0)}_2(t,\rho^\text{ss})$ is at $t = (2k+1)\pi/\Omega$ with $k=0,1,\cdots$. If we have the equilibrium environments, the steady state coherence in the energy basis vanishes, i.e., $\rho^\text{ss}_{23}=0$. One can verify that the steady state $\rho^\text{ss}$ either in the bosonic case or in the fermionic case has the relation $\rho^\text{ss}_{33}<\rho^\text{ss}_{22}$. Therefore, the detuning angle $\theta=-\pi/2$ gives the maximum of $\mathcal I^{(0)}_2(t,\rho^\text{ss})$ when $t = (2k+1)\pi/\Omega$. Based on the above optimal choices of the parameters, we have
\begin{multline}
    \mathcal I^{(0)}_2(t=\pi/\Omega,\rho^\text{ss}(\theta=-\pi/2))  \\
    =2(\rho^\text{ss}_{11}-\rho^\text{ss}_{44})+2(\rho^\text{ss}_{22}+\rho^\text{ss}_{33})-1
\end{multline}
It is easy to see that the two-time LGI function in the zeroth order $\mathcal I^{(0)}_2(t,\rho^\text{ss})$ does not violate the classical bound:
\begin{equation}
    \mathcal I^{(0)}_2(t,\rho^\text{ss}) \leq 1
\end{equation}
Single qubit with Rabi oscillation can violate the two-time LGIs dependent on the initial state \cite{Halliwell16,MHGJR19}. Here the two-time LGIs are not violated (two qubits coupled with the equilibrium environments) because of the nature of the steady state and the specific dynamics of the system.  

Given by the zeroth order correlation function $C^{(0)}_\text{q}(t,\rho)$ in Eq. (\ref{def c q 0}), the three-time LGI functions $\mathcal I_\pm(t,\rho^\text{ss})$ defined in Eq. (\ref{LGI steady state}) have the forms
\begin{multline}
\label{LGI 0}
\mathcal I_+^{(0)}(t,\rho^\text{ss})\\
=1+\left(2\cos(\Omega t)-\cos(2\Omega t)-1\right)\sin^2\theta(\rho^\text{ss}_{22}+\rho^\text{ss}_{33})
\end{multline}
\begin{multline}
\label{LGI 0 -}
\mathcal I_-^{(0)}(t,\rho^\text{ss})\\
=-3+(3-2\cos(\Omega t)-\cos(2\Omega t))\sin^2\theta (\rho^\text{ss}_{22}+\rho^\text{ss}_{33})
\end{multline}
It is easy to find that the zeroth order of three time LGI functions $\mathcal I_+^{(0)}(t,\rho^\text{ss})$ and $\mathcal I_-^{(0)}(t,\rho^\text{ss})$ have the first maximums at $t=\pi/(3\Omega)$ and $t=2\pi/(3\Omega)$ respectively. And we have
\begin{align}
\label{def I + 0}
    &\mathcal I_+^{(0)}(\pi/(3\Omega),\rho^\text{ss}) =  \frac 1 2 \sin^2\theta(\rho^\text{ss}_{22}+\rho^\text{ss}_{33})+1 \\
\label{def I - 0}
    &\mathcal I_-^{(0)}(2\pi/(3\Omega),\rho^\text{ss}) =  \frac 9 2 \sin^2\theta(\rho^\text{ss}_{22}+\rho^\text{ss}_{33})-4
\end{align}
Obviously we have 
\begin{equation}
\label{eq I + -}
\mathcal I_+^{(0)}(\pi/(3\Omega),\rho^\text{ss})\geq \mathcal I_-^{(0)}(2\pi/(3\Omega),\rho^\text{ss})
\end{equation}
Only at $\theta=-\pi/2$ and $\rho^\text{ss}_{22}+\rho^\text{ss}_{33}=1$ we have the equal sign. The classical description of the LGI functions defined in (\ref{def LGI ss}) is bounded by 1. We have $\mathcal I_+^{(0)}(t,\rho^\text{ss})>1$ during the time $0<t<\pi/(2\Omega)$ and $(3/2+2k)\pi/\Omega<t<(5/2+2k)\pi/\Omega$ with $k=0,1,\cdots$, as long as we have (nonzero) steady state population $(\rho_{22}+\rho_{33})$ and nonzero inter-qubit coupling $\lambda\neq0$.

According to the maximum of the three-time LGI functions $\mathcal I_\pm^{(0)}(t,\rho^\text{ss})$ in Eqs. (\ref{def I + 0}) and (\ref{def I - 0}), we have the zeroth order of the MLGI function defined in (\ref{def I max}):
\eq
\label{MLGI 0}
 \mathcal I_\text{max}^{(0)} (\rho^\text{ss})=\frac 1 2 \sin^2\theta(\rho^\text{ss}_{22}+\rho^\text{ss}_{33})
\en
Note that only the populations $\rho_{22}$ and $\rho_{33}$ contribute to the violation of LGIs, because states $|1\rangle$ and $|4\rangle$ defined in (\ref{def nonlocal basis}) are product states and their evolution admits classical descriptions (no off-diagonal terms). Also when $\lambda=0$ (the two-qubit systems are decoupled), eigenstates are all product states and the time-evolution operator for the local states is diagonalized (classical probabilistic descriptions). {When we have two identical qubits ($\omega_1=\omega_2$), if the population for states $|2\rangle$ and $|3\rangle$ is maximal, i.e., $\rho^\text{ss}_{22}+\rho^\text{ss}_{33}=1$,} the violation of LGI is saturated \cite{ELN13}. Notice that the symmetric qubits have eigenstates $|2\rangle$ and $|3\rangle$ as maximal entangled two qubit states (Bell-type states). We can see that the coherent evolution of the spatial maximal entangled states also has the maximal violation of LGIs. Then it is reasonable to view the maximum of the LGI functions defined in Eq. (\ref{def I max}) as a quantitative measure for quantum evolution. 

The zeroth order of LGI functions $\mathcal I_\pm^{(0)}(t,\rho^\text{ss})$ in Eqs. (\ref{def I + 0}) and (\ref{def I - 0}) oscillate without decay, which implies the information of the system is preserved (the LGI violations can occur at any long time interval between the two measurements). See FIG. \ref{fig_LGI} for the comparison between the zeroth order, the first order and the numerical LGI function $\mathcal I_+(t,\rho^\text{ss})$. The physical meaning of the zeroth order of LGI functions $\mathcal I_2^{(0)}(t,\rho^\text{ss})$ and $\mathcal I_\pm^{(0)}(t,\rho^\text{ss})$ is that the steady state (coupled with the environments) evolves coherently (decoupled from the environments) after the first measurement. The zeroth order LGI functions have the same form both in equilibrium and nonequilibrium cases (and the same for bosonic and fermionic baths), being different only in the steady state $\rho^\text{ss}$.

\begin{figure}
	\includegraphics[width=0.85\columnwidth]{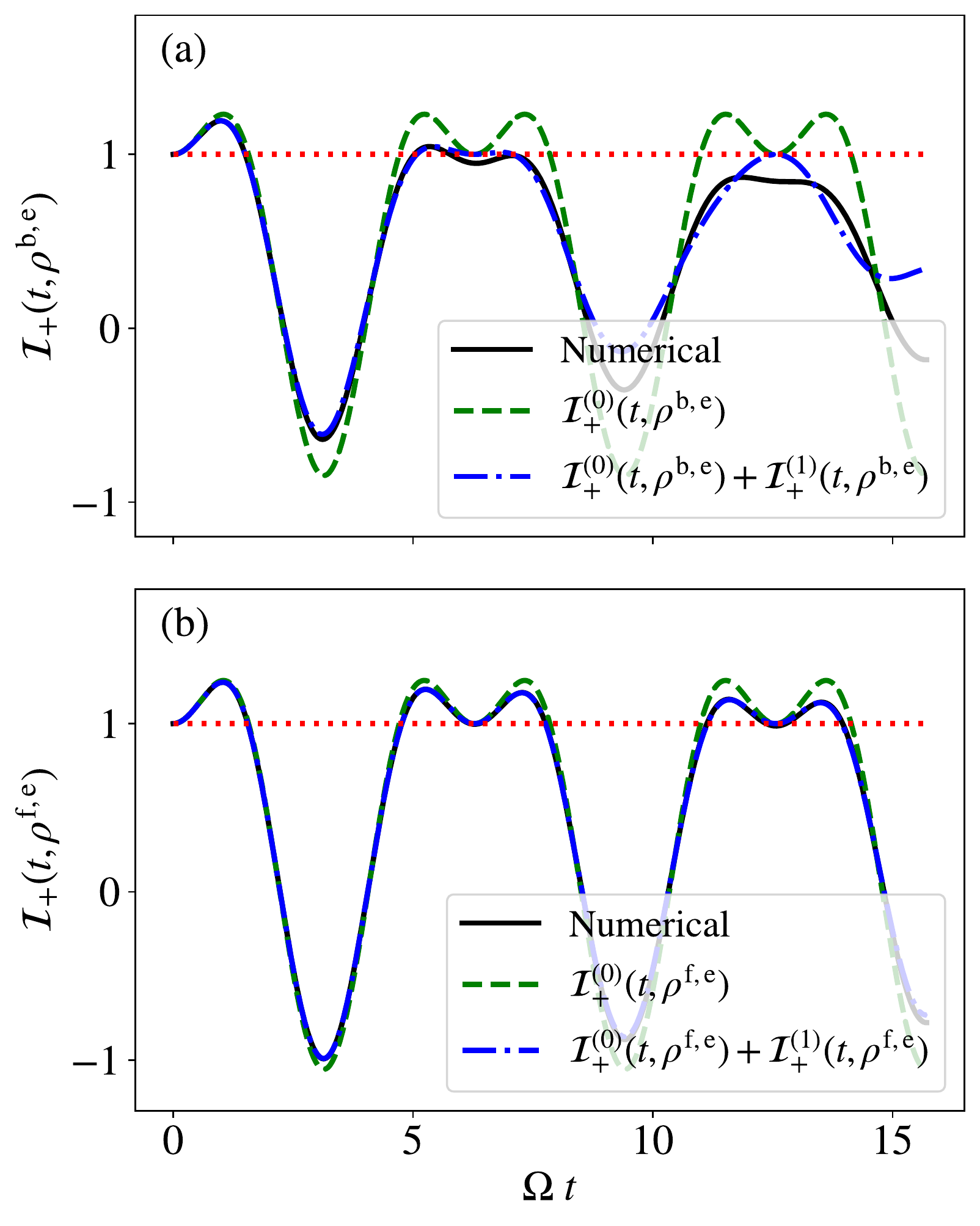}
	\caption{\label{fig_LGI} LGI function $\mathcal I_+(t,\rho^\text{ss})$ defined in (\ref{LGI steady state}) for the two-qubit system coupled with two (a) bosonic or (b) fermionic baths. The oscillating frequency $\Omega$ is defined in Eq. (\ref{def Omega}). Parameters are set as $\lambda=\bar\omega$, $\theta=-\pi/2$ and $J=0.005\bar\omega$. The solid line is the numerical results based on the Bloch-Redfield equation (\ref{QME}); the dashed line is the zeroth order $J$ of the LGI function $\mathcal I_+^{(0)}(t,\rho^\text{ss})$ in (\ref{LGI 0}); the dotted-dashed line is up to the first order $J$ of the LGI function: $\mathcal I_+^{(0)}(t,\rho^\text{ss})+\mathcal I_+^{(1)}(t,\rho^\text{ss})$; analytical expression of $\mathcal I_+^{(1)}(t,\rho^\text{b,e})$ and $\mathcal I_+^{(1)}(t,\rho^\text{f,e})$ can be found in (\ref{LGI 1}) and (\ref{LGI 1 fermi}) respectively. Above the red line suggests the violation of LGIs (beyond the classical description). (a) Two bosonic baths have the same temperature $T=1.5\bar\omega$. (b). Two fermionic baths have the same chemical potentials $\mu=\bar\omega$ and temperatures $T=1.5\bar\omega$.}
\end{figure}

\subsection{Leggett-Garg Inequalities in the First Order Coupling}

The superoperator $\mathcal W$ characterizing the time evolution of the system has the matrix form $\mathcal M$ (Liouville space), see (\ref{QME matrix}). Then we have the time evolution:
\begin{equation}
\label{e M t}
     |\rho_S(t)\rangle = e^{\mathcal M t}|\rho_S\rangle
\end{equation}
Note that the evolution operator $e^{\mathcal M t}$ is defined in the energy basis. The analytical form of $e^{\mathcal M t}$ is complicated, even when the matrix $\mathcal M$ can be diagonalized. Note that the real part of $\mathcal M$ is related to the overall coupling constant $J$ and the imaginary part of $\mathcal M$ is related to the coherent oscillation only, namely we have
\begin{equation}
\label{decompose M}
    \mathcal M=\mathcal M_0+\mathcal M_J
\end{equation}
with
\begin{equation}
    \mathcal M_J=\re \mathcal M,\quad \mathcal M_0=i\im \mathcal M,
\end{equation}
{see the matrix elements of $\mathcal M$ in (\ref{M})-(\ref{M_end}).}

We can apply the Zassenhaus formula \cite{Kimura07} in the first order of the coupling constant $J$:
\begin{align}
\label{e Mt approximation}
e^{\mathcal M t}=\left(1\!\!1+\sum_{n=1}^\infty \frac{t^n}{n!}\mathcal L_{\mathcal M_0}^{n-1}\mathcal M_J+\mathcal O(\mathcal M^2_J)\right)e^{\mathcal M_0 t}
\end{align}
where $\mathcal L$ is the commutator operator
\begin{equation}
    \mathcal L_{\mathcal M_0}\mathcal M_J=[\mathcal M_0,\mathcal M_J]
\end{equation}
Evolution $\exp(\mathcal M_0t)$ is the coherent part discussed before. The above expansion is based on the small coupling $J$. More specifically, the matrix elements of $\mathcal M_Jt$ should be much less than 1, formally 
\begin{equation}
\label{approximate condition}
     tJ n_l(\omega'_{1,2})\ll 1
\end{equation}
It suggests that we have good agreements for the first-order LGI functions when $t$ is small, for example see FIG. \ref{fig_LGI}. The approximation condition in Eq. (\ref{approximate condition}) requires the low temperature: $T< \omega'_{1,2}$ for bosonic reservoirs. Since the mean particle occupation number $n_l(\omega'_{1,2})$ is bounded in fermionic baths due to the exclusion principle, we have the approximation condition $t\ll 1/J$. The bounded mean particle occupation number in fermionic cases also suggests we have better approximation in the first order LGI function, see FIG. \ref{fig_LGI}. The systems are supposed to be ``classical'' and all LGIs are preserved when the environments are at high temperatures. We will numerically check that the LGIs are not violated in the high temperature regime.

\begin{figure}
	\includegraphics[width=0.75
	\columnwidth]{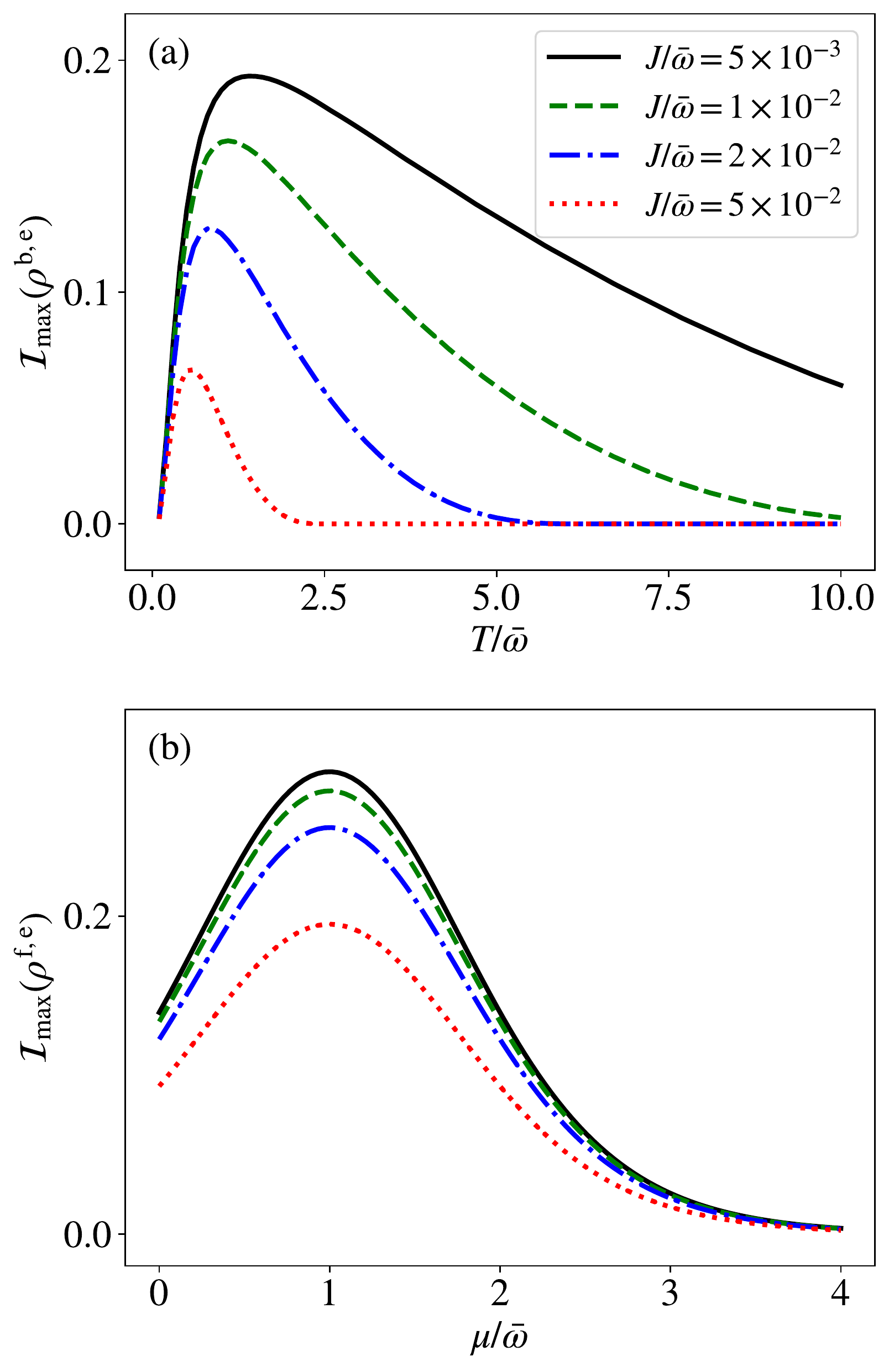}
	\caption{\label{MGLI_T} MLGI function {$\mathcal I_\text{max}(\rho^\text{b,e})$ defined in (\ref{def I max})} in terms of system-bath coupling $J$ and (a) equilibrium temperature $T=T_1=T_2$ for bosonic bath and (b) equilibrium chemical potential $\mu=\mu_1=\mu_2$ for fermionic bath with temperature $T=0.5\bar\omega$. Parameters are set as $\lambda=\bar\omega$ and $\theta=-\pi/2$. }
\end{figure}

\subsubsection{Equilibrium bosonic Bath}

The equilibrium steady state does not have the coherence terms in the energy basis, i.e., the population space and coherence space are decoupled in the Bloch-Redfield equation, see (\ref{QME}). However, the coherence in the localized state representation always survives if we have non-vanishing inter-qubit coupling $\lambda$. Matrix elements of $\mathcal M$ in (\ref{QME matrix}) is block diagonalized in the energy basis. Moreover, we can check that the real (dissipation) and imaginary (coherent evolution) parts in matrix $\mathcal M$ commute, i.e., 
\begin{equation}
\label{commutative M}
    [\mathcal M_0,\mathcal M_J]=0,\quad \text{if~} T_1=T_2,~\mu_1=\mu_2
\end{equation}
The expansion of the exponential (\ref{e Mt approximation}) is then greatly simplified as:
\eq
e^{\mathcal Mt}=\left(1\!\!1+\mathcal M_Jt+\mathcal O(\mathcal M_J)\right)e^{\mathcal M_0t},\quad \text{if~} T_1=T_2
\en
The above relation also holds for the equilibrium fermionic bath ($\mu_1=\mu_2=\mu$). 

We know that the zeroth order of the LGI functions oscillates without decay. Numerical results in FIG. \ref{fig_LGIs} show that the LGI functions are damped when the nonunitary evolution is included. Therefore, it is expected that the MLGI function in Eq. (\ref{def I max}) is equivalent to the maximum of the LGI function $I_+(t,\rho^\text{ss})$. Recall that the two-time LGIs are not violated with the unitary evolution. In the following, we concentrate on the first order LGI function $I_+(t,\rho^\text{ss})$ which gives the first order of the MLGI function. It is straightforward to find that
\begin{multline}
\label{LGI 1}
\mathcal I_+ ^{(1)}(t,\rho^\text{b,e})= 4tJ\sin^2\theta (\rho^\text{b,e}_{22}+\rho^\text{b,e}_{33})\left(n(\omega_1')+n(\omega_2')+1)\right) \\
 \times (\cos(2\Omega t)-\cos(\Omega t))
\end{multline}
with notation $n(\omega_{1,2}')=n_1(\omega_{1,2}')=n_2(\omega_{1,2}')$ (because we have $T_1=T_2=T$). We omit the superscript ``ss'' for simplicity.

The first-order correction $\mathcal I_+^{(1)}(t,\rho^\text{b,e})$ is proportional to $tJn(\omega'_{1,2})$ and oscillates in period $2\pi/\Omega$. We know that the zeroth-order LGI function $\mathcal I_+^{(0)}(t,\rho^\text{b,e})$ in (\ref{LGI 0}) has the maximums at $(1/(3\Omega)+2k)\pi$ and $(5/(3\Omega)+2k)\pi$ with the integer number $k\geq0$. We can check that those extreme points always give negative $\mathcal I_+^{(1)}(t,\rho^\text{b,e})$ in (\ref{LGI 1}). Also note that $\mathcal I_+^{(1)}(t,\rho^\text{b,e})$ linearly increases with time $t$. Therefore, the LGI function $\mathcal I_+(t,\rho^\text{b,e})$ is decaying due to the incoherent evolution (system coupled with the environments). After a threshold time {(the interval time between two measurements)}, the inequality will not be violated. The first order correction is proportional to $\sin^2\theta$ and the population sum $\rho^\text{b,e}_{22}+\rho^\text{b,e}_{33}$. It is interesting to see that when $\lambda=0$ or $\rho^\text{b,e}_{22}+\rho^\text{b,e}_{33}=0$, the first order correction is also zero. Although the two qubits are decoupled $\lambda=0$, the LGI functions are still expected to decay due to coupling with the environments {(information leaked into environments)}. However such effects are beyond the first order $J$, which means that coupled two-qubit system is more easily affected by the environments (more fragile than the local states). See FIG. \ref{fig_LGI} for the comparison among numerical calculations, its zeroth order and first order of LGI function $\mathcal I_+(t,\rho^\text{e,b})$.

\begin{figure}
	\includegraphics[width=0.75
	\columnwidth]{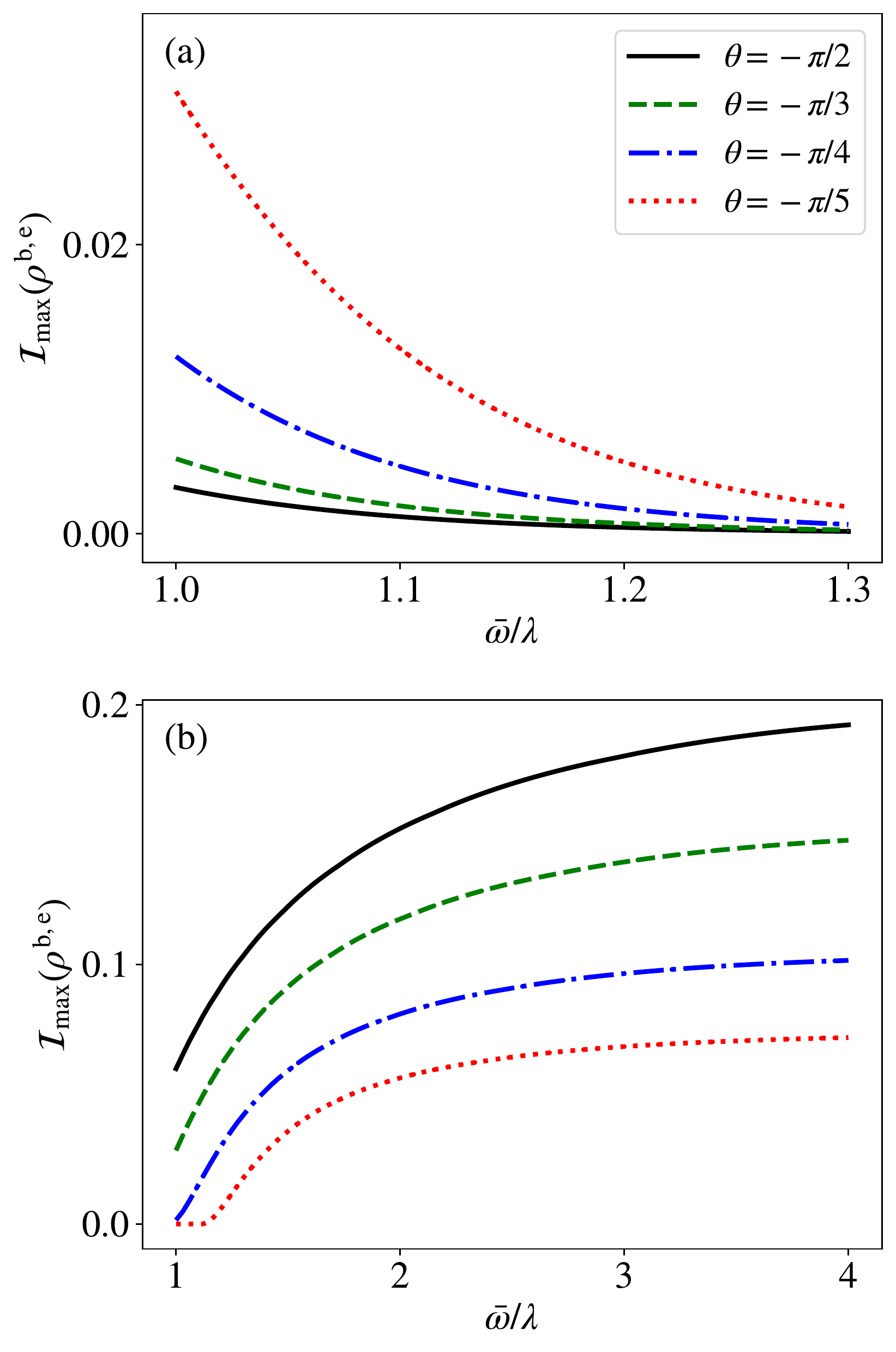}
	\caption{\label{MLGI_theta_boson} MLGI function $\mathcal I_\text{max}(\rho^\text{b,e})$ defined in (\ref{def I max}) in terms of the mean energy splitting $\bar\omega$ and the detuning angle (with fixed $\lambda=1$) characterizing the asymmetry of the two-qubit system. The coupling is set as $J=0.005$. The two bosonic bath have the same (a) temperatures $T=0.1$ or (b) temperatures $T=10$.}
\end{figure}

The zeroth order of LGI function $\mathcal I_+^{(0)}(t,\rho^\text{b,e})$ in (\ref{LGI 0}) has the maximum at $t=\pi/(3\Omega)$. And the zeroth order MLGI function is equivalent to the maximum of the zeroth order of LGI function $\mathcal I_+(t,\rho^\text{b,e})$. We can safely approximate the first order of MLGI function $\mathcal I_\text{max}(\rho^\text{b,e})$ defined in (\ref{def I max}) by the first order of the LGI function $\mathcal I_+(t,\rho^\text{b,e})$ at $t=\pi/(3\Omega)$:
\begin{equation}
    \mathcal I^{(1)}_\text{max}(\rho^\text{b,e}) \approx \mathcal I_+^{(1)}(t=\pi/(3\Omega),\rho^\text{b,e}),
\end{equation}
{which gives}
\begin{multline}
\label{MLGI 1}
\mathcal I^{(1)}_\text{max}(\rho^\text{b,e})\\
=-\frac{4\pi J}{3\Omega}\sin^2\theta (\rho^\text{b,e}_{22}+\rho^\text{b,e}_{33})\left(n(\omega_1')+n(\omega_2')+1)\right)
\end{multline}
The first order $\mathcal I^{(1)}_\text{max}(\rho^\text{b,e})$ is always negative irrespective to the bath parameters. Increasing the coupling $J$ will always decrease the MLGI function, namely moving the system towards the classical description. 

To better see how the temperature $T$ affects $\mathcal I^{(1)}_\text{max}(\rho^\text{b,e})$, we can approximate the mean particle occupation number as
\begin{equation}
   n(\omega'_{1,2}) \approx e^{-\omega'_{1,2}/T}
\end{equation} 
{if $T\ll \omega'_{1,2}$. Then we have}
\eq
\label{MLGI 0 1}
 \mathcal I^{(0)}_\text{max}(\rho^\text{b,e})+\mathcal I^{(1)}_\text{max}(\rho^\text{b,e})\approx\alpha\sin^2\theta\left(\frac 1 2 -\frac{4\pi J}{3\Omega}\right)
\en
with 
\begin{equation}
\label{def alpha}
    \alpha=2e^{-\bar\omega/T}\cosh\left(\frac{\Omega}{2T}\right)
\end{equation}
{Here, the parameter $\alpha$ has the physical meaning: proportional to the population term, i.e., $\alpha\propto (\rho^\text{b,e}_{22}+\rho^\text{b,e}_{33})$. Increasing the temperature $T$ (when $T\ll \omega'_{1,2}$) gives the LGI violation enhancement. Approximately zero temperature does not have LGI violation} because the system will always be in the ground state $|1\rangle$, and the time evolution of the ground state is trivial. The population of excited states is the key for the LGIs violations. If we do not have the non-local ground state, but have non-local excited state, then the bath temperature, which gives the excitation, is beneficial for enhancing the LGIs violations. 

High temperature environment deteriorates the quantumness of the system and the LGIs are expected to be preserved. For example, if the temperature is around $T\sim 10\omega'_{1,2}$ (approximation condition in (\ref{approximate condition}) can still be valid), this gives almost even distribution of the populations, see (\ref{eqa rho high T}). Further increasing the temperature does not increase the population sum $\rho^\text{b,e}_{22}+\rho^\text{b,e}_{33}$. However, increasing the temperature {(when $T\sim 10\omega'_{1,2}$)} will dramatically increase $n(\omega_1')+n(\omega_2')$ and therefore {the first order correction $\mathcal I^{(1)}_\text{max}(\rho^\text{b,e})$ in (\ref{MLGI 1}) will decrease.} At intermediate temperature $T\sim \omega'_{1,2}$, the MLGI function $\mathcal I_\text{max}(\rho^\text{b,e})$ is compromised between the non-local state population and the decoherence. We can numerically explore the non-monotonic relation between $\mathcal I_\text{max}(\rho^\text{b,e})$ and temperature $T$, see FIG. \ref{MGLI_T}. We can also check the monotonic relation between $\mathcal I_\text{max}(\rho^\text{b,e})$ and the coupling $J$, see FIG. \ref{MGLI_T}. Stronger coupling $J$ leads to that the environment destroys the coherence (in energy basis) faster. Increasing the temperature can lead to both the excited nonlocal states and the decoherence effect, giving rise to a non-monotonic behavior. 

When the two qubits have different transition frequencies $\Delta \omega \neq0$, the eigenstates $|2\rangle$ and $|3\rangle$ (defined in Eq. (\ref{def nonlocal basis})) will become less entangled. The detuning angle $\theta$ in (\ref{def theta}) is $\theta=-\pi/2$ when $\Delta \omega =0$. Turning off the coupling of the two qubit systems ($\lambda=0$) leads to the classical descriptions (probabilistic equation) of the time evolution of the system. The inter-qubit coupling strength $\lambda$ has the monotonic relation with the MLGI function $\mathcal I_\text{max}(\rho^\text{b,e})$ (smaller $\lambda$ gives smaller $\mathcal I_\text{max}(\rho^\text{b,e})$). However, $\mathcal I_\text{max}(\rho^\text{b,e})$ does not have a monotonic relation with $\Delta \omega$ (changing $\theta$ with fixed $\lambda$), see FIG. \ref{MLGI_theta_boson}. 

Recall that in the low temperature regime ($T\ll \omega'_{1,2}$), the MLGI function up to the first order of $J$ has the form (\ref{MLGI 0 1}), where $\alpha$ is proportional to the population term. If we increase the transition frequency difference $\Delta\omega$ (with fixed average $\bar\omega$), the oscillation frequency $\Omega$ in (\ref{def Omega}) will increase. Then $\alpha$ is larger with larger $\Omega$. In other words, we can increase the population sum $\rho^\text{b,e}_{22}+\rho^\text{b,e}_{33}$ by increasing the frequency difference $\Delta\omega$. In a more intuitive understanding, if the ground state is dominated, population $\rho^\text{b,e}_{22}$ will increase if we lower the energy level of state $|2\rangle$. Note that the eigenstate $|2\rangle$ has the energy $\omega_1'=\bar\omega-\sqrt{\Delta\omega^2+\lambda^2}/2$. Besides, we can enhance the MLGI function in the low temperature regime by lowering the average frequency $\bar\omega$ (which is physically equivalent to increase the equilibrium temperature of the two baths). See the numerical results in FIG. \ref{MLGI_theta_boson} which are consistent with the analytical arguments.

If the equilibrium temperature $T$ is high ($T\gg \omega'_{1,2}$), the population sum $\rho^\text{b,e}_{22}+\rho^\text{b,e}_{33}$ is saturated around $1/2$. The overall factor $\sin^2\theta$ is dominated in $\mathcal I_\text{max}^{(1)}(\rho^\text{b,e})$ with the expression in (\ref{MLGI 1}). Decreasing $\Delta\omega$ with fixed $\lambda$ or increasing $\lambda$ with fixed $\Delta\omega$ will both enhance $\mathcal I_\text{max}(\rho^\text{b,e})$. The more physical intuitive explanation is that the maximal entangled nonlocal states violate the local realism maximally (because the local states have the classical time evolution). Therefore, we want both qubits to have the same frequencies in order to form the Bell-type eigenstates. Also in high temperature case, lowering temperature is beneficial to test LGI violation, see FIG. \ref{MGLI_T}. Then we can enhance $\mathcal I_\text{max}(\rho^\text{b,e})$ by increasing the mean energy splitting $\bar\omega$. We numerically checked the above arguments, see FIG. \ref{MLGI_theta_boson}.

\subsubsection{Equilibrium fermionic Bath}

\begin{figure}
	\includegraphics[width=0.75
	\columnwidth]{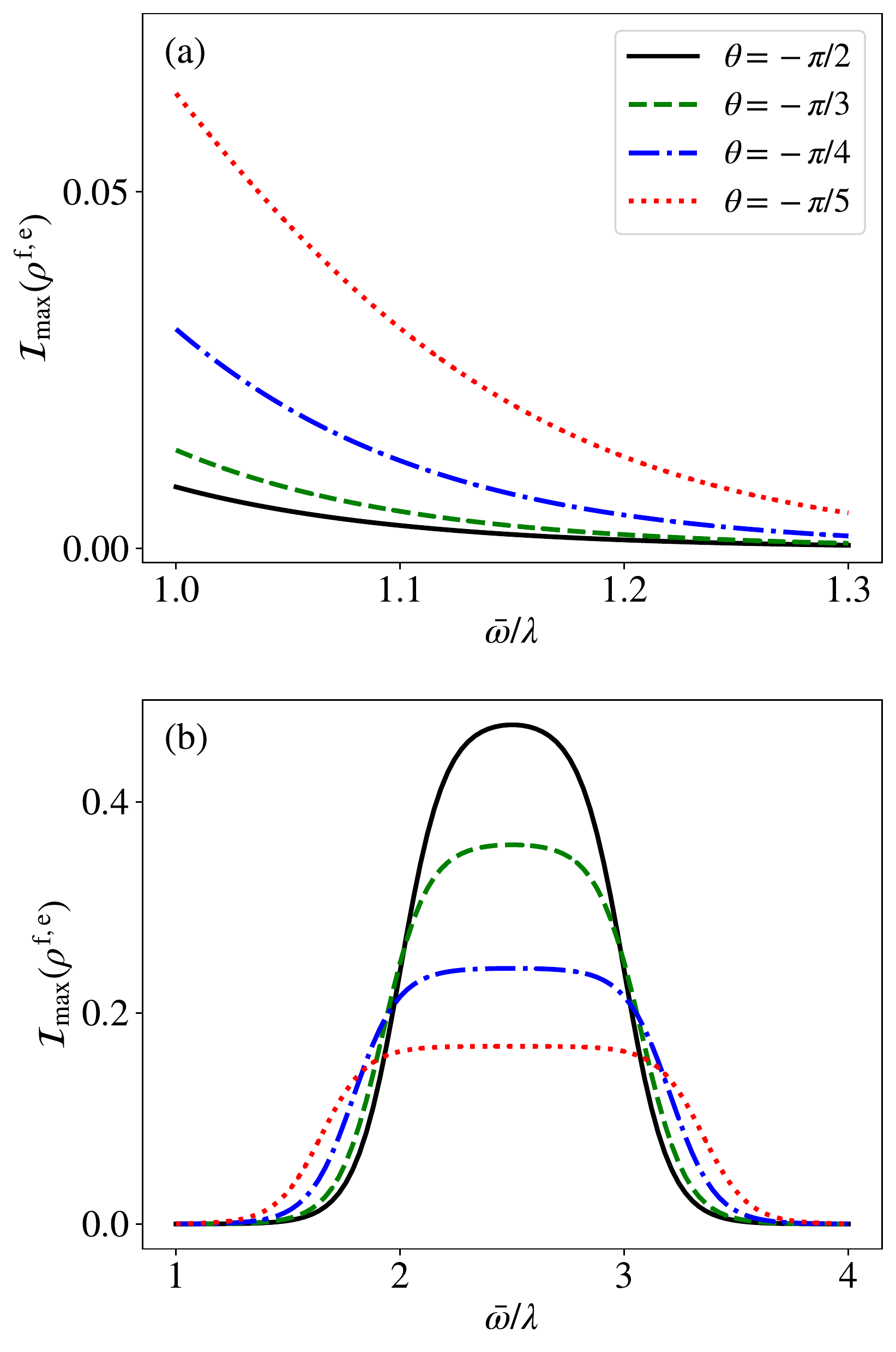}
	\caption{\label{MLGI_theta_fermi} The MLGI function $\mathcal I_\text{max}(\rho^\text{b,e})$ defined in (\ref{def I max}) in terms of the mean energy splitting $\bar\omega$ and the detuning angle (with fixed $\lambda=1$) characterizing the asymmetry of the two-qubit system. The coupling is set as $J=0.005$. And the temperature is $T=0.1$. The two fermionic baths have the same (a) chemical potential $\mu=0.1$ or (b) $\mu=2.5$.}
\end{figure}

We study the LGIs violations in two-qubit system (toy model of double quantum dots) coupled with equilibrium fermionic environments. We limit in the low-temperature regimes in following parts (for the fermionic environments). Similar to the equilibrium bosonic case, population space and coherent space in energy basis are decoupled {in the Bloch-Redfield equation (\ref{QME})}, which implies {the commutative relation (\ref{commutative M})}. {Note that the time-evolution matrix $\mathcal M$ defined in (\ref{QME matrix}) has different matrix elements for bosonic and fermionic setups, because of difference in Bose-Einstein distribution and Fermi-Dirac distribution.}

In the first order coupling constant $J$, the LGI function $\mathcal I_+(t,\rho^\text{f,e})$ (\ref{LGI steady state}) has the form
\eq
\label{LGI 1 fermi}
 \mathcal I_+^{(1)}(t,\rho^\text{f,e})= 4tJ\sin^2\theta (\rho^\text{f,e}_{22}+\rho^\text{f,e}_{33})(\cos(2\Omega t)-\cos(\Omega t))
\en
The steady state $\rho^\text{f,e}$ has the form in (\ref{Fermi ss rho}). We can safely approximate the first order MLGI function $\mathcal I_\text{max}(\rho^\text{f,e})$ defined in (\ref{def I max}) by 
\begin{align}
\label{MLGI 1 fermi}
    \mathcal I^{(1)}_\text{max}(\rho^\text{f,e}) & \approx  \mathcal I_+ ^{(1)}(t=\pi/(3\Omega),\rho^\text{f,e}) \nonumber\\
    &  =-\frac{4\pi J}{3\Omega}\sin^2\theta (\rho^\text{f,e}_{22}+\rho^\text{f,e}_{33})
\end{align}
Note that the zeroth order of LGI function $\mathcal I^{(0)}_+(t,\rho^\text{f,e})$ has a maximum at $t=\pi/(3\Omega)$ and the zeroth order of MLGI function is equivalent to the maximum of the zeroth order of LGI function $\mathcal I_+(t,\rho^\text{f,e})$.

The first order correction $\mathcal I^{(1)}_\text{max}(\rho^\text{f,e})$ is always negative. Both the zeroth and the first order of MLGI function are even functions of $\bar\omega-\mu$, indicating local minimum or maximum at $\mu=\bar\omega$. If $T\ll\bar\omega$ (the low temperature regime), we have the population sum
\eq
\rho^\text{f,e}_{22}+\rho^\text{f,e}_{33} \approx \frac{1}{\exp\left(\frac{2|\bar\omega-\mu|-\Omega}{2T}\right)+1}
\en
The resonant point $\bar\omega=\mu$ gives the maximal population of the nonlocal states (eigenstates $|2\rangle$ and $|3\rangle$ are entangled states). The ground state has the particle occupation number 0; the first and the second excited states have the particle occupation number 1 and the highest excited state has the particle occupation number 2. At the resonant point $\bar\omega=\mu$, the system is driven into the first and the second excited states. {Therefore the zeroth order of the MLGI function (\ref{MLGI 0}) has the global maximum at $\bar\omega=\mu$.} Low chemical potential reservoir can not {pump electrons into dots (described by the state $|1\rangle$); but high chemical potential reservoir makes all dots occupied (described by the state $|4\rangle$). Both the evolutions of states $|1\rangle$ and $|4\rangle$ admit local realism descriptions.} See FIG. \ref{MGLI_T} for the numerical results of $\mathcal I_\text{max}(\rho^\text{f,e})$ in terms of the coupling constant and the chemical potential $\mu$.

When the ground state is dominated (two empty sites), we can increase the population sum of the first and the second excited states by lowering the mean energy splitting $\bar\omega$ or increasing the local energy level difference $\Delta\omega$. We can reversely argue the high chemical potential case. The above arguments are similar to the bosonic case. Consider the intermediate chemical potential regime $\mu\sim\bar\omega$: population sum of the first and the second excited state is almost saturated. Then changing the energy difference $\Delta\omega$ does not affect the population sum of states $|2\rangle$ and $|3\rangle$. However, states $|2\rangle$ and $|3\rangle$ will become less entangled (more local). Then we always have larger MLGI function $\mathcal I_\text{max}(\rho^\text{b,e})$ in (\ref{def I max}) when $\theta=-\pi/2$, see the numerical results in FIG. \ref{MLGI_theta_fermi}. On the contrary, away from the resonant point $\mu\sim\bar\omega$, we always have MLGI $\mathcal I_\text{max}(\rho^\text{b,e})$ enhancement by increasing the local energy level difference $\Delta\omega$, although we have less entangled excited states. 

%Consider the intermediate chemical potential $\mu\sim\bar\omega$: numerical results, see Fig. \ref{MLGI_theta_fermi}, suggest that when $\bar\omega<\mu$, we have threshold $\omega_1'\sim\mu$ to determine that the detuning $|\theta+\pi/2|$ is beneficial for MLGI or not; when $\bar\omega<\mu$, the threshold becomes $\omega_2'\sim\mu$. This results from the competition between forming the maximal excited states and more excited state population. More specifically, when $\bar\omega\sim\bar \mu$, the population sum of the first and the second excited state is almost saturated. Then the detuning angle cannot change the population sum much. Instead, it can determinine the entangled degree of the first and the second excited states. Therefore, if we want $\sin^2\theta\approx 1$ then $\bar\omega\sim \bar\mu$. Away from the resonant point,  detuning the qubit energy levels can give arise to more nonlocal excited state population although they are not maximal entangled states.

%\section{MLGI enhanced by nonequilibrium condition}
\section{Legget-Garg inequalities violations in the nonequilibrium cases}

\label{sec:nonequ_LGI}

In this section, we explore how the nonequilibrium condition, characterized by the temperature difference $\Delta T=T_2-T_1$ for the bosonic bath or chemical potential difference $\Delta\mu=\mu_2-\mu_1$ for the fermionic bath, contributes to the violation of LGIs. The nonequilibrium environment suggests the heat or particle current flowing through the system \cite{CRQ13,ZW14,WWCW18,QRRP07,WS11,DSH12,ZW15,WWW18,WW19}. Consequently, there will be nonzero thermodynamic dissipation characterized by the entropy production rate \cite{ZW14,ZW15}. Violations of the LGIs given by the nonequilibrium steady state also imply the quantum transport phenomenon \cite{LECN10}. Firstly, we show that the heat or particle current and entropy production rate through the system increases monotonically in terms of the environment bias (temperature difference $\Delta T$ or chemical potential difference $\Delta \mu$). Then analytically, we derive the LGI function $\mathcal I_+(t,\rho^\text{ss})$ defined in (\ref{LGI steady state}) in the first order of coupling constant $J$ (approximation valid by the condition (\ref{approximate condition})). Based on that, we give analytical analysis on the first order of the MLGI function $\mathcal I_\text{max}(\rho^\text{ss})$ in (\ref{def I max}) in the nonequilibrium cases. We numerically check the analytical results and numerically explore the relative high-temperature cases.

\subsection{Heat/Particle Current and Entropy Production Rate}

\begin{figure}
	\includegraphics[width=0.75\columnwidth]{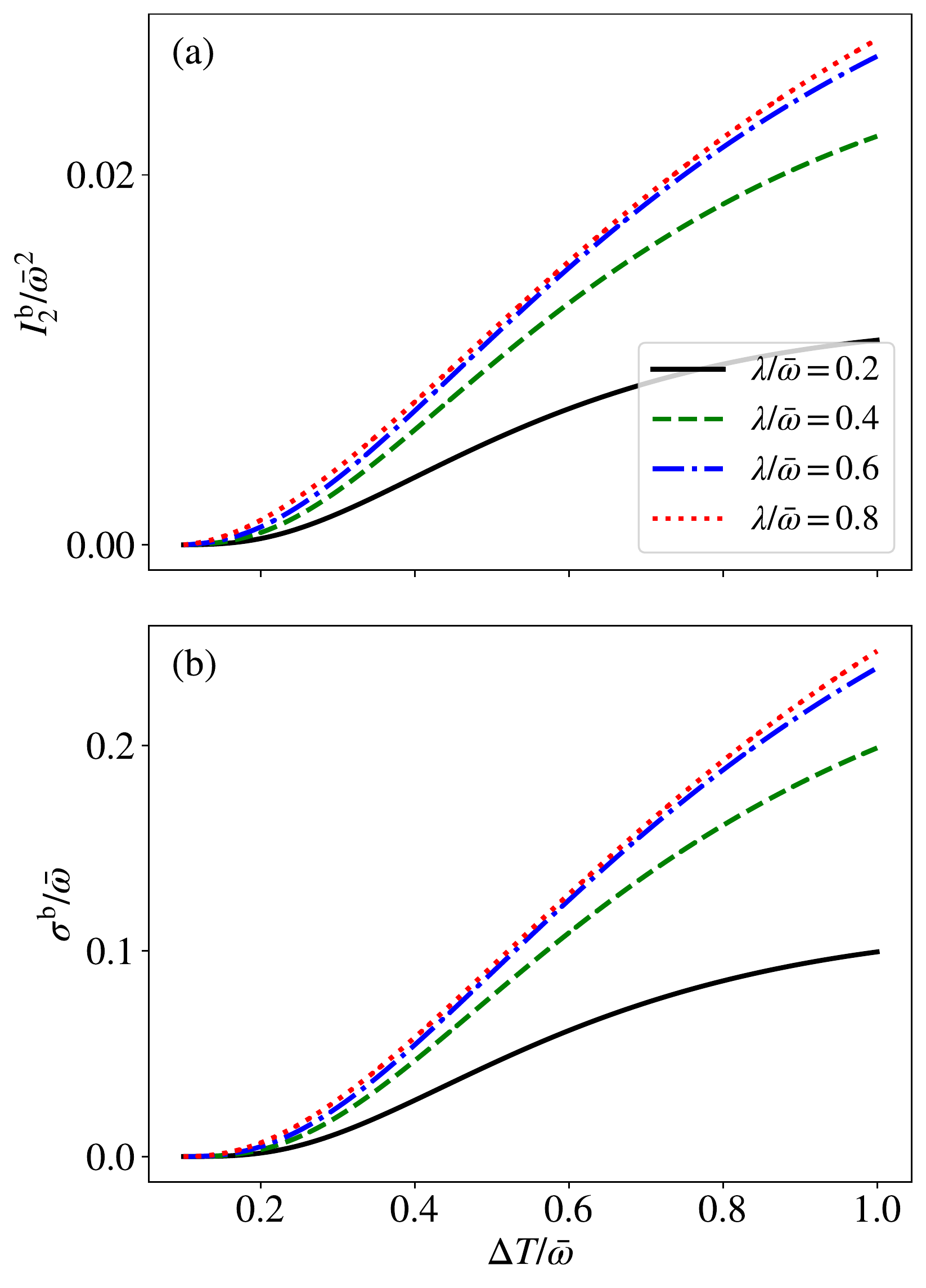}
	\caption{\label{FIG b epr} (a) Heat current $I_2^\text{b}$ defined in (\ref{def I b}) and (b) entropy production rate $\sigma^\text{b}$ defined in (\ref{def S b}) at steady state in terms of the nonequilibrium condition $\Delta T=T_2-T_1$ with fixed $T_1=0.1\bar\omega$. Parameters are set as $\theta=-\pi/2$ and $J=0.1\bar\omega$.}
\end{figure}

When the system reaches the steady state (under nonequilibrium environments), there is a constant heat or particle current flowing through the system (from the high temperature or chemical potential bath to the low temperature or chemical potential bath). The heat current is characterized by the energy change with the two environments. We have
\begin{equation}
    \tr\left(\frac{d\rho}{dt} H_S\right) = \sum_{l=1}^2 I^\text{b}_l
\end{equation}
where $I^\text{b}_l$ with $l=1,2$ are the heat currents through the bath 1 or 2. Superscript b reminds that we have bosonic environments. Positive $I^\text{b}_l$ means that the heat current is flowing into the system. According to the Bloch-Redfield equation (\ref{QME}), the heat currents are related with dissipators:
\begin{equation}
\label{def I b}
    I^\text{b}_l = \tr(\mathcal D_l[\rho]H_S)
\end{equation}
Steady state means that the energy of the system does not change. Therefore the two currents have the same magnitude but different directions:
\begin{equation}
    I^\text{b}_1+I^\text{b}_2=0
\end{equation}
If $T_2>T_1$, we have $I^\text{b}_2>0$ and $I^\text{b}_1<0$.

Given by the analytical results of the nonequilibrium steady state $\rho^\text{b}$, see (\ref{Boson rho 11})-(\ref{Boson rho 44}) and (\ref{def rho 23}), we can find the steady state heat current. We give the analytical steady state heat current in Appendix \ref{appen C}. We plot the heat current $I^\text{b}_2$ in terms of the nonequilibrium condition $\Delta T=T_2-T_1$ (with fixed $T_1$) in FIG. \ref{FIG b epr}. The current magnitude increases monotonically with the temperature bias $\Delta T$. The inter-qubit coupling (characterized by strength $\lambda$) plays an important role in heat transport \cite{WWCW18,WWW18,WW19}. As $\lambda =0$, the two-qubit system is decoupled and the two environments are separated. No heat current flows through the system. Increasing the inter-qubit coupling strength $\lambda$ can enhance the heat flow, see FIG. \ref{FIG b epr}.

%At far from equilibrium case, for example $T_1\rightarrow 0$ and $T_2\rightarrow \infty$, we have almost equally mixed populations $\rho^\text{b}_{jj}\approx1/4$ with $j=1,2,3,4$. And the heat current can be approximated as
%\begin{equation}
%    I_2^\text{b} \approx  J \bar\omega(\rho^\text{b}_{23}+\rho^\text{b}_{32}+2)
%\end{equation}
%assuming $\omega_1=\omega_2$ and constant symmetric coupling spectral $J$ (\ref{constant J}). The coherence terms $\rho^\text{b}_{23}$ and $\rho^\text{b}_{32}$ in energy basis are generated from nonequilibrium conditions $\Delta T$. At far from equilibrium case, the heat current is proportional to the energy basis coherence. Note that Lindblad form of the description does not give energy basis coherence in nonequilibrium environments. 

Since the two reservoirs are assumed to be infinitely large compared with system, the steady state can be maintained for a very long time. We assume that the equilibrium temperatures of the two baths are always constant. The nonzero constant heat transfer at steady state implies that we have constant entropy production rate defined as
\begin{equation}
\label{def S b}
 \sigma^\text{b} = -\frac{I_1^\text{b}}{T_1}-\frac{I_2^\text{b}}{T_2} = I_2^\text{b}\left(\frac 1 {T_1}-\frac 1 {T_2}\right)
\end{equation}
The same magnitude $|\Delta T|$ gives the same entropy production rate $\sigma^\text{b}>0$. The entropy production rate also monotonically increases with nonequilibrium condition, see FIG. \ref{FIG b epr}. The stronger inter-qubit coupling gives larger entropy production rate, since we have larger steady-state heat current. At far from equilibrium case, for example $T_1\rightarrow 0$ and $T_2\rightarrow \infty$, the entropy production rate is proportional to the heat current:
\begin{equation}
    \sigma^\text{b} \approx  \frac{I_2^\text{b}}{T_1}
\end{equation}

\begin{figure}
	\includegraphics[width=0.75\columnwidth]{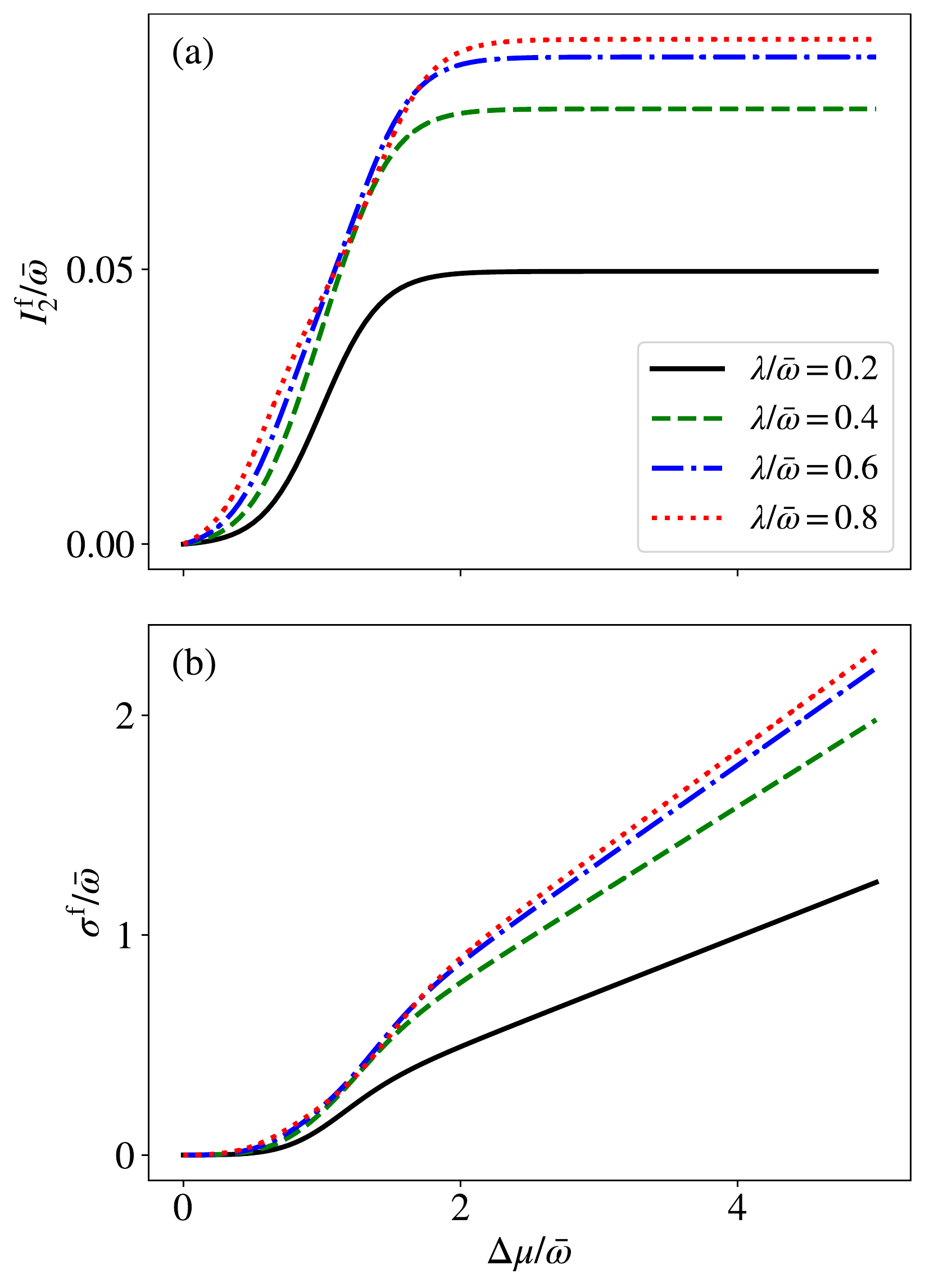}
	\caption{\label{FIG f epr} (a) Particle current $I_2^\text{f}$ defined in (\ref{def I f}) and (b) entropy production rate $\sigma^\text{f}$ defined in (\ref{def S f}) at steady state in terms of the nonequilibrium condition $\Delta \mu=\mu_2-\mu_1$ with fixed $\mu_1=0$. The two baths have the same temperatures $T_1=T_2=0.2\bar\omega$. Parameters are set as $\theta=-\pi/2$ and $J=0.1\bar\omega$. }
\end{figure}

Nonequilibrium fermionic environments suggest particle current flowing through the system. We can keep track of the particle number change in the system to reveal the particle current. Similarly with heat current, we define the particle current with respect to the two baths as
\begin{equation}
\label{def I f}
    I^\text{f}_l = \tr(\mathcal D_l[\rho]N_S)
\end{equation}
with the particle operator $N_S$ in energy basis $N_S=|2\rangle\langle 2|+|3\rangle\langle 3|+2|4\rangle\langle 4|$. The steady state gives
\begin{equation}
    I^\text{f}_1+I^\text{f}_2=0
\end{equation}
The positive current means that the electron flows into the system. The analytical expression of the particle current is given in Appendix \ref{appen C}. We plot the particle current in terms of the chemical potential difference $\Delta \mu =\mu_2-\mu_1$ with fixed $\mu_1$ in FIG. \ref{FIG f epr}. Due to the Pauli exclusion principle (particle occupation number $n(\omega)$ (\ref{def n i omega}) is less than 1), the particle current becomes saturated when we have large chemical potential bias. For example, at relative low temperature regime $T\ll \bar\omega$, if $\mu_1\rightarrow 0$ and $\mu_2\rightarrow \infty$, we have the particle current
\begin{equation}
    I_2^\text{f} \approx J(1+\rho^\text{f}_{23}+\rho^\text{f}_{32}) = J\left(1-\frac{1}{1+\left(\lambda/2J\right)^2}\right)
\end{equation}
assuming $\omega_1=\omega_2$ and constant symmetric coupling spectral $J$ (\ref{constant J}). The current is proportional to the coherence in the energy basis. The particle current is bounded by $J$. Increasing the inter-qubit coupling gives larger particle current as expected, see FIG. \ref{FIG f epr}.

The two fermionic environments are characterized by the (different) equilibrium chemical potentials and the temperatures respectively at the steady state. The entropy production rate of the environments is
\begin{equation}
\label{def S f}
    \sigma^\text{f} = \frac{\mu_1 I_1^\text{f}}{T}+\frac{\mu_2 I_2^\text{f}}{T} = \frac{\mu_2-\mu_1}{T}I_2^\text{f}
\end{equation}
Here for simplicity, we assume the same temperatures $T = T_1 = T_2$ of the two
baths to explore how chemical potential difference influences the dynamics and thermodynamic dissipation. At far from equilibrium case, the particle current is saturated. Therefore the entropy production rate increases linearly with the nonequilibrium condition $\Delta \mu=\mu_2-\mu_1$, see FIG. \ref{FIG f epr}.

\subsection{MLGI Enhanced by the Nonequilibrium bosonic Environments}

In Liouville space, the time-evolution operator is given by $e^{\mathcal Mt}$ with the matrix elements in (\ref{M})-(\ref{M_end}), see (\ref{e M t}). When we have nonzero nonequilibrium condition $\Delta T$, the density matrix in population space and coherence space is not decoupled, unlike the equilibrium case. Therefore, the steady state coherence in the eigenstate representation can survive in nonequilibrium environments \cite{LCS15,WWCW18,WWW18}. The time-evolution matrix $\mathcal M$ can be decomposed into coherent evolution part $\mathcal M_0$ and the dissipation part $\mathcal M_J$, see (\ref{decompose M}). These two parts do not commute in the nonequilibrium cases. As a result, the Zassenhaus formula in the first order $\mathcal M_J$ in (\ref{e Mt approximation}) will have summation for infinite series. Fortunately, the sum has the closed form due to the simple structure of $\mathcal M_0$. For simplicity, we consider the symmetric qubit $\omega_1=\omega_2$ in the following analytical study. 

The zeroth order of the LGI function $\mathcal I_2(t,\rho^\text{ss})$ has the contribution from the steady state coherence. Thus the nonzero steady state coherence may suggest that the two-time LGIs are violated. In the symmetric qubit setting, the inequality  $\mathcal I^{(0)}_2(t,\rho^\text{ss})>1$ can be rewritten into
\begin{equation}
\label{rho 44 23}
    \rho_{44}<\re\rho_{23},
\end{equation}
which is possible in some nonequilibrium cases. Note that the nonequilibrium steady state coherence is proportional to the coupling constant $J$, see Eq. (\ref{def rho 23}) and Eq. (\ref{def rho 23 f}). Therefore, the inequality in Eq.(\ref{rho 44 23}) also implies that the possible violation of two-time LGIs is in the order $J$. We numerically check to see that the possible violation from the coherent evolution is wiped out when the nonunitary evolution is included. In the following, we concentrate on the LGI function $\mathcal I_+(t,\rho^\text{ss})$ defined in (\ref{LGI steady state}). Note that the violation by the LGI function $\mathcal I_+(t,\rho^\text{ss})$ is always smaller than the LGI function $\mathcal I_-(t,\rho^\text{ss})$ in the coherent evolution, see Eq.(\ref{def I + 0})-(\ref{def I - 0}).

We obtain the first order LGI function $\mathcal I_+(t,\rho^\text{b})$ with the nonequilibrium corrections:
\begin{widetext}
	\eq
	\mathcal I_+^{(1)}(t,\rho^\text{b})= 4tJ (\rho^\text{b}_{22}+\rho^\text{b}_{33})\left(\tilde n_1+\tilde n_2+1)\right)(\cos(2\lambda t)-\cos(\lambda t)) + \frac {2J} \lambda \left(\Delta n_1+\Delta n_2\right)(\sin(2\lambda t)-2\sin(\lambda t))
	\en
\end{widetext}
{The steady state $\rho^\text{b}$ has the analytical solution in (\ref{Boson rho 11})-(\ref{Boson rho 44}). The first term is the equilibrium correction (with $\sin^2\theta=1$), see (\ref{LGI 1}).} To better see how the nonequilibrium condition affects the LGIs violations, we approximate the MLGI function $\mathcal I_\text{max}(\rho^\text{b})$ as (same as the equilibrium case):
\begin{align}
\label{MLGI 1 none}
    \mathcal I^{(1)}_\text{max}(\rho^\text{b}) & \approx  \mathcal I_+^{(1)}(t=\pi/(3\lambda),\rho^\text{b}) \nonumber\\
    &  =-\frac{4\pi J}{3\lambda}(\rho^\text{b}_{22}+\rho^\text{b}_{33})-\frac{\sqrt 3 J}{\lambda}(\Delta n_1+\Delta n_2)
\end{align}
with $\lambda\neq 0$. Note that $\lambda=0$ (decoupled two qubit system) is trivial for the LGIs violations. The first order $\mathcal I^{(1)}_\text{max}(\rho^\text{b})$ is always negative irrespective to either $T_2>T_1$ or $T_1<T_2$. The second term in $\mathcal I^{(1)}_\text{max}(\rho^\text{b})$ is the corresponding nonequilibrium correction. Note that we also have nonequilibrium corrections in steady state solution (\ref{Boson rho 11})-(\ref{Boson rho 44}), which is proportional to $J^2$ order.

\begin{figure}
	\includegraphics[width=0.75
	\columnwidth]{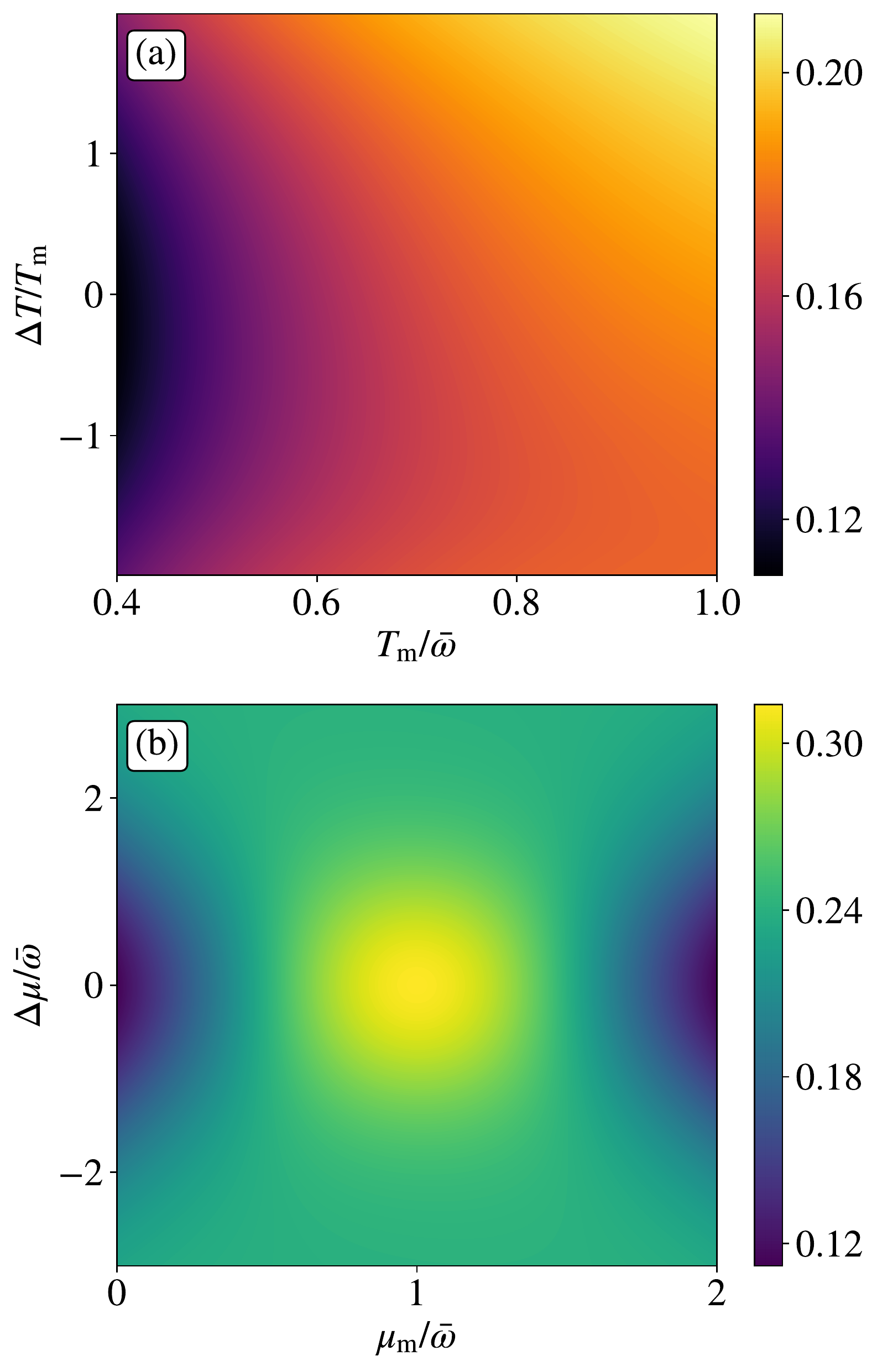}
	\caption{\label{MLGI phase 1} {The contour plot diagram of MLGI function $\mathcal I_\text{max}(\rho^\text{ss})$ defined in (\ref{def I max})} in terms of (a) the mean temperature $T_\text{m}$ and the temperature difference $\Delta T=T_2-T_1$ for the bosonic bath and (b) the mean chemical potential $\mu_\text{m}$ and the chemical potential difference $\Delta\mu=\mu_2-\mu_1$ for the fermionic bath with $T=0.4\bar\omega$. Other parameters are set as $\lambda=\bar\omega$, $\theta=-\pi/2$ and $J=0.005\bar\omega$. {Red-based color is for the bosonic environment and green-based color is for the fermionic environment.} }
\end{figure}

\begin{center}
\begin{figure}
	\includegraphics[width=0.9
	\columnwidth]{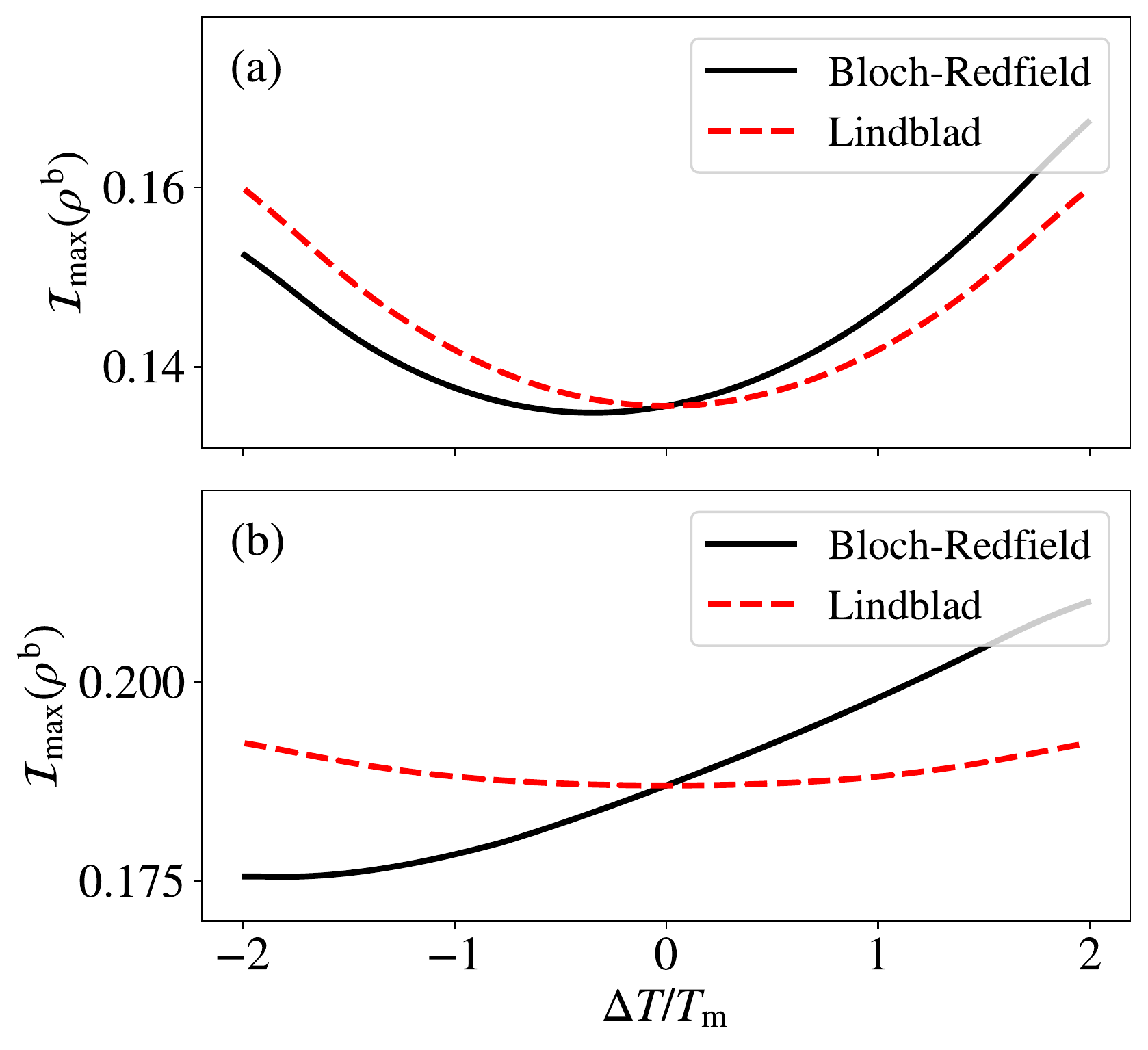}
	\caption{\label{fig_compare} Comparison between the MLGI functions given by the Lindblad and the Bloch-Redfield equation. The parameters are set as $\lambda=\bar\omega$, $\theta=-\pi/2$ and $J=0.005\bar\omega$. (a) The mean temperature is relatively low $T_\text{m}=0.5\bar\omega$. (b) The mean temperature is relatively high $T_\text{m}=\bar\omega$.}
\end{figure}
\end{center}

The nonequilibrium correction in $\mathcal I^{(1)}_\text{max}(\rho^\text{b})$ is not a symmetric function of temperatures $T_1$ and $T_2$. Simple argument shows that $\Delta n_{1,2}<0$ with $T_2>T_1$. Therefore we can expect MLGI function $\mathcal I^{(1)}_\text{max}(\rho^\text{b})$ enhancement with $T_2>T_1$. The two-qubit system is symmetric with $T_1$ and $T_2$ if the two qubits are identical ($\omega_1=\omega_2$). In other words, the identical two qubit system has the spatial asymmetry which comes from the nonequilibrium environments. In the LGIs tests, we assume that the one-qubit system is measured locally, which introduces another asymmetry. Therefore, we expect that measuring qubit 1 will have different results with measuring qubit 2, if the two qubits are surrounded by the nonequilibrium environments $\Delta T\neq0$. If we tune the two-qubit coupling $\lambda$ to be small, the qubit 2 coupled bath with temperature $T_2$ becomes another environment in terms of qubit 1. If bath 1 has high temperature, then qubit 1 becomes classical.

In the equilibrium scenario, the MLGI function $\mathcal I_\text{max}(\rho^\text{b,e})$ has a nonmonotonic relationship with the temperatures $T=T_1=T_2$, see FIG. \ref{MGLI_T}, resulting from the competition between the nonlocal state population and the decoherence effects on the system. In the following, we fix the mean temperature and study the relationship between the MLGI function $\mathcal I_\text{max}(\rho^\text{b})$ and nonequilibrium condition $\Delta T=T_2-T_1$. If we have relative low mean temperature ($T_\text{m}\ll \bar\omega$ with $T_\text{m}=1/2(T_1+T_2)$), we get the approximation:
\eq
\label{MLGI 0 1 none}
 \mathcal I_\text{max}^{(0)}(\rho^\text{b})+\mathcal I_\text{max}^{(1)}(\rho^\text{b}) \approx \alpha'\left(\frac 1 2 +\frac{J}{\lambda}\left(\pm\sqrt 3-\frac{4\pi}{3}\right)\right)
\en
The plus sign is for $T_2>T_1$ and minus sign is for $T_2<T_1$. Parameter $\alpha'$ is defined as
\begin{equation}
    \alpha'=e^{-\omega'_1/T_\text{eff}}+e^{-\omega'_2/T_\text{eff}}
\end{equation}
{with the effective temperature}
\begin{equation}
     T_\text{eff}=2T_\text{m}^2/(2T_\text{m}-|\Delta T|)
\end{equation}
With $\Delta T=0$, the effective temperature is the equilibrium temperature $T_\text{eff}=T_1=T_2=T$, and $\alpha'$ is back to the equilibrium parameter $\alpha$ defined in (\ref{def alpha}). The parameter $\alpha'$ is approximated from the nonlocal state population sum $\rho^\text{b}_{22}+\rho^\text{b}_{33}$. We can increase $\alpha'$ by increasing the mean temperature $\bar T$ or the temperature difference $\Delta T$. Therefore the increased MLGI function $\mathcal I_\text{max}(\rho^\text{b,e})$ via the nonequilibrium condition can be understood as the increasing nonlocal state population. Besides, the temperature difference at $T_2>T_1$ is more favorable for LGI violations than the temperature difference at $T_2<T_1$ (magnitude in the order of $J$). The asymmetric roles of $T_1$ and $T_2$ can only be characterized by the Bloch-Redfield equation. We compare the difference of the MLGI functions given by the Bloch-Redfield equation and the Lindblad respectively, see FIG. \ref{fig_compare}. Note that the Lindblad form always treats the (strongly coupled) two-qubit system collectively and therefore misses the genuine nonequilibrium effects \cite{WW19}.

When the mean temperature is relatively large, say $T_\text{m}\sim 10\bar\omega$, we know that the states are almost evenly distributed i.e., $\rho^\text{b}_{22}+\rho^\text{b}_{33}\approx 1/2$. Up to the first order $J$, we have
\begin{multline}
\label{MLGI 0 1 none 2}
 \mathcal I_\text{max}^{(0)}(\rho^\text{b})+\mathcal I_\text{max}^{(1)}(\rho^\text{b}) \approx \\
 \frac 1 4 +\frac{J}{\lambda}\left(\left(\frac 1 {\omega'_1}+\frac 1 {\omega'_2}\right)\left(\frac{\sqrt 3}{2}\Delta T-\frac{2\pi}{3}T_\text{m}\right)-\frac{2\pi}{3}\right)
\end{multline}
We can see that $\Delta T>0$ can give LGIs violations however $\Delta T<0$ can not, see FIG. \ref{MLGI phase 1}. Such asymmetric contribution is beyond the Lindblad description \cite{CRQ13}. See FIG. \ref{fig_compare} for the comparison between the Lindblad and the Bloch-Redfield equation. Note that we have the same MLGI functions given by the Lindblad or the Bloch-Redfield equation if $\Delta T=0$.  

\begin{center}
\begin{figure}
	\includegraphics[width=0.75
	\columnwidth]{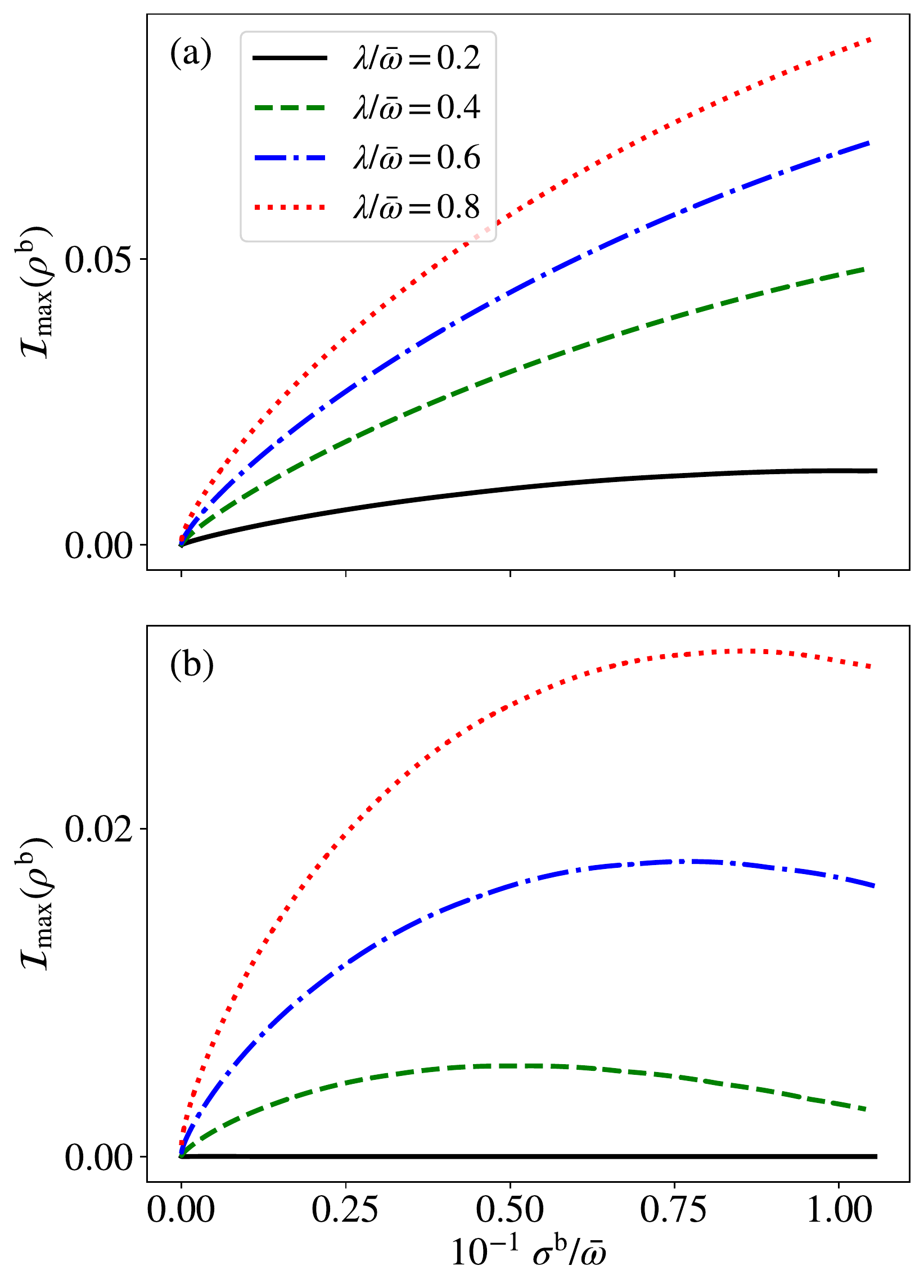}
	\caption{\label{FIG MLGI epr} The MLGI function $\mathcal I_\text{max}(\rho^\text{b})$ defined in (\ref{def I max}) in terms of the entropy production rate $\sigma^\text{b}$ defined in (\ref{def S b}) with the different inter-qubit coupling strength $\lambda$. (a) The temperature $T_1$ is fixed with $T_1=0.1\bar\omega$ or (b) the temperature $T_2$ is fixed with $T_2=0.1\omega$. Other parameters are set as $\theta=-\pi/2$ and $J=0.05\bar\omega$.}
\end{figure}
\end{center}

The nonequilibrium bosonic environments have the thermodynamic cost characterized by the entropy production rate $\sigma^\text{b}$ defined in (\ref{def S b}). We know that the steady state gives the constant entropy production rate which monotonically increases with the temperature difference of the two baths, see FIG. \ref{FIG b epr}. The entropy production rate $\sigma^\text{b}$ can be viewed as another thermodynamic nonequilibrium measure, aside from the temperature difference $\Delta T$. {Suppose that we fix the temperature $T_1$ at relatively low regime $T_1\ll \bar\omega$. The system will stay at the ground state at the equilibrium case $T_1=T_2$. Increasing the temperature $T_2$ will induce the heat current (at steady state) flowing from the bath 2 to the bath 1. We will have non-zero entropy production rate $\sigma^\text{b}$ generated from the nonequilibrium environments. The consumption in the environments characterized by the entropy production rate $\sigma^\text{b}$ can enhance the LGI violation if the temperature $T_1$ is relatively low, see FIG. \ref{FIG MLGI epr}. If we fix the temperature $T_2$, increasing the temperature $T_1$ will induce the heat current (at steady state) flowing from the bath 1 to the bath 2. The same magnitude $|\Delta T|$ gives the same entropy production rate. However, we have the stronger enhancement for the MLGI function $\mathcal I_\text{max}(\rho^\text{b})$ consumed by the thermodynamic cost (the nonzero entropy production rate) if $T_1<T_2$, see FIG. \ref{FIG MLGI epr}. With the same entropy production rate $\sigma^\text{b}$, larger inter-qubit coupling $\lambda$ gives larger MLGI function $\mathcal I_\text{max}(\rho^\text{b})$ either $T_1>T_2$ or $T_1<T_2$. In other words, we have more efficient thermodynamic cost (to enhance the LGI violations) if we have stronger inter-qubit coupling $\lambda$. Intuitively, stronger inter-qubit coupling $\lambda$ gives more population on the first excited state $|2\rangle$, which is nonlocal. }

\begin{figure}
	\includegraphics[width=0.75
	\columnwidth]{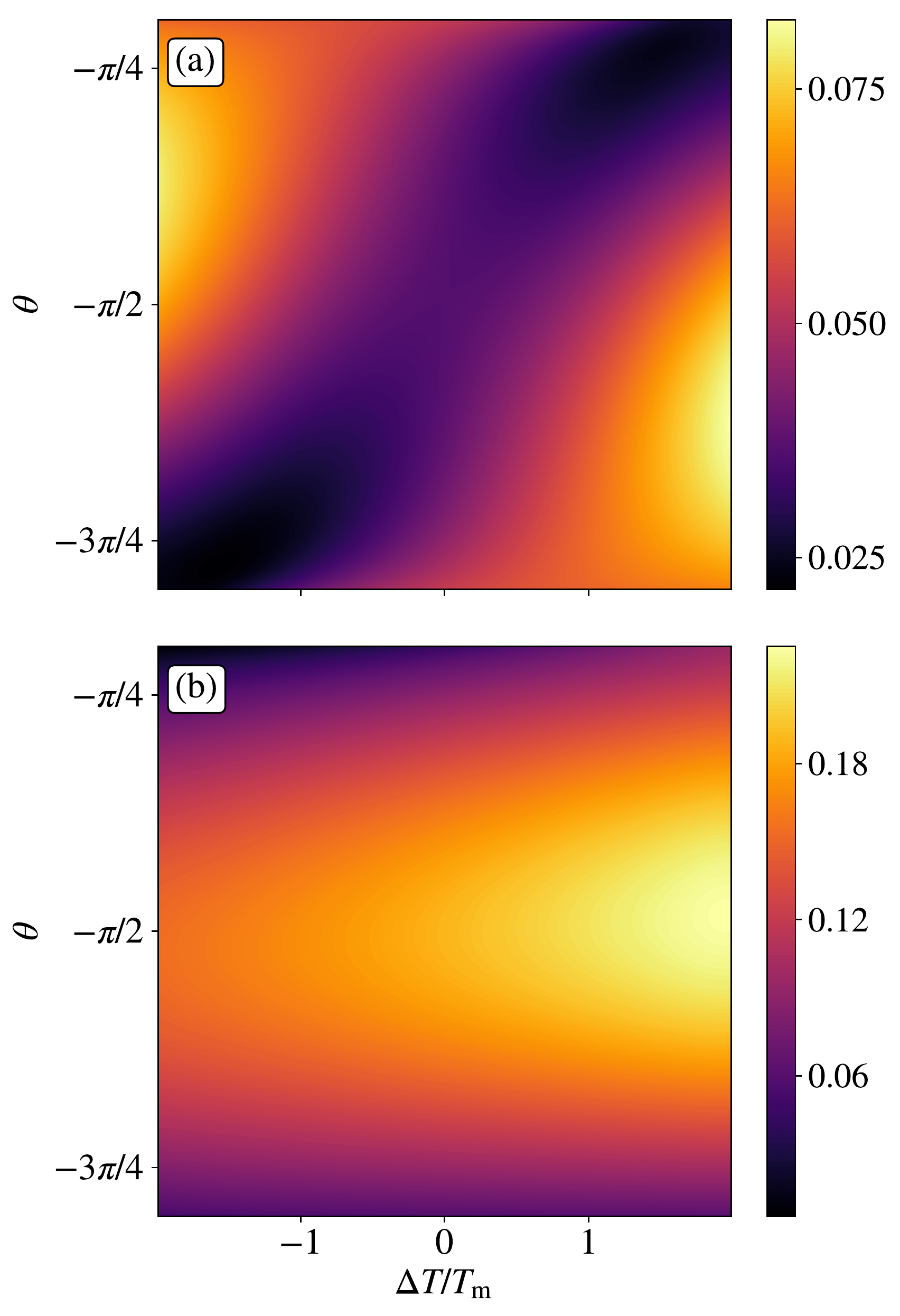}
	\caption{\label{MLGI phase theta boson} {The MLGI function $\mathcal I_\text{max}(\rho^\text{ss})$ defined in (\ref{def I max})} in terms of the nonequilibrium condition $\Delta T=T_2-T_1$ and the detuning angle $\theta$ defined in (\ref{def theta}) with fixed $\lambda=\bar\omega$. (a) Mean temperature is relatively low $T_\text{m}=0.2\bar\omega$ or (b) high $T_\text{m}=2\bar\omega$. The coupling constant is set as $J=0.005\bar\omega$.}
\end{figure}

%We fixed the temperature of one bath (the bath 1 with the temperature $T_1$ or the bath 2 with the temperature $T_2$), and plot the MLGI function $\mathcal I_\text{max}(\rho^\text{b})$ in terms of the entropy production rate $\dot S^\text{b}$ in FIG. \ref{FIG MLGI epr}. The higher inter-qubit coupling $\lambda$ gives both the higher entropy production rate $\dot S^\text{b}$ and the MLGI function $\mathcal I_\text{max}(\rho^\text{b})$. Since increasing the inter-qubit coupling $\lambda$ gives the lower first excited state energy $\omega_1'$, see Eq. (\ref{def nonlocal basis}), we will have more population in the nonlocal state $|2\rangle$. Although the same temperature difference magnitude $|\Delta T|$ gives the same entropy production rate $\dot S^\text{b}$, we know that choosing the $T_1$ smaller than the $T_2$ is better than the $T_2$ smaller than the $T_1$ because we only measure the local observables of qubit 1. We can see the obvious difference with fixing the $T_1$ or the $T_2$ in FIG. \ref{FIG MLGI epr}. For example, with the small inter-qubit coupling strength $\lambda$, we can have the LGI violation for $T_2>T_1$, but the LGI is preserved in $T_1>T_2$ with the same entropy production rate $\dot S^\text{b}$ or temperature difference magnitude $|\Delta T|$. The nonequilibrium thermodynamic cost, such as entropy production rate $\dot S^\text{b}$, is the necessary but not the sufficient condition for the enhancement of LGI violation via the nonequilibrium conditions. 

{Detuning the qubit frequency $\Delta \omega$ can increase the MLGI function $\mathcal I_\text{max}(\rho^\text{b,e})$ (with equilibrium environments) if the temperature $T=T_1=T_2$ is relatively low, see FIG. \ref{MLGI_theta_boson}. The relationship between the frequency difference $\Delta \omega$ and MLGI function $\mathcal I_\text{max}(\rho^\text{b})$ is more complicated in nonequilibrium cases. We know that the LGI function $\mathcal I_\text{max}(t,\rho^\text{b})$ is asymmetric in terms of temperature difference $\Delta T$.} We plot the MLGI function $\mathcal I_\text{max}(\rho^\text{b})$ with respect to detuning angle $\theta$ defined in (\ref{def theta}) (with fixed $\lambda$) and nonequilibrium condition $\Delta T$ in FIG. \ref{MLGI phase theta boson}. When the mean temperature is relatively low $T_\text{m}<\bar\omega$, detuning $\Delta \omega>0$ or $\Delta \omega<0$ can increase MLGI function $\mathcal I_\text{max}(\rho^\text{b})$ with $\Delta T<0$ or $\Delta T>0$ respectively, see FIG. \ref{MLGI phase theta boson} (a). In other words, the qubit with smaller frequency should couple with higher temperature bath and vice versa. Recall that in the low mean temperature regime, the MLGI function $\mathcal I_\text{max}(\rho^\text{b})$ is enhanced from increasing the nonlocal state population. One can check that the nonlocal state population sum is also increased by detuning $\Delta \omega>0$ or $\Delta \omega<0$ if $\Delta T<0$ or $\Delta T>0$. Intuitively we can understand that the low frequency qubit coupled with the high temperature bath can help the excitation, since the mean temperature is relatively low and the system is at the ground state with high probability. When the mean temperature is relatively large, i.e., $T_\text{m}\approx\bar\omega$, from FIG. \ref{MLGI phase theta boson} (b), we see that detuning $\Delta \omega$ will not enhance the violations of LGIs significantly anymore for either $\Delta T>0$ or $\Delta T<0$. The asymmetric temperature contribution is obvious: $\Delta T>0$ always gives stronger enhancement of MLGI function $\mathcal I_\text{max}(\rho^\text{b})$. {In equilibrium environments, around $T_\text{m}\approx\bar\omega$, the MLGI function $\mathcal I_\text{max}(\rho^\text{b,e})$ has the maximal value, see FIG. \ref{MGLI_T}. Such maximal point is the result of the competition between the excitation of nonlocal states and the decoherence effect. In nonequilibrium case around $T_\text{m}\approx\bar\omega$, detuning the two qubits will not help the excitation. In fact, detuning the two qubits leads to the nonlocal state to be less entangled and therefore does not boost the LGI violation.}

\subsection{MLGI Enhanced by Nonequilibrium fermionic Environments}

\begin{figure}
	\includegraphics[width=0.75
	\columnwidth]{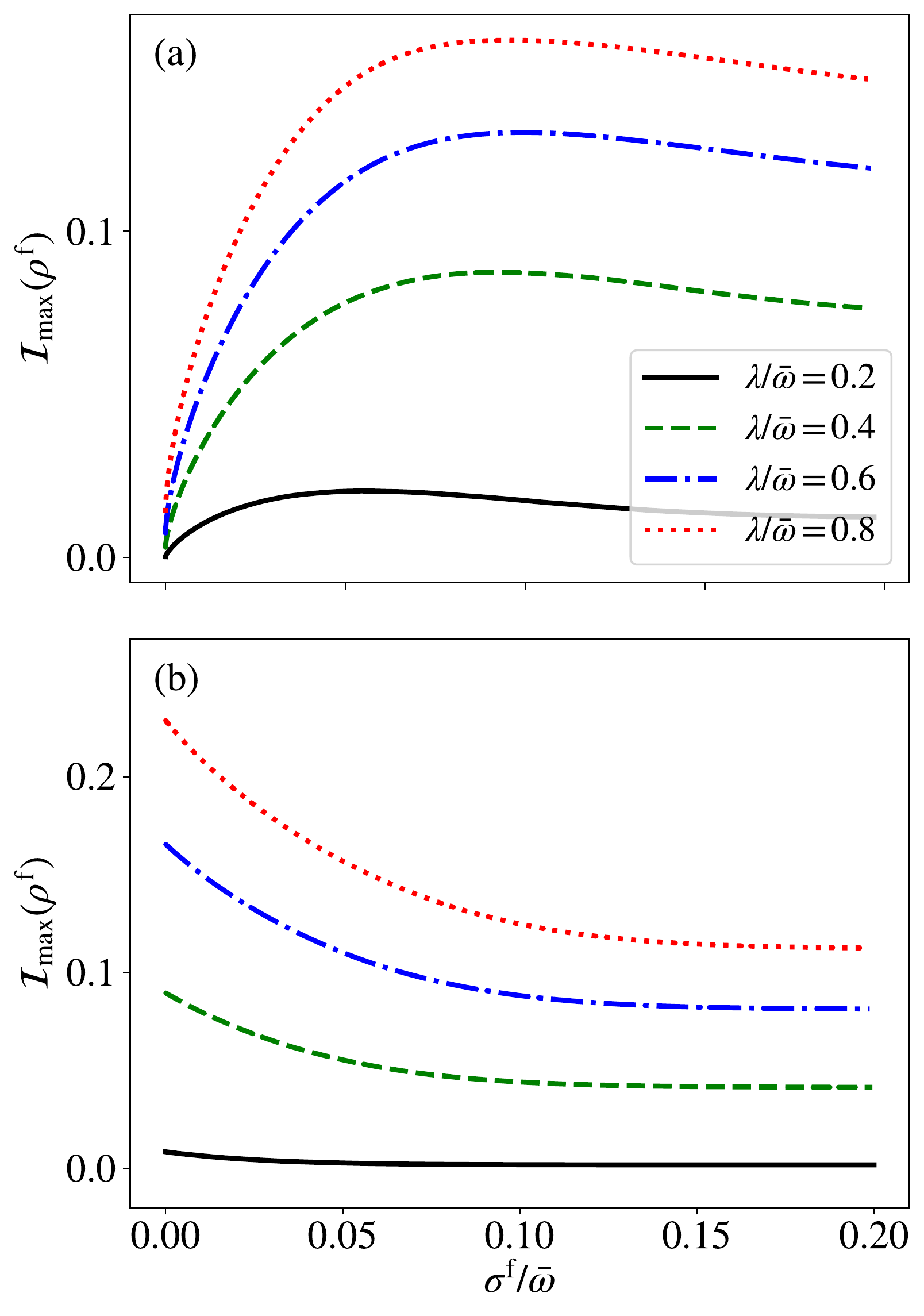}
	\caption{\label{FIG MLGI epr f} The MLGI function $\mathcal I_\text{max}(\rho^\text{f})$ defined in (\ref{def I max}) in term of the entropy production rate $\sigma^\text{f}$ defined in (\ref{def S f}) with the different inter-qubit coupling strength $\lambda$. (a) The chemical potential $\mu_1$ is fixed with $\mu_1=0$ or (b) fixed with $\mu_1=\bar\omega$. Other parameters are set as $\theta=-\pi/2$, $T_1=T_2=0.2\bar\omega$ and $J=0.05\bar\omega$.}
\end{figure}

When the two-qubit system is coupled with the two fermionic baths, we consider how the nonequilibrium condition given by $\Delta \mu=\mu_2-\mu_1$ contributes to the violations of LGIs. Same as bosonic nonequilibrium setup, {the coherent evolution $\mathcal M_0$ and dissipator $\mathcal M_J$ do not commute with non-vanishing $\Delta \mu\neq 0$.} However, we still can have closed form of Zassenhaus formula (\ref{e Mt approximation}) in the first order of the coupling constant $J$. {We have the first order LGI function $\mathcal I_+(t,\rho^\text{f})$ with the nonequilibrium corrections:}
\begin{widetext}
	\eq
	 \mathcal I_+^{(1)}(t,\rho^\text{f})= 4tJ (\rho^\text{f}_{22}+\rho^\text{f}_{33})(\cos(2\lambda t)-\cos(\lambda t)) + \frac {2J} \lambda \left(\Delta n_1+\Delta n_2\right)\left(\rho^\text{f}_{11}-\rho^\text{f}_{44}\right)(\sin(2\lambda t)-2\sin(\lambda t))
	\en
\end{widetext}
{with $\omega_1=\omega_2$ and $\lambda\neq 0$.} The nonequilibrium correction term is proportional to $(\Delta n_1+\Delta n_2)$. {We can approximate the first order MLGI function $\mathcal I_\text{max}(\rho^\text{b})$ by}
\begin{equation}
    \mathcal I^{(1)}_\text{max}(\rho^\text{f})\approx  \mathcal I^{(1)}(t=\pi/(3\lambda),\rho^\text{f}),
\end{equation}
{which gives} 
\begin{multline}
\label{MLGI 1 none f}
\mathcal I^{(0)}_\text{max}(\rho^\text{f})+\mathcal I^{(1)}_\text{max}(\rho^\text{f})=\left(\frac 1 2-\frac{4\pi J}{3\lambda}\right)(\rho^\text{f}_{22}+\rho^\text{f}_{33})\\
+\frac{\sqrt 3 J}{\lambda}(\Delta n_1+\Delta n_2)(\rho^\text{f}_{44}-\rho^\text{f}_{11})
\end{multline}
The nonequilibrium term $\Delta n_l$ (\ref{delta n 1})-(\ref{delta n 2}) with $l=1,2$ is bounded $|\Delta n_l|<1/2$ in fermionic case. {Nonequilibrium conditions $\Delta \mu>0$ and $\Delta \mu<0$ only have difference in $\mathcal I^{(1)}_\text{max}(\rho^\text{f})$ with magnitude in order $J$. In other words, the asymmetric nonequilibrium condition $\Delta\mu$ does not give significantly asymmetric MLGI function $\mathcal I_\text{max}(\rho^\text{f})$,} unlike the bosonic nonequilibrium case.

In the equilibrium setup, we know that the resonant point $\bar\mu=\bar\omega$ gives the maximal MLGI. Analytically, with $\bar\mu=\bar\omega$, the population sum $\rho^\text{f}_{22}+\rho^\text{f}_{33}$ (up to first order coupling $J$) has the expression:
\begin{multline}
\rho^\text{f}_{22}+\rho^\text{f}_{33}=\frac{\cosh\left(\frac{\lambda}{2T}\right)}{\cosh\left(\frac{\Delta u}{2T}\right)+\cosh\left(\frac{\lambda}{2T}\right)} \\
+\frac{\sinh^2\left(\frac{\Delta \mu}{2T}\right)}{2\left(\cosh\left(\frac{\Delta \mu+\lambda}{2T}\right)+1\right)\left(\cosh\left(\frac{\Delta \mu-\lambda}{2T}\right)+1\right)}
\end{multline}
{It is easy to see that the equilibrium setup $\Delta\mu=0$ gives the maximal of population sum $\rho^\text{f}_{22}+\rho^\text{f}_{33}$. Therefore nonequilibrium condition does not give enhancement of MLGI $\mathcal I_\text{max}(\rho^\text{f})$ when $\bar\mu=\bar\omega$. See FIG. \ref{MLGI phase 1} for numerical results.}

%especially we have $\rho^\text{f}_{22}+\rho^\text{f}_{33}\rightarrow 1$ as $|\Delta\mu|<\lambda$ and $\rho^\text{f}_{22}+\rho^\text{f}_{33}\rightarrow 1/2$ as $|\Delta\mu|>\lambda$ at low temperature. As for the nonequilibrium correction for first order $J$ of MLGI (second term in (\ref{MLGI 1 none f})), the prefactor $(\rho^\text{f}_{44}-\rho^\text{f}_{11})$ at resonant point $\bar\omega=\bar\mu$ has the analytical form
%\begin{multline}
%\rho^\text{f}_{44}-\rho^\text{f}_{11}=\frac{\cosh\left(\frac{\lambda}{2T}\right)}{\cosh\left(\frac{\Delta u}{2T}\right)+\cosh\left(\frac{\lambda}{2T}\right)} \\
%+\frac{\cosh\left(\frac{\Delta\mu}{2T}\right)\left(\cosh\left(\frac{\Delta\mu}{2T}\right)+\cosh\left(\frac{\lambda}{2T}\right)\right)}{\left(\cosh\left(\frac{\Delta \mu+\lambda}{2T}\right)+1\right)\left(\cosh\left(\frac{\Delta \mu-\lambda}{2T}\right)+1\right)}-1
%\end{multline}
%which will converge to 0 both at $|\Delta\mu|<\lambda$, $|\Delta\mu|>\lambda$ and $|\Delta\mu|=\lambda$ at low temperature regime. Therefore, the nonequilibrium term (as well as asymmetric contribution from $\Delta \mu$) is surpassed around the resonant point $\bar\omega=\bar\mu$. And MLGI will not be enhanced by the nonequilibrium condition $\Delta \mu$ around the resonant point $\bar\mu=\bar\omega$, since the resonant point gives arise to almost Bell-like mixed state which has maximally violated LGI already. See Fig. \ref{MLGI phase 1} for numerical confirmation.

{The major contribution up to the first order MLGI function in (\ref{MLGI 1 none f}) is from $\rho^\text{f}_{22}+\rho^\text{f}_{33}$, which is the population sum of the non-local eigenstates.} Away from the resonant point $\bar\mu=\bar\omega$, we can find that the population sum increases with the nonequilibrium condition $\Delta \mu$. Therefore, MLGI $\mathcal I_\text{max}(\rho^\text{f})$ can be enhanced by the nonequilibrium condition away from the resonant point. See FIG. \ref{MLGI phase 1} for numerical results that MLGI $\mathcal I_\text{max}(\rho^\text{f})$ can be enhanced by $\Delta \mu$ away from $\bar\mu=\bar\omega$.

%Another way to look at the approximated MLGI (\ref{MLGI 1 none f}) is that nonequilibrium term $\Delta n_i$ is bounded by $|\Delta n_i|<1/2$ for Fermionic distribution. Then the major contribution to the MLGI from the nonequilibrium condition is  $\rho^\text{f}_{22}+\rho^\text{f}_{33}$, which is the population sum of the non-local eigenstates.   Note that the asymmetric nonequilibrium condition $\Delta\mu$ contributes to the MLGI in the order of coupling constant $J$. See Fig. \ref{MLGI phase 1} for numerical results that MLGI can be enhanced by $\Delta \mu$ away from $\bar\omega=\bar\mu$.

\begin{figure}
	\includegraphics[height=9.5cm]{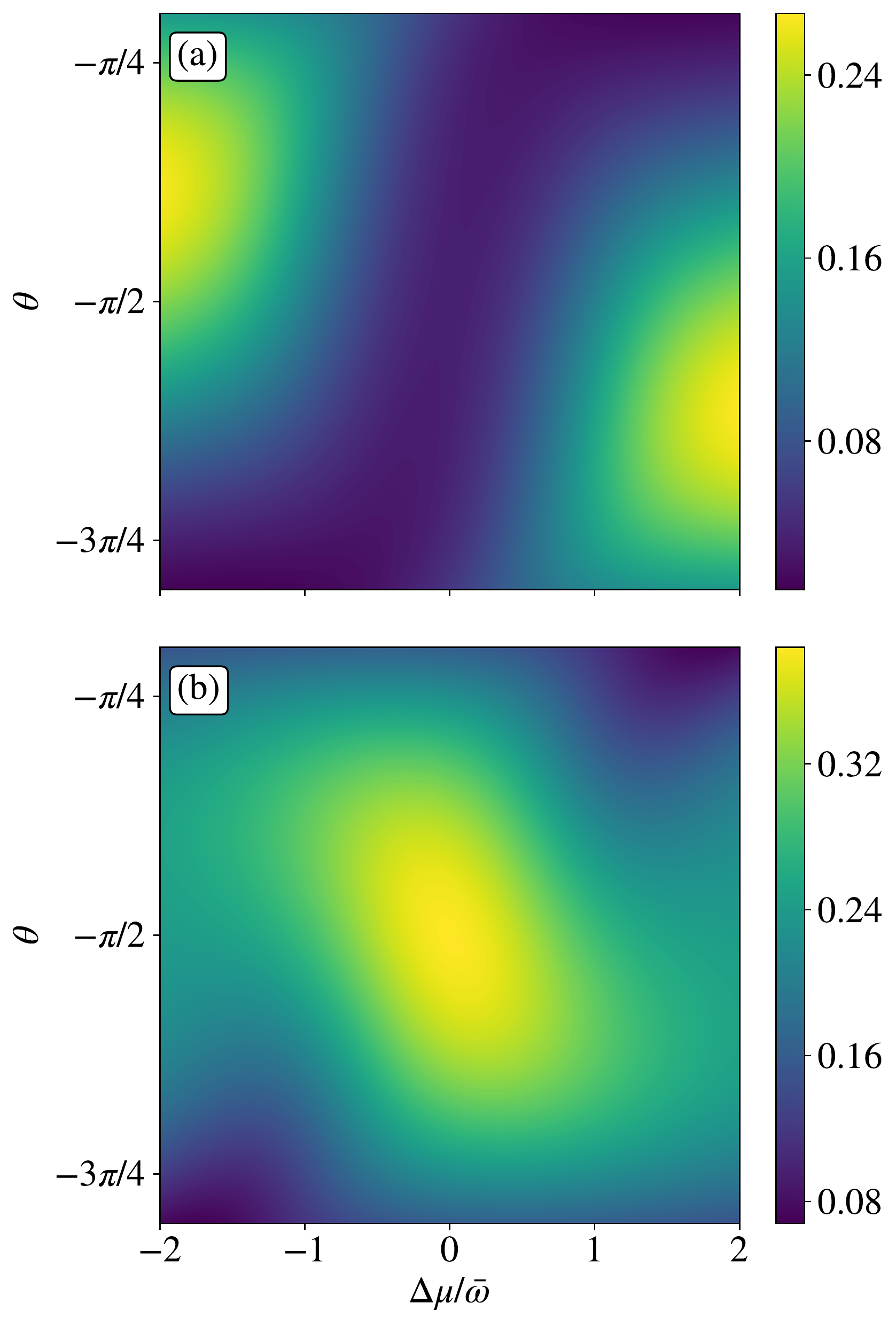}
	\caption{\label{MLGI phase theta fermi} {The MLGI function $\mathcal I_\text{max}(\rho^\text{ss})$ defined in (\ref{def I max})} in terms of nonequilibrium condition $\Delta\mu=\mu_2-\mu_1$ (nonequilibrium fermionic bath with $T=0.2\bar\omega$) and detuning angle $\theta$ defined in (\ref{def theta}) with fixed $\lambda=\bar\omega$. (a) Small mean chemical potential $\mu_\text{m}=0.2\bar\omega$ and (b) large mean chemical potential  $\mu_\text{m}=0.8\bar\omega$. The coupling constant is set as $J=0.005\bar\omega$.}
\end{figure}

{
The fermionic environments with the different chemical potentials lead to the particle current flowing through the system. The nonequilibrium environments have the dissipation characterized by the entropy production rate $\sigma^\text{f}$ defined in (\ref{def I f}), see FIG. \ref{FIG f epr}.} {The MLGI function $\mathcal I_\text{max}(\rho^\text{f})$ can be enhanced by consuming the nonequilibrium environments (with nonzero entropy production rate $\sigma^\text{f}$ if the chemical potentials $\mu_1$ and $\mu_2$ are away from the resonant point $\bar\omega$. For example, if we fix the chemical potential $\mu_1$ at $\mu_1=0$, increasing the entropy production rate $\sigma^\text{f}$ gives the enhancement of LGI violation, see FIG. \ref{FIG MLGI epr f}. However, if we fix the $\mu_1$ around the value $\bar\omega$, the nonequilibrium thermodynamic cost gives the smaller or zero MLGI function $\mathcal I_\text{max}(\rho^\text{f})$, see FIG. \ref{FIG MLGI epr f}. The equilibrium case $\mu_1=\mu_2\approx \bar\omega$ with zero entropy production rate $\sigma^\text{f}$ already gives almost saturated populations $\rho_{22}^\text{f}+\rho_{33}^\text{f}$, therefore the nonequilibrium cost $\sigma^\text{f}$ does not enhance the LGI violation. Similarly with bosonic environments, stronger inter-qubit coupling $\lambda$ always increases the LGI violation either $\mu_1$ and $\mu_2$ away from or around the value $\bar\omega$.}

%Suppose that we fix the chemical potential $\mu_1$. When $\mu_1$ is away from the resonant point $\bar\omega$, the system stays at ground state or highest excited state with high probability, in which the ground state and the highest exited state are local.

%We show the relationship between the entropy production rate $\dot S^\text{f}$ and the MLGI function $\mathcal I_\text{max}(\rho^\text{f})$ in FIG. \ref{FIG MLGI epr f}. We fix the chemical potential $\mu_1$. The asymmetric properties about the MLGI function $\mathcal I_\text{max}(\rho^\text{f})$ in terms of the asymmetric nonequilibrium condition $\Delta\mu$ is suppressed due to the Pauli exclusion principle (the occupation particle number difference $|\Delta n_l|<1/2$ is bounded). When the chemical potential $\mu_1$ is away from the resonant value $\bar\omega$, the nonequilibrium cost characterized by the entropy production rate $\dot S^\text{f}$ can increases the violation of LGI function. 

We also studied how the detuning frequency $\Delta \omega$ combined with nonequilibrium condition $\Delta\mu$ contributes to the violations of LGIs. Numerically, {we separately contour plot the two-dimensional diagram of the MLGI function $\mathcal I_\text{max}(\rho^\text{f})$ in terms of the nonequilibrium condition $\Delta\mu$ and the detuning angle $\theta$ (defined in (\ref{def theta})) with fixed tunneling strength $\lambda$, when the system is away from the resonant point or around the resonant point $\bar\mu=\bar\omega$,} see FIG. (\ref{MLGI phase theta fermi}). When the system is away $\bar\mu=\bar\omega$, the lower frequency qubit should couple the higher chemical potential reservoir and higher frequency qubit should couple the lower chemical potential reservoir, in order to reach the larger violations of LGIs. However, around the resonant point, detuning the system $\omega_1\neq\omega_2$ with nonequilibrium condition $\Delta\mu$ does not give larger violations of LGIs. The explanations are still based on the population sum $\rho^\text{f}_{22}+\rho^\text{f}_{33}$ in terms of the detuning angle $\theta$ and the nonequilibrium condition $\Delta \mu$. Away from the resonant point, the system (with high probability) stays at ground states or highest excited states which has classical description. Then detuning the two qubits by $\Delta\omega$ or by nonequilibrium condition $\Delta \mu>0$ (or $\Delta \mu<0$), we can increase the nonlocal state (the first and the second excited states) population. On the contrary, around the resonant point, the nonequilibrium condition always gives less population of nonlocal states.

\section{Conclusion}

\label{sec:conclusion}

We have studied the LGIs violations in a quantum system (two interacting qubits) coupled with equilibrium or nonequilibrium environments. Each qubit is coupled with one (bosonic or fermionic) bath. We have derived the evolution equation of the reduced density matrix beyond the secular approximation (Bloch-Redfield equation). There are analytical solutions of the reduced density matrix of steady states. We calculate the heat current (the system coupled with the nonequilibrium bosonic environments) or the particle current (the system coupled with the nonequilibrium fermionic environments) in the steady state regime. Correspondingly, we study the entropy production rate as the nonequilibrium thermodynamic cost. We use the maximum to LGI functions, which is called MLGI function $\mathcal I_\text{max}(\rho^\text{ss})$ defined in (\ref{def I max}), to quantify the degree of the LGIs violations. We have obtained the analytical form of the two- and three-time LGI functions $\mathcal I_2(t,\rho^\text{ss})$ and $\mathcal I_\pm(t,\rho^\text{ss})$. Based on that, we analyze the MLGI function $\mathcal I_\text{max}(\rho^\text{ss})$ in the zeroth order and the first order of coupling $J$. The zeroth order represents the coherent evolution and the first order describes the non-unitary part (due to coupled with the environments). We also analytically separate the equilibrium and nonequilibrium effects in the first order LGI function $\mathcal I_+(t,\rho^\text{ss})$ and MLGI function $\mathcal I_\text{max}(\rho^\text{ss})$.

In the equilibrium set up, LGIs violations are caused by the unitary evolution of the first and the second excited states (entangled states in two qubit system). In the bosonic case, the MLGI function $\mathcal I_\text{max}(\rho^\text{b,e})$ has a non-monotonic relationship with the equilibrium temperature. The analytical results reveal that the environment can give excitation and therefore the nonlocal eigenstates (enhance the LGIs violations) but can also have decoherence effect (reduce the LGIs violations). The fermionic bath has similar results.

The nonequilibrium condition can be quantified by the temperature difference or chemical potential difference for the bosonic or fermionic reservoir respectively. The entropy production rate also characterizes the thermodynamic nonequilibrium cost. In the bosonic case, the nonequilibrium environment can magnify the violations of LGIs (increasing the MLGI function $\mathcal I_\text{max}(\rho^\text{b})$ by the temperature difference), if the mean temperature is relatively low. Correspondingly, the LGIs violations enhancement monotonically increases with the entropy production rate. In particular, the MLGI function $\mathcal I_\text{max}(\rho^\text{b})$ can be enhanced more if $T_2>T_1$ (we measure the local observable of qubit 1). Such asymmetric result is beyond the Lindblad description (wiped out by the secular approximation). In the fermionic bath setups, we have the MLGI function $\mathcal I_\text{max}(\rho^\text{b})$ enhancement by the nonequilibrium condition $\Delta\mu$ (chemical potential difference) if the system is {\textit away} from the resonant point $\bar\mu=\bar\omega$. Such enhancement is realized by the thermodynamic nonequilibrium cost (nonzero entropy production rate). When we have the detuned two qubits (different frequencies), the low (high) frequency qubit should couple to high (low) temperature or chemical potential (with low temperature) bath in order to enhance the LGIs violations.

Whether we can realize the macroscopic coherence or Shr\"odinger's cat in the experiments is a very important fundamental question in quantum mechanics. The inevitable coupling with environments leads to the decoherence (classical macrorealism description). Our study suggests new ways to test the macrorealism via the LGIs. We can take the advantage from coupling with the environments by designing nonlocal excited states of the system. Furthermore, with the detuning of the two-qubit system, the nonequilibrium environments (with different temperatures or chemical potentials) can significantly enhance the degree of the violations of LGIs. In our study, we describe quite general environments. Our results are not based on the sophisticated designed interaction between the system and the environments. It is much easier to create a system coupled with the nonequilibrium environments than a completely decoupled system. It is expected that our results can be generalized into multi-qubit system coupled with multi-nonequilibrium environments.

%Our study suggests new approach for maintaining the quantumness in terms of temporal correlations of open quantum system and sheds light on quantum information processing based on open quantum system.

%The nonequilibrium condition is directly quantified by the temperature difference or chemical potential difference for Bosonic or Fermionic reservoir respectively. The dynamic of system is described by the {Bloch-Redfield} equation beyond the secular approximation. We {have obtained} the analytical form of LGI in terms of the first order coupling constant at both equilibrium and nonequilibrium conditions. 

%The maximal violation of LGI (MLGI) can be enhanced by the nonequilibrium condition. For Bosonic baths, the enhancement can be achieved at low mean temperature regime and greater enhancement of MLGI can be realized for $\Delta T>0$ if we choose qubit 1 observable. For Fermionic baths, MLGI can be enhanced by $\Delta\mu$ if the system is away from the resonant point $\bar\mu=\bar\omega$. Moreover, we find that detuned system $\omega_1\neq\omega_2$ can achieve larger MLGI by coupling with nonequilibrium environment, i.e., low frequency qubit should couple to high temperature or chemical potential (with low temperature) bath. 

\begin{acknowledgments}
	 K.Z. and J.W. appreciate the supports from Grant No. NSF-PHY 76066. W.W. is supported by the National Natural Science Foundation of China under Grant No. 11905222. We thank Professor J. Kestner, Professor V. Korepin, and X. Wang for helpful discussions.
\end{acknowledgments}

\appendix

\section{Transition Matrix Elements for Time-Evolution Equation}

\label{Appendix:matrix}

    Matrix $\mathcal M$ defined in (\ref{QME matrix}) describing the time evolution of the system based on Bloch-Redfield equation (\ref{QME}) has the matrix elements $\mathcal M_{ab,cd}$ (with row index $ab$ and column index $cd$):
	\begin{align}
	\label{M}
	\mathcal M_{11,11} =& -2\left(\cos^2\frac \theta 2\left(\alpha_1(\omega_1')+\alpha_2(\omega_2')\right)\right.\nonumber\\
	&+\left.\sin^2\frac \theta 2\left(\alpha_1(\omega_2')+\alpha_2(\omega_1')\right)\right); \\
	\mathcal M_{11,22} =& \mathcal M_{33,44} = 2\left(\cos^2\frac \theta 2\beta_1(\omega_1')+\sin^2\frac \theta 2\beta_2(\omega_1')\right); \\
	\mathcal M_{11,33} =& \mathcal M_{22,44} = 2\left(\sin^2\frac \theta 2\beta_1(\omega_2')+\cos^2\frac \theta 2\beta_2(\omega_2')\right); \\
	\mathcal M_{22,11} =& \mathcal M_{44,33} = 2\left(\cos^2\frac \theta 2\alpha_1(\omega_1')+\sin^2\frac \theta 2\alpha_2(\omega_1')\right); \\
	\mathcal M_{22,22} =& -2\left(\cos^2\frac \theta 2\left(\beta_1(\omega_1')+\alpha_2(\omega_2')\right)\right.\nonumber\\
	&\left.+\sin^2\frac \theta 2\left(\alpha_1(\omega_2')+\beta_2(\omega_1')\right)\right); \\
	\mathcal M_{33,11} =& \mathcal M_{44,22} =2\left(\sin^2\frac \theta 2\alpha_1(\omega_2')+\cos^2\frac \theta 2\alpha_2(\omega_2')\right); \\
	\mathcal M_{33,33} =& -2\left(\cos^2\frac \theta 2\left(\alpha_1(\omega_1')+\beta_2(\omega_2')\right)\right.\nonumber \\
	&\left.+\sin^2\frac \theta 2\left(\alpha_2(\omega_1')+\beta_1(\omega_2')\right)\right); \\
	\mathcal M_{44,44} =& -2\left(\cos^2\frac \theta 2\left(\beta_1(\omega_1')+\beta_2(\omega_2')\right)\right.\nonumber \\
	&\left.+\sin^2\frac \theta 2\left(\beta_1(\omega_2')+\beta_2(\omega_1')\right)\right); \\
	\mathcal M_{11,23} =& \mathcal M_{11,32} = - \mathcal M_{23,44} = - \mathcal M_{32,44} \nonumber \\
	 =& \frac 1 2 \sin\theta\left(\beta_2(\omega_1')+\beta_2(\omega_2')-\beta_1(\omega_1')-\beta_1(\omega_2')\right); \\
	\mathcal M_{22,23} =& \mathcal M_{22,32} = \mathcal M_{23,33} = \mathcal M_{32,33} \nonumber \\
	=& \frac 1 2 \sin\theta\left(\alpha_2(\omega_1')+\beta_1(\omega_2')-\alpha_1(\omega_1')-\beta_2(\omega_2')\right); \\
	\mathcal M_{33,23} =& \mathcal M_{33,32} = \mathcal M_{23,22} = \mathcal M_{32,22} \nonumber \\
	=& \frac 1 2 \sin\theta\left(\beta_1(\omega_1')+\alpha_2(\omega_2')-\alpha_1(\omega_2')-\beta_2(\omega_1')\right); \\
	\mathcal M_{44,23} =& \mathcal M_{44,32} = -\mathcal M_{23,11} = -\mathcal M_{32,11} \nonumber \\
	=& \frac 1 2 \sin\theta\left(\alpha_1(\omega_1')+\alpha_1(\omega_2')-\alpha_2(\omega_1')-\alpha_2(\omega_2')\right); \\
	\mathcal M_{23,23} =& \mathcal M^*_{23,23}=i\Omega \nonumber \\ &-(\cos^2\frac \theta 2\left(\alpha_1(\omega_1')+\alpha_2(\omega_2')+\beta_1(\omega_1')+\beta_2(\omega_2')\right)\nonumber \\
	&+\sin^2\frac \theta 2\left(\alpha_2(\omega_1')+\alpha_1(\omega_2')+\beta_2(\omega_1')+\beta_1(\omega_2')\right)) \\
	\label{M_end}\mathcal M_{11,44} =&\mathcal M_{44,11}=\mathcal M_{22,33} \nonumber \\
	&=\mathcal M_{33,22}=\mathcal M_{23,32}=\mathcal M_{32,23}=0
	\end{align}
	{Parameters $\alpha_l(\omega)$ and $\beta_l(\omega)$ with $l=1,2$ are defined in (\ref{def alpha beta}). Angle $\theta$ defined in (\ref{def theta}) tells the detuning frequency $\Delta \omega$ of the two qubits.  }
	
\section{Elements of Steady State Population Matrix}

\label{appen B}

Steady state population matrix defined in (\ref{def A}) has the matrix elements when the two baths are bosonic or fermionic (with symmetric constant coupling $J_1(\omega)=J_2(\omega)=J$)
\eq
\label{def A aa bb}
\mathcal A_{aa,bb}=\mathcal M_{aa,bb}+\frac{\kappa^\text{b,f}}{J}\mathcal M_{aa,23}\mathcal M_{23,bb}
\en
with {coefficient $\kappa^\text{b}$ for bosonic bath defined in (\ref{kappa b}) and coefficient $\kappa^\text{f}$ for fermionic bath defined in (\ref{kappa f}).}

\subsection{Elements of Steady State Population Matrix for bosonic Bath}

The explicit expressions for the matrix elements $\mathcal A$, when the two baths are bosonic, are:
\begin{align}
\label{A Bosonic}
&\mathcal A^\text{b}_{11,11}=-2\left(\tilde n_1+\tilde n_2\right)+\kappa_b\left(\Delta n_1+\Delta n_2\right)^2 \\
&\mathcal A^\text{b}_{11,22}=2\left(\tilde n_1+1\right)+\kappa_b\left(\Delta n_2^2-\Delta n_1^2\right) \\
&\mathcal A^\text{b}_{11,33}=2\left(\tilde n_2+1\right)+\kappa_b\left(\Delta n_1^2-\Delta n_2^2\right) \\
&\mathcal A^\text{b}_{11,44}=-\kappa_b\left(\Delta n_1+\Delta n_2\right)^2 \\
&\mathcal A^\text{b}_{22,11}=2\tilde n_1+\kappa_b\left(\Delta n_1^2-\Delta n_2^2\right) \\
&\mathcal A^\text{b}_{22,22}=-2\left(\tilde n_1+\tilde n_2+1\right)-\kappa_b\left(\Delta n_1-\Delta n_2\right)^2  \\
&\mathcal A^\text{b}_{22,33}=\kappa_b\left(\Delta n_1-\Delta n_2\right)^2 \\
&\mathcal A^\text{b}_{22,44}=2\left(\tilde n_2+1\right)-\kappa_b\left(\Delta n_1^2-\Delta n_2^2\right) \\
&\mathcal A^\text{b}_{33,11}=2\tilde n_2-\kappa_b\left(\Delta n_1^2-\Delta n_2^2\right) \\
&\mathcal A^\text{b}_{33,22}=\kappa_b\left(\Delta n_2-\Delta n_1\right)^2  \\
&\mathcal A^\text{b}_{33,33}=-2\left(\tilde n_1+\tilde n_2+1\right)-\kappa_b\left(\Delta n_1-\Delta n_2\right)^2 \\
&\mathcal A^\text{b}_{33,44}=2\left(\tilde n_1+1\right)+\kappa_b\left(\Delta n_1^2-\Delta n_2^2\right) \\
&\mathcal A^\text{b}_{44,11}=-\kappa_b\left(\Delta n_1+\Delta n_2\right)^2 \\
&\mathcal A^\text{b}_{44,22}=2\tilde n_2+\kappa_b\left(\Delta n_1^2-\Delta n_2^2\right) \\
&\mathcal A^\text{b}_{44,33}=2\tilde n_1-\kappa_b\left(\Delta n_1^2-\Delta n_2^2\right) \\
\label{A Bosonic_end}&\mathcal A^\text{b}_{44,44}=-2\left(\tilde n_1+\tilde n_2+2\right)+\kappa_b\left(\Delta n_1+\Delta n_2\right)^2
\end{align}
{Notations $\tilde n_l$ and $\Delta$ with $l=1,2$ are defined in (\ref{def tilde n1})-(\ref{delta n 1}).} Note that the overall constant $J$ in matrix $\mathcal A$ has omitted here.

\subsection{Matrix Elements of Steady State Population Matrix for fermionic Bath}

The explicit expressions for the matrix elements $\mathcal A$, when the two bath are fermionic, are:	
\begin{align}
\label{A Fermionic}
&\mathcal A^\text{f}_{11,11}=-2\left(\tilde n_1+\tilde n_2\right)+\kappa_f\left(\Delta n_1+\Delta n_2\right)^2 \\
&\mathcal A^\text{f}_{11,22}=2\left(1-\tilde n_1\right)-\kappa_f\left(\Delta n_1+\Delta n_2\right)^2 \\
&\mathcal A^\text{f}_{11,33}=2\left(1-\tilde n_2\right)-\kappa_f\left(\Delta n_1+\Delta n_2\right)^2 \\
&\mathcal A^\text{f}_{11,44}=-\kappa_f\left(\Delta n_1+\Delta n_2\right)^2 \nonumber \\
&\mathcal A^\text{f}_{22,11}=2\tilde n_1+\kappa_f\left(\Delta n_1+\Delta n_2\right)^2 \\
&\mathcal A^\text{f}_{22,22}=2\left(\tilde n_1-\tilde n_2-1\right)+\kappa_f\left(\Delta n_1+\Delta n_2\right)^2   \\
&\mathcal A^\text{f}_{22,33}=+\kappa_f\left(\Delta n_1+\Delta n_2\right)^2 \\
&\mathcal A^\text{f}_{22,44}=2\left(1-\tilde n_2\right)+\kappa_f\left(\Delta n_1+\Delta n_2\right)^2 \\
&\mathcal A^\text{f}_{33,11}=2\tilde n_2+\kappa_f\left(\Delta n_1+\Delta n_2\right)^2 \\
&\mathcal A^\text{f}_{33,22}=\kappa_f\left(\Delta n_1+\Delta n_2\right)^2  \nonumber \\
&\mathcal A^\text{f}_{33,33}=2\left(\tilde n_2-\tilde n_1-1\right)+\kappa_f\left(\Delta n_1+\Delta n_2\right)^2 \\
&\mathcal A^\text{f}_{33,44}=2\left(1-\tilde n_1\right)+\kappa_f\left(\Delta n_1+\Delta n_2\right)^2  \\
&\mathcal A^\text{f}_{44,11}=-\kappa_f\left(\Delta n_1+\Delta n_2\right)^2 \\
&\mathcal A^\text{f}_{44,22}=2\tilde n_2-\kappa_f\left(\Delta n_1+\Delta n_2\right)^2  \\
&\mathcal A^\text{f}_{44,33}=2\tilde n_1-\kappa_f\left(\Delta n_1+\Delta n_2\right)^2  \\
\label{A Fermionic_end}&\mathcal A^\text{f}_{44,44}=2\left(\tilde n_1+\tilde n_2-2\right)-\kappa_f\left(\Delta n_1+\Delta n_2\right)^2
\end{align}
{Notations $\tilde n_l$ and $\Delta$ with $l=1,2$ are defined in (\ref{def tilde n1})-(\ref{delta n 1}).} 

\section{Steady State Heat/Particle Current}

\label{appen C}

Given the steady state of nonequilibrium bosonic baths in (\ref{Boson rho 11})-(\ref{Boson rho 44}), the heat current $I^\text{b}_2$ defined in (\ref{def I b}) has the form
\begin{align}
    I^\text{b}_2 = & -2(\cos^2\frac \theta 2 \alpha_1(\omega_1')\omega_1'+\sin^2\frac \theta 2 \alpha_1(\omega_2')\omega_2')\rho_{11}^\text{b} \nonumber \\
    &+2(\cos^2\frac \theta 2 \beta_1(\omega_1')\omega_1'-\sin^2\frac \theta 2 \alpha_1(\omega_2')\omega_2')\rho_{22}^\text{b} \nonumber \\
    &+2(\sin^2\frac \theta 2 \beta_1(\omega_2')\omega_2'-\cos^2\frac \theta 2 \alpha_1(\omega_1')\omega_1')\rho_{33}^\text{b} \nonumber \\
    &+2(\cos^2\frac \theta 2 \beta_1(\omega_1')\omega_1'+\sin^2\frac \theta 2 \beta_1(\omega_2')\omega_2')\rho_{44}^\text{b} \nonumber \\
    &-\frac 1 2\sin\theta (\beta_1(\omega_2')+\alpha_1(\omega_2'))\omega_1'(\rho^\text{b}_{23}+\rho^\text{b}_{32})\nonumber \\
    &-\frac 1 2\sin\theta (\beta_1(\omega_1')+\alpha_1(\omega_1'))\omega_2'(\rho^\text{b}_{23}+\rho^\text{b}_{32})
\end{align}
where the parameters $\alpha_1(\omega)$ and $\beta_1(\omega)$ are defined in (\ref{def alpha beta}). Note that the occupation particle number obeys the Bose-Einstein distribution.

Given the steady state of nonequilibrium fermionic baths in (\ref{Fermi rho 11})-(\ref{Fermi rho 44}), the particle current $I^\text{f}_2$ defined in (\ref{def I f}) has the form
\begin{align}
    I^\text{f}_2 = & -2(\cos^2\frac \theta 2 \alpha_1(\omega_1')+\sin^2\frac \theta 2 \alpha_1(\omega_2'))\rho_{11}^\text{f} \nonumber \\
    &+2(\cos^2\frac \theta 2 \beta_1(\omega_1')-\sin^2\frac \theta 2 \alpha_1(\omega_2'))\rho_{22}^\text{f} \nonumber \\
    &+2(\sin^2\frac \theta 2 \beta_1(\omega_2')-\cos^2\frac \theta 2 \alpha_1(\omega_1'))\rho_{33}^\text{f} \nonumber \\
    &+2(\cos^2\frac \theta 2 \beta_1(\omega_1')+\sin^2\frac \theta 2 \beta_1(\omega_2'))\rho_{44}^\text{f} \nonumber \\
    &-\frac 1 2\sin\theta (\beta_1(\omega_2')+\alpha_1(\omega_2'))(\rho^\text{f}_{23}+\rho^\text{f}_{32})\nonumber \\
    &-\frac 1 2\sin\theta (\beta_1(\omega_1')+\alpha_1(\omega_1'))(\rho^\text{f}_{23}+\rho^\text{f}_{32})
\end{align}
where the parameters $\alpha_1(\omega)$ and $\beta_1(\omega)$ are defined in (\ref{def alpha beta}). Note that the occupation particle number obeys the Fermi-Dirac distribution.

%\bibliographystyle{apsrev4-2}
%\bibliographystyle{unsrt}
%\bibliography{LGI_bio}


\end{document}